\documentclass[rmp,twocolumn,aps]{revtex4-2}
\RequirePackage{amssymb}
\RequirePackage{amsmath}
\RequirePackage{gensymb}
\RequirePackage{graphicx}
\RequirePackage{natbib}
\RequirePackage{nicefrac}
\RequirePackage{multirow}
\RequirePackage{epstopdf}
\RequirePackage{float}
\RequirePackage{physics}
\RequirePackage{braket}
\RequirePackage[linktocpage=true, colorlinks, allcolors=blue]{hyperref}
\RequirePackage{color} 
\RequirePackage{epigraph}
\RequirePackage{breqn}  
\RequirePackage{bbold}
\RequirePackage[utf8]{inputenc}
\RequirePackage{newunicodechar} 
\RequirePackage[table]{xcolor}
\RequirePackage{setspace}
\RequirePackage{hhline} 

\def\vecsign{\mathchar"017E}
\def\dvecsign{\smash{\stackon[-1.95pt]{\vecsign}{\rotatebox{180}{$\vecsign$}}}}
\def\dvec#1{\def\useanchorwidth{T}\stackon[-4.2pt]{#1}{\,\dvecsign}}
\usepackage{stackengine}
\stackMath

\newunicodechar{ }{ }  
\newunicodechar{−}{\ensuremath{-}}        
\newunicodechar{∕}{\ensuremath{/}}        
\newunicodechar{×}{\ensuremath{\times}}   
\newunicodechar{°}{\degree}               

\newboolean{ShowComments}
\setboolean{ShowComments}{true}  

\begin{document}


\newcommand{\te}[1]{\dvec{#1}}
\ifthenelse{\boolean{ShowComments}}%
{
	\newcommand{\ColorComment}[3]{%
		{\colorbox{#1}{\color{white}   \textsf{\textbf{#2}}} \textcolor{#1}{#3}}}
	\newcommand{\nyacite}[1]{[#1]}
}%
{
	\newcommand{\ColorComment}[3]{}
	\newcommand{\nyacite}[1]{}
}%

\definecolor{outlinecolor}{rgb}{1,0,0}\newcommand{\OutlineInfo}[1]{\ColorComment{outlinecolor}{Outline}{#1}}
\definecolor{tdlcolor}{rgb}{0,0.7,0}\newcommand{\tdl}[1]{%
	\ColorComment{tdlcolor}{tdl}{\textsf{#1}}}
\definecolor{tdltextcolor}{rgb}{0,0.4,0}
\definecolor{gbtextcolor}{rgb}{1,0.6,0}
\definecolor{asptextcolor}{rgb}{0,0.5,0.5}
\newcommand{\tdlsuggests}[1]{%
	\ColorComment{tdltextcolor}{PROPOSED:}{\textsf{#1}}}
\newcommand{\gbsuggests}[1]{%
	\ColorComment{gbtextcolor}{PROPOSED:}{\textsf{#1}}}
\definecolor{jpcolor}{rgb}{0,0,1}\newcommand{\jp}[1]{\ColorComment{jpcolor}{JP}{#1}}
\definecolor{jmncolor}{rgb}{0.5,0,0.5}
\newcommand{\jmn}[1]{\ColorComment{jmncolor}{jmn}{#1}}
\newcommand{\jmnsuggests}[1]{%
	\ColorComment{jmncolor}{PROPOSED:}{\textsf{#1}}}
\definecolor{gbcolor}{rgb}{1,0.6,0}\newcommand{\gb}[1]{\ColorComment{gbcolor}{gb}{#1}}
\definecolor{aspcolor}{rgb}{0,0.5,0.5}\newcommand{\asp}[1]{%
	\ColorComment{aspcolor}{ap}{#1}}
\newcommand{\aspsuggests}[1]{%
	\ColorComment{asptextcolor}{PROPOSED:}{\textsf{#1}}}
\definecolor{oldcolor}{rgb}{0.5,0.5,0.5}\newcommand{\old}[1]{%
	\ColorComment{oldcolor}{PREVIOUSLY:}{#1}}
\definecolor{toreconcile}{rgb}{1,0,0}\newcommand{\recon}[1]{%
	\ColorComment{toreconcile}{ToReconcile}{#1}}

\newcommand{\concernaddressed}[1]{}

\newcommand{\isotope}[2]{$^{#2}$#1}
\newcommand{\Si}{\isotope{Si}{29}}
\newcommand{\Ge}{\isotope{Ge}{73}}
\newcommand{\Pspin}{\isotope{P}{31}}

\newcommand{\aver}[1]{\mbox{$\langle\!#1\!\!\rangle$}}
\renewcommand{\vec}[1]{\mathbf{#1}}

\def\ua{\uparrow}
\def\da{\downarrow}
\def\be{\begin{equation}}
\def\ee{\end{equation}}
\def\bea{\begin{eqnarray}}
\def\eea{\end{eqnarray}}
\def\bfig{\begin{figure}[htbp]}
\def\efig{\end{figure}}
\def\<{\langle}
\def\>{\rangle}
\def\up{\uparrow}
\def\down{\downarrow}
\newcommand{\ts}[2]{#1_{\text{#2}}}
\newcommand{\killblock}[1]{}  
\newcommand{\ketbraproj}[1]{\ketbra{#1}{#1}}

\newcommand{\refsec}[1]{Sec.~\ref{#1}}
\newcommand{\refeq}[1]{Eq.~\eqref{#1}}
\newcommand{\reffig}[1]{Fig.~\ref{#1}}
\newcommand{\reftable}[1]{Table~\ref{#1}}

\newcommand{\vac}{\ket{\text{vac}}}
\newcommand{\nodag}{\phantom{\dag}}

\title{Semiconductor Spin Qubits}

\newcommand{\Konstanz}{Department of Physics, University of Konstanz, D-78457 Konstanz, Germany}
\newcommand{\HRL}{HRL Laboratories LLC, 3011 Malibu Canyon Road, Malibu, California 90265, USA}
\newcommand{\Rochester}{Department of Physics and Astronomy, University of Rochester, Rochester, New York 14627, USA}
\newcommand{\Princeton}{Department of Physics, Princeton University, Princeton, New Jersey 08544, USA}

\author{Guido Burkard}\affiliation\Konstanz
\author{Thaddeus D. Ladd}\affiliation\HRL
\author{John M. Nichol}\affiliation\Rochester
\author{Andrew Pan}\affiliation\HRL
\author{Jason R. Petta}\affiliation\Princeton

\begin{abstract}
%
%
The spin degree of freedom of an electron or a nucleus is one of the most basic properties of nature and functions as an excellent qubit, as it provides a natural two-level system that is insensitive to electric fields, leading to long quantum coherence times.
This coherence survives when the spin is isolated and controlled within nanometer-scale, lithographically fabricated semiconductor devices, enabling the existing microelectronics industry to help advance spin qubits into a scalable technology.  
Driven by the burgeoning field of quantum information science, worldwide efforts have developed semiconductor spin qubits to the point where quantum state preparation, multiqubit coherent control, and single-shot quantum measurement are now routine.  
The small size, high density, long coherence times, and available industrial infrastructure of these qubits provide a highly competitive candidate for scalable solid-state quantum information processing. We review the physics of semiconductor spin qubits, focusing not only on the early achievements of spin initialization, control, and readout in GaAs quantum dots, but also on recent advances in Si and Ge spin qubits, including improved charge control and readout, coupling to other quantum degrees of freedom, and scaling to larger system sizes. We begin by introducing the four major types of spin qubits: single spin qubits, donor spin qubits, singlet-triplet spin qubits, and exchange-only spin qubits. 
We then review the mesoscopic physics of quantum dots, including single-electron charging, valleys, and spin-orbit coupling. 
We next give a comprehensive overview of the physics of exchange interactions, a crucial resource for single- and two-qubit control in spin qubits.  
The bulk of this review is centered on the presentation of results from each major spin qubit type, the present limits of fidelity, and a brief overview of alternative spin qubit platforms. 
We then give a physical description of the impact of noise on semiconductor spin qubits, aided in large part by an introduction to the filter function formalism. 
Lastly, we review recent efforts to hybridize spin qubits with superconducting systems, including charge-photon coupling, spin-photon coupling, and long-range cavity-mediated spin-spin interactions. 
Cavity-based readout approaches are also discussed. 
This review is intended to give an appreciation for the future prospects of semiconductor spin qubits, while highlighting the key advances in mesoscopic physics over the past two decades that underlie the operation of modern quantum-dot and donor spin qubits.
\end{abstract}

\maketitle
\tableofcontents

\newpage
\newpage

\section{Introduction} \label{Sec:Intro}
Quantum computers are fundamentally capable of vastly outperforming all classical computers for a growing list of problems \cite{Feynman1982,Shor1997,DiVincenzo255,Ekert1996,Childs2010,Montanaro2016,JordanZoo,Nielsen2000}.
In order to perform a quantum computation, the information to be processed must be represented in a suitable physical form
\cite{Landauer1991}. Semiconductor spin qubits are one platform that has fulfilled the main criteria for the implementation of quantum computation. 

\begin{figure*}[t]
    \centering
	\includegraphics[width=2\columnwidth]{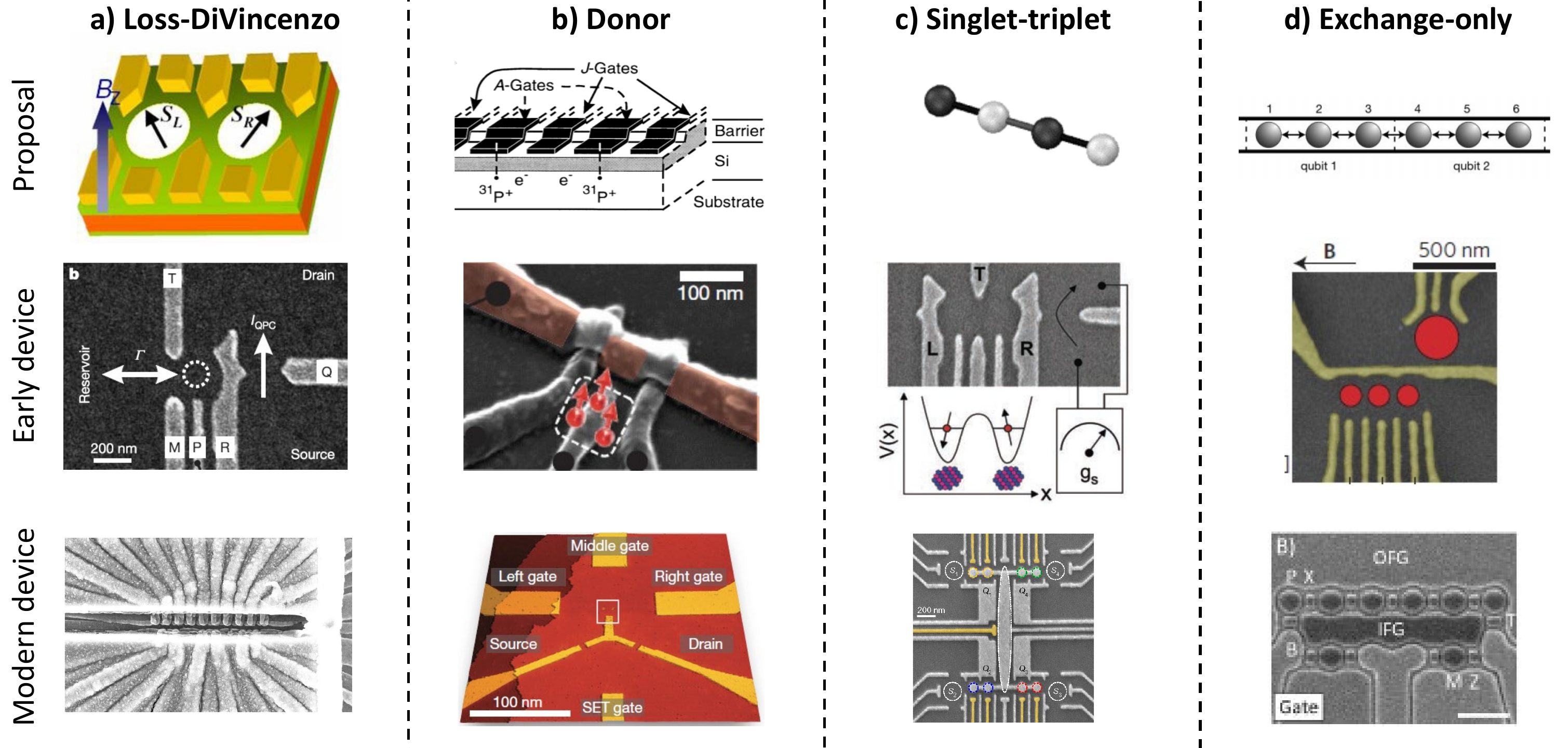}
	\caption{The four major qubit types covered in this review, with images depicting the original proposals, early devices, and modern devices. a) Loss-Divincenzo (LD) single spin qubits \cite{Loss1998},\cite{Elzerman2004},\cite{Mills2019}. (b) Donor spin qubits \cite{Kane1998},\cite{Morello2010},\cite{He_2donorexchange_2019}. (c) Singlet-triplet (ST) spin qubits \cite{Levy2002},\cite{Petta2005},\cite{Fedele_PRXQ_2021}. (d) Exhange-only (EO) spin qubits \cite{DiVincenzo2000ExchangeQC},\cite{Medford2013b},\cite{Ha_archive}.
	}
	\label{fig:intro}
\end{figure*}

The requirements for quantum computation can be stated as follows \cite{DiVincenzo1998RoySoc,DiVincenzo2000}: 1) The elementary units of information need to be stored in a scalable quantum register.  In analogy to  binary logic where bits take on the value of 0 or 1, quantum information is typically stored in the form of quantum bits (qubits).  A qubit is a quantum two-level system with orthogonal, i.e. distinguishable, basis states $\ket{0}$ and $\ket{1}$. Systems with spin-1/2 are perhaps the simplest example of this encoding, although other spin-based possibilities exist, as we will discuss.  2) A further requirement is that the qubits can be prepared in a fiducial state, e.g. $\ket{00\ldots 0}$.  3) The quantum system must remain coherent for times much longer than the duration of elementary logic gates, since decoherence causes computational errors. 4) Along with maintaining coherence, a high-fidelity gate set (single qubit and two qubit gates) must be attainable. 5) Finally, it is required that a sufficiently large part of the quantum register can be read out at the end of a computation.

The spin degree of freedom quite naturally defines a qubit, as spin-up or spin-down in the case of one electron \cite{Loss1998}, or as two distinct nuclear spin states \cite{Kane1998}. As we will show, spin qubits have satisfied the DiVincenzo criteria. Electron spins can be electrically initialized and read out with high fidelity using energy dependent tunneling or the Pauli exclusion principle \cite{Elzerman2004,Petta2005}. While coupling of the charge to electric fields allows for electrical control of spin states, the small magnetic moment of the electron spin is weakly coupled to the environment leading to long spin coherence times. Semiconductors may be ideal hosts for solid state qubits, as materials such as Si can be chemically and isotopically purified to extremely high levels. As Kane pursuasively points out \cite{Kane1998}, ``Because of the advanced state of Si materials technology and the tremendous effort currently underway in Si nanofabrication, Si is the obvious choice for the semiconductor host.'' Experiments on large spin ensembles demonstrating seconds-long electron spin coherence times and hours-long nuclear spin coherence times in isotopically enriched silicon give credence to Kane's statement \cite{Tyryshkin2012,Saeedi2015}.

Single spins have been controlled with electron spin resonance \cite{Koppens2006} and two-electron spin states with exchange coupling \cite{Petta2005}. Silicon quantum devices have achieved high fidelity single qubit \cite{Yoneda2018} and two-qubit gates \cite{Veldhorst2015,Zajac2017CNot,Watson2018}, and recent advances have pushed the fidelity beyond the thresholds required to enter a regime for fault-tolerant operation \cite{xue_2Q_2021,noiri_2Q_2021,mills_2Q_2021}.

Another motivation for harnessing the spin degree of freedom is scale. Given that a fully-error corrected quantum computer is likely to require at least one million physical qubits \cite{Fowler2012}, the small $\sim$ 100 nm intrinsic scale of quantum dots (QDs) lends itself to the creation of a dense quantum computing architecture that could be mass-produced by the semiconductor microelectronics industry \cite{Vandersypen2017}. At the same time, the small size scale of a spin qubit can lead to engineering challenges associated with addressing each qubit and achieving sufficient connectivity for quantum error correction. Indeed, many recent exciting physics results from the QD community have shown that spins can be coherently coupled to microwave photons \cite{Mi2018,Samkharadze2018,Landig2018}, providing tantalizing opportunities for long-range coupling of spin qubits and readout \cite{Borjans2020,Petersson2012,Mi2018,zheng_rapid_2019,Borjans_sensing_2021}.

The scope of this review is limited to semiconductor spin qubits in shallow donors and gate-defined QDs. Electronic and nuclear spins of point defects in wide-bandgap semiconductors such as diamond or SiC are outside the scope of this review,  and we refer the interested reader to \onlinecite{Doherty2013b,childress_hanson_2013}.  Optically addressable and self-assembled QDs have provided seminal studies toward semiconductor spin qubits, including early measures of semiconductor spin decoherence rates, but are more relevant for photonic implementations of quantum information systems that are not the focus of this review \cite{Imamoglu1999,Kroutvar2004,Bracker2005,de_greve_ultrafast_2011,Warburton2013}. Topological quantum computation,  both with anyons in quantum Hall systems \cite{DasSarma2006} and with Majorana fermions in superconductor-semiconductor hybrid systems \cite{Mourik1003,DasSarma2015} will not be covered. 

The following Sec.~\ref{Sec:Basics} will introduce the four major types of spin qubits, namely the single-spin qubit, donor spin qubit, singlet-triplet spin qubit, and exchange-only spin qubit. Figure \ref{fig:intro} gives an overview of the four qubit types, with images illustrating the theoretical proposals, early devices, and modern devices. Readers familiar with the basic spin qubit types can skip ahead to Sec.~\ref{Sec:Principles}, which covers the mesoscopic physics underpinning the operation of semiconductor spin qubits. The initiated reader may want to directly delve into the subsequent sections for selected topics. Details regarding the control of spin-spin interactions, in particular exchange, can be found in \ref{Sec:2spins}.  The implementation of quantum gates and circuits for the various spin qubit flavors is discussed in Sec.~\ref{Sec:Gates}.  Dephasing and decoherence of spin qubits due to  uncontrolled interactions with their environment is covered in Sec.~\ref{Sec:Decoherence}. Hybrid systems consisting of semiconductor spin qubits embedded into superconducting circuits can be found in Sec.~\ref{Sec:Hybrid}. We conclude by commenting on future directions for the field (Sec.~\ref{Sec:Outlook}).

\section{Basics of spin qubits} \label{Sec:Basics}
\begin{figure*}[t]
    \centering
    \includegraphics[width=2\columnwidth]{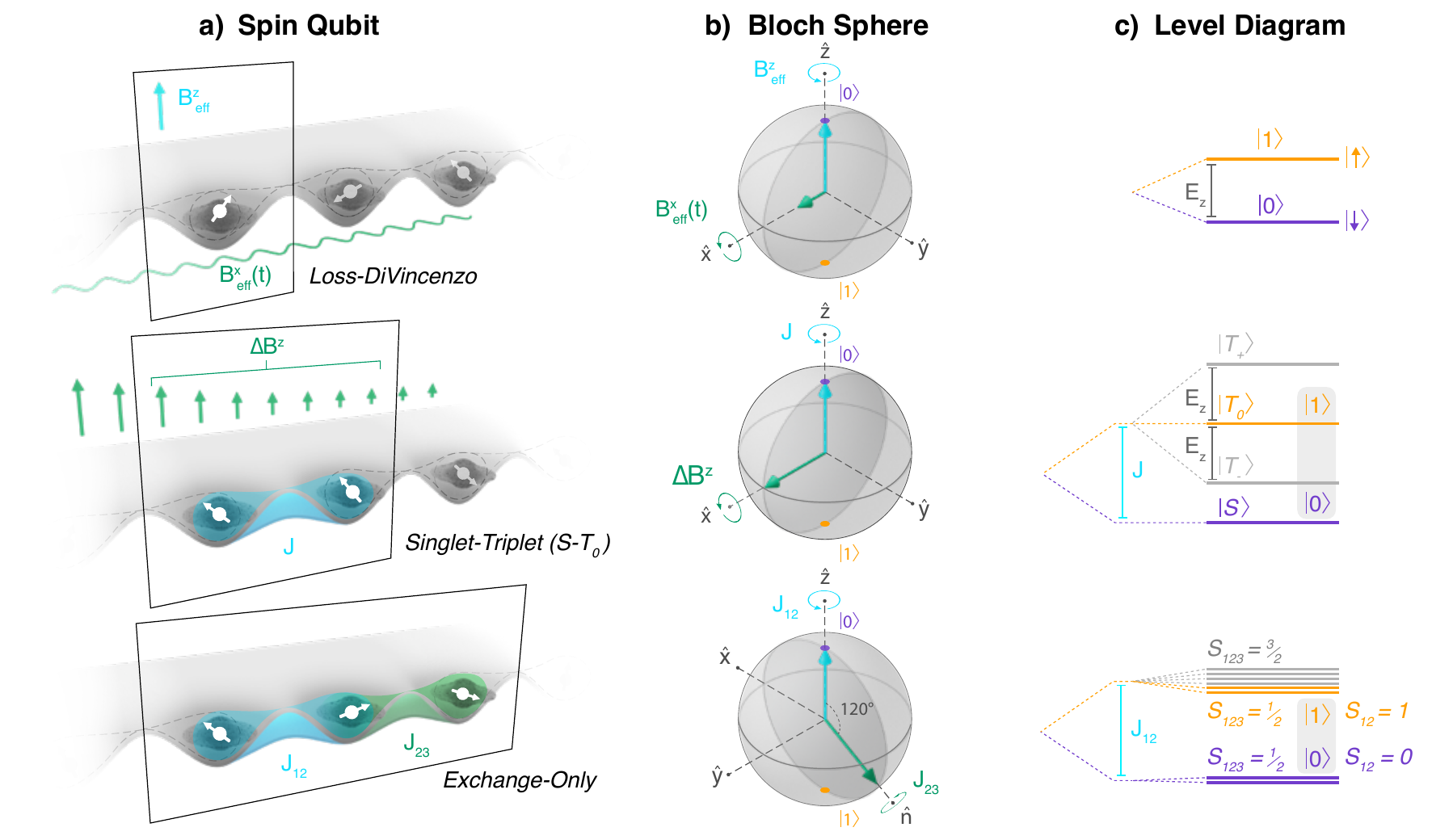}
    \caption{a) Spin configurations, b) Bloch spheres, and c) energy level diagrams associated with Loss-DiVincenzo (LD) single spin qubits, two-spin singlet-triplet (ST$_0$) qubits, and three-spin exchange-only (EO) spin qubits. Donor spin qubits also rely on single-spins, similar to the LD case.  We conventionally identify the north pole of the Bloch sphere with the qubit $|0\rangle$ state and the south pole with $|1\rangle$, irrespective of which state is lower in energy. For the LD qubit, a static magnetic field $B^z_{\rm eff}$ defines the quantization axis of the single spin, while a transverse (and smaller) ac magnetic field $B^x_{\rm eff}(t)$ drives coherent spin rotations between spin-up and spin-down.
    We identify $|0\rangle =|\downarrow\rangle$ and $|1\rangle =|\uparrow\rangle$ and note that the level ordering in c) holds for $g>0$ e.g. for Si. For the ST$_0$ qubit,
    exchange coupling $J$ and a longitudinal magnetic field gradient $\Delta B^z$ provide two orthogonal control axes.  For the EO spin qubit, nearest-neighbor exchange couplings $J_{12}$ and $J_{23}$ provide two control axes that are separated by 120\degree\ on the Bloch sphere.}
    \label{fig:blochspheres}
 \end{figure*}

In this Section, we introduce the various kinds of spin qubits. At the most basic level, we can classify spin qubit types based on the number of spins used to encode the qubit. Figure \ref{fig:blochspheres} shows the Bloch spheres and control axes for single spin qubits, two-spin singlet-triplet qubits, and three-spin exchange only qubits. For example, the single spin Loss-DiVincenzo qubit encodes quantum information in the spin state of a single electron. A static magnetic field lifts the degeneracy between the spin-up and spin-down states of the electron, while a transverse ac magnetic field drives coherent rotations between spin-up and spin-down \cite{Loss1998}. 

At a more detailed level (see Table \ref{table:layouts}), the different types of spin qubits are distinguished by how they encode spins into qubits; by the number and species of particle that carries the spin (atomic nucleus, electron, hole); by their placement in a single-site or multi-site arrangement, where a site can be a QD or a donor atom; and by their initialization, measurement, and control methods, all of which we elaborate on in this section.

\begin{table*}[t]
	\centering
	\caption{Spin qubit configurations grouped by the number of spin-1/2 particles per qubit (rows) and number of sites -- usually QDs (columns).  Spins are indicated by grey dots (electrons/holes) or white dots (nuclei), numbered to adhere to the basis description of Table~\ref{table:basis}. Sites are indicated by pink circles; their overlap indicates ``always-on exchange," meaning that the spins contained are somewhat delocalized across the site even for the idle qubit.}
	\includegraphics[width=2\columnwidth]{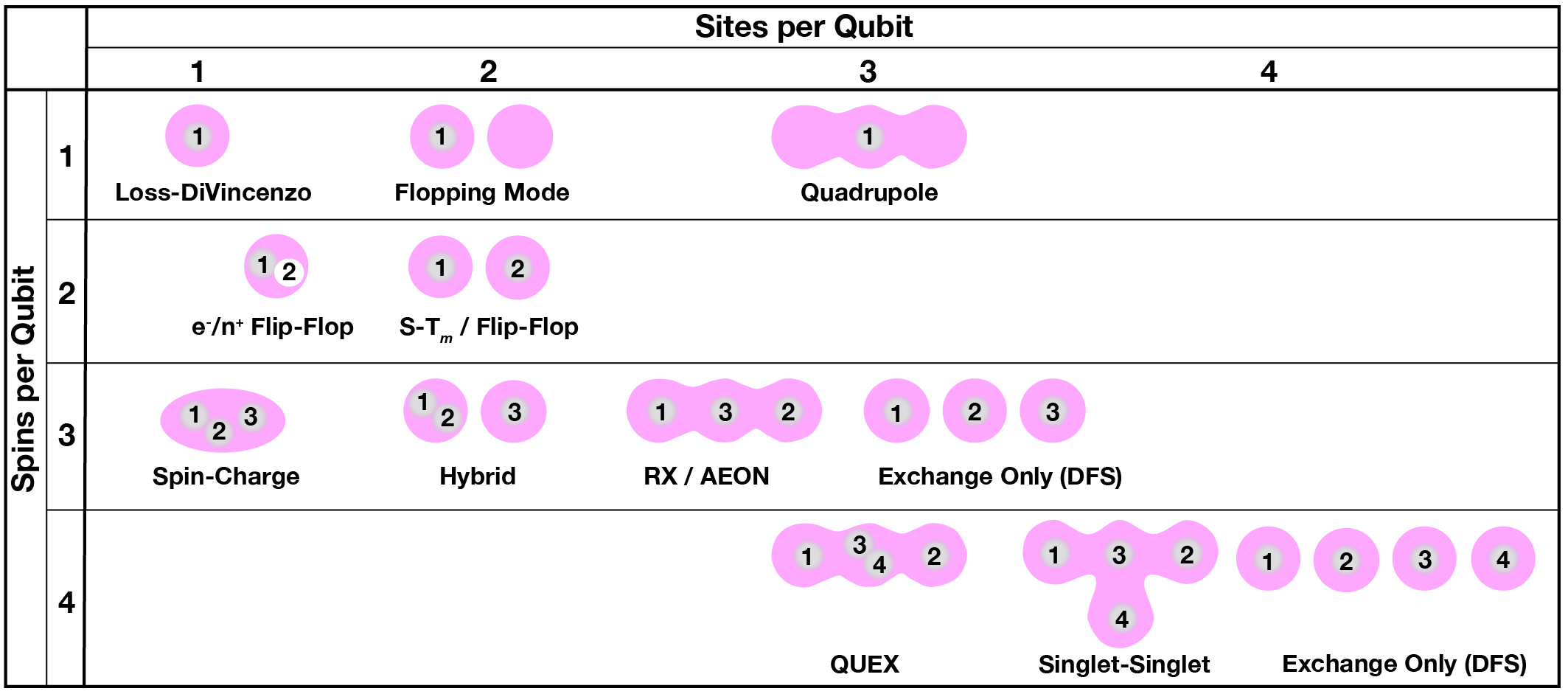}
	\label{table:layouts}
\end{table*}

Common to all semiconductor spin qubits is the confinement of spin to isolated sites.  In semiconductors, in contrast to metals, the density of conduction electrons can be depleted to be arbitrarily low.  The density may in fact be engineered, starting from zero in an intrinsic semiconductor at low temperature.  This allows for the restriction of electron motion to two dimensions (2D) in quantum wells (QWs) or at interfaces between two materials \cite{ando_electronic_1982}, and further to one or even zero dimensions (1D or 0D) with electrostatic tailoring of the potential landscape \cite{Wiel2002,Kouwenhoven2001}. Confinement in all spatial dimensions is achieved in QDs which localize electrons and act as artificial atoms \cite{KastnerRMP1992}.  A collection of electrons, each of which is confined to one such QD, provides a nearly ideal arena for the realization of spin-based quantum information processing~\cite{Loss1998}.

Another commonality to all flavors of semiconductor spin qubits is some use of the exchange interaction.  The physics of exchange will be discussed in more detail in Secs. \ref{Subsec:Principles:Coulomb} and \ref{Sec:2spins}, but the basic principle is that when the wavefunctions of two electrons in two distinguishable locations overlap, the energy of the spin-singlet state is lowered relative to the three spin triplet states by an amount called the exchange coupling $J$.  This effect (sometimes referred to as pseudo-exchange or kinetic exchange) occurs due to the ability of electrons in the spin-asymmetric singlet state to move to and from the same location (while maintaining a totally antisymmetric wavefunction, as per the Pauli exclusion principle), whilst such motion is forbidden for the symmetric triplet states.  The lowered energy of the singlet relative to the triplets for spins $i$ and $j$ is captured by the Heisenberg exchange Hamiltonian $H=J_{ij}\vec{S}_i\cdot\vec{S}_j$, where $\mathbf{S}_i$ denotes the quantum operator for the spin of the electron residing in the $i$-th site.  From a quantum control perspective, an appeal of spin qubits is that $J_{ij}$ can typically be tuned over many orders of magnitude by adjusting gate voltages \cite{Petta2005}.  Depending on the type of spin qubit, the exchange interaction may be used for both single \cite{Levy2002,Petta2005,Eng2015} and two-qubit gates \cite{Nowack2011,Veldhorst2015,Watson2018,Zajac2017CNot}.

\subsection{Loss-DiVincenzo (LD) spin qubit}  \label{Subsec:Basics:LDintro}
The spin-1/2 of an electron represents a natural realization of a qubit.  The encoding for a single electron spin `Loss-DiVincenzo' qubit is a direct mapping $\mathbf{S}_i={\boldsymbol\sigma}_i/2$ between spin operators and encoded Pauli operators. In the limit of tight electronic confinement, with one electron per dot, the electron spin dynamics are governed by the Heisenberg exchange Hamiltonian (as discussed above) and the single-electron Zeeman Hamiltonian, leading to a total Hamiltonian of the form:
\begin{equation}
\label{eq:Heisenberg}
    H(t)=\frac{1}{4}\sum_{\langle i,j\rangle} J_{ij}(t){\boldsymbol\sigma}_i\cdot{\boldsymbol\sigma}_j 
    +\frac{1}{2}\sum_i g_i \mu_B \mathbf{B}_i\cdot {\boldsymbol\sigma}_i,
\end{equation}
where $\mathbf{B}_i$ and $g_i$ are the (effective) magnetic field and g-factor at site $i$.

The Loss-DiVincenzo qubit requires a method of initialization and measurement of single electron spin states.  The original proposal~\cite{Loss1998} suggested spin-selective ferromagnetic elements in the device, however actual practice has employed spin-selective tunneling to a Fermionic bath of electrons \cite{Elzerman2004}, in which a large static magnetic field $B \gg k_BT_e/g\mu_B$ enables tunneling of the higher energy QD spin-state to the Fermi sea, while tunneling from the lower energy spin state is energetically forbidden. Here $k_B$ is Boltzmann's constant and $T_e$ is the electron temperature. The presence or absence of a tunneling event, as measured using sensitive charge detectors (see \refsec{Subsec:Principles:Dots_and_sensors}), is then used to infer the orientation of the electron spin. This spin readout protocol is commonly refered to as `Elzerman readout' and it requires relatively large magnetic fields, which in turn sets the Larmor frequency for spins in the tens of GHz range.  

For this qubit type, the single-spin $B$-dependent (Zeeman) terms provide single-qubit control. Time-dependent control of $\mathbf{B}_i$ or $g_i$ is required for the implementation of single-qubit gates; this has been realized using a combination of static and oscillatory magnetic fields within the framework of electron spin resonance (ESR) \cite{Koppens2006,Veldhorst2015,Pla2012}, or using oscillatory electric fields in combination of spin-orbit coupling \cite{Nowack2007,Nadj2010} or magnetic field gradients \cite{Brunner2011,Zajac2017CNot,Yoneda2018} by applying electric dipole spin resonance (EDSR).

The exchange coupling, which can be adjusted with gate voltages \cite{Petta2005}, allows for time-dependent two-qubit control and hence the realization of entangling two-qubit gates between nearest-neighbor spins \cite{Nowack2011}. Recent implementations of Loss-DiVincenzo qubits use static field gradients for $\vec{B}$, pulsed or ac-driven exchange for $J_{ij}(t)$, and oscillatory electric fields \cite{Zajac2017CNot,Watson2018} to achieve full control of a two-qubit system.

\subsection{Donor spin qubits and Kane's proposal}

Shortly after the publication of the Loss-DiVincenzo proposal on quantum computation with QDs, Bruce Kane published a proposal to use the nuclear spins of \Pspin\ donor atoms in silicon to construct a quantum computer \cite{Kane1998}. Nuclear spins are highly coherent since the nuclear gyromagnetic ratio, $\gamma_n/2\pi$~=~17.2~MHz/T for \Pspin, is nearly 2,000 times smaller than the electron gyromagnetic ratio $\gamma_e/2\pi$~$\approx$~28~GHz/T, and their lack of mobility in a solid-state host inhibits charge-hybridizing or spin-orbit-related decoherence mechanisms (which are discussed in detail in section \ref{Sec:Decoherence}).  

Kane proposed using the $I$ = 1/2 nuclear spin of a \Pspin\ donor in Si as a quantum bit. \Pspin\ is a shallow donor in Si with a 45 meV ionization energy \cite{Feher1959,Wilson1961}. The donor electron has a hydrogenic s-like ground state with an effective Bohr radius of 1.8 nm \cite{SmithSciRep2017}. To maintain a high degree of nuclear spin coherence, the donor nuclear spins would ideally be embedded in a host material composed of $I$ = 0 isotopes as background nuclear spins can lead to decoherence. Despite their small effective mass and widespread use in mesoscopic physics, common III-V semiconductors such as GaAs and InAs only have stable isotopes with $I \neq 0$. In contrast, Si is primarily composed of $I$ = 0 nuclear spin isotopes $^{28}$Si and $^{30}$Si. The remaining 5\% of $I$~=~1/2~$^{29}$Si can be removed through isotopic enrichment. 

Gate voltage control of the donor-bound electronic wavefunction is a crucially important aspect of the Kane quantum computer. Kane proposed using an array of $^{31}$P donor atoms placed $\approx$ 200 \AA~beneath the Si surface as the register of qubits. By adjusting the voltage $V_g$ on an $A$-gate placed above each donor, the donor electron can be pulled away from the donor towards the Si/SiO$_2$ interface to reduce the hyperfine interaction $A(V_g)$ and control the nuclear spin resonance frequency. Nuclear spin exchange is mediated by electrons achieved using gates called $J$-gates, which are located between adjacent donor sites. The $J$-gate voltage influences the overlap between adjacent donor electron wavefunctions, and through the hyperfine interaction, the nuclear spin exchange coupling. Measurements of the nuclear spin state are performed by again leveraging the tunability of the electronic wavefunction using gates. Nuclear spin initialization can be achieved using the same steps for nuclear spin state readout, with an additional radio-frequency driven rotation to the desired starting spin state if required. 

Since Kane's proposal, many elements of this qubit type have been demonstrated, and in so doing many critical variations on the donor-qubit concept have emerged.  \Pspin\ nuclei have been placed in isotopically enhanced silicon substrates using both masked ion-implantation methods \cite{Morello2010} and scanning tunneling microscopy \cite{Fuechsle2012Transistor}. Control of the exchange interaction between \Pspin\ donor-bound electrons has been demonstrated using both fabrication methods \cite{He_2donorexchange_2019,madzik_controllable_2020}.  The initialization and readout of a single \Pspin\ nuclear spin has been performed with over 99\% fidelity \cite{Pla2013}, the $A$-gate-modulated hyperfine interaction has been used as envisioned by Kane to tune electron and nuclear Larmor resonances \cite{madzik_controllable_2020},  and multi-qubit electron and nuclear processes have been characterized with gate-set-tomography for total single and two-qubit gate fidelities  exceeding 99\% \cite{Nielsen2021,madzik_arxiv}. A key challenge of the Kane proposal is that the required exchange interaction is highly sensitive to the \Pspin\ donor placement \cite{Koiller2001Exchange}, requiring either impeccable fabrication tolerance or more tolerant forms of two-qubit gates, several of which have been proposed \cite{tosi_silicon_2017,Broome2018Correlations}.

\begin{table*}[t]

\setlength{\tabcolsep}{1pt}
\setlength{\arrayrulewidth}{1pt}

\renewcommand{\arraystretch}{1.0}
\newcommand{\atb}[1]{\mbox{$\begin{aligned}#1\end{aligned}$}}
\newcommand{\qkets}{\atb{&\ket{0}\\
				&\ket{1}}}
\newcommand{\lb}[1]{\parbox{0.85in}{\centering #1}} 

\newcommand{\tblue}{ebf1f2}
\newcommand{\tgreen}{f1f7f0}
\newcommand{\tpink}{ededdd}
\newcommand{\twhite}{FFFFFF}

\newcolumntype{N}{@{}m{0pt}@{}}

	\caption{Spin qubit encodings: The first column $N$ is the number of spin-1/2 particles per qubit, followed by a named ``Type" of qubit discussed in this review.  The two qubit states $\ket{0}$ and $\ket{1}$ are then specified in terms of both conserved and qubit-dependent ``q-number" describing the total angular momentum; here $m$ always refers to the \emph{total} spin projection, whereas $S_{jk\cdots}$ refers to the combined total spin angular momentum of spins $j,k,\ldots$.  Clebsch-Gordan coefficients translate these spin angular momentum combinations into ``States."  For the three-spin case, $m$ may take either value $\pm 1/2$ in the encoded subspace.  The final column shows the encoded Pauli operators $\boldsymbol\sigma$ of the \emph{qubit} in terms of the spin operators $\vec{S}_j$ of each spin-1/2 particle $j$.  The qubit states are the $\pm 1$ eigenstates of $\sigma^z$; degeneracies in these eigenstates indicate gauge freedom, and the null space of these operators are leakage states.  For the LD qubit, the constant relating the logical qubit to the spin changes with the $g$-factor; the minus value shown here is consistent with the Si $g>0$ choice used in \reffig{fig:blochspheres}. \label{table:basis}}
\begin{center}
  
  \begin{tabular}{
>{\arrayrulecolor[HTML]{\twhite}}
>{\columncolor[HTML]{\tblue}\centering\arraybackslash}p{5mm}
>{\columncolor[HTML]{\tblue}\centering\arraybackslash}p{20mm}
>{\columncolor[HTML]{\tgreen}\centering\arraybackslash}p{5mm}
>{\columncolor[HTML]{\tgreen}\centering\arraybackslash}p{24mm}
>{\columncolor[HTML]{\tgreen}\centering\arraybackslash}p{37mm}
>{\columncolor[HTML]{\tpink}\centering\arraybackslash}p{84mm}N}

\rule{0pt}{3Ex}\textbf{\textit{N}} & 
\textbf{Type} &   & 
\textbf{q-numbers}  & 
\textbf{States} & 
\textbf{Encoded Qubit Pauli Operators} &\\[2pt] \hline\hline
  \textbf{1} 
& \lb{Loss-DiVincenzo}  & \qkets &  \atb{m &= -1/2\\
				m &= +1/2} & \rule{0ex}{4ex}\atb{&\ket\downarrow \\
                               &\ket\uparrow}
            & 
			\rule{1.5ex}{0ex}\atb{\boldsymbol{\sigma}&=-2\vec{S}_1}
                             &\\  \hline\hline
%
&
			\lb{Singlet-Triplet \\ (ST$_0$)} 
			& \qkets & \atb{S_{12}&=m=0 \\
				S_{12}&=1, m=0} & 
			\atb{{ 
					\ket{\text{S}}}   & 
				= (\ket{\up\down}\!-\!\ket{\down\up})/\sqrt{2} \\
				{\ket{\text{T}_0}} & 
				= (\ket{\up\down}\!+\!\ket{\down\up})/\sqrt{2} }
			&                  
			\rule{0in}{5.8Ex}
			\atb{\sigma^x &= S_1^z-S_2^z \\
				\sigma^y &= 2\hat{z}\cdot \vec{S}_2\times\vec{S}_1 \\
				\sigma^z &= 2(S_1^zS_2^z-\vec{S}_1\cdot\vec{S}_2)}          	
    \\ \hline
	\multirow{1}{*}{\textbf{2}}			
& \lb{Flip-Flop} & \qkets & $m=0$ & \atb{&\ket{\up\down}\\&\ket{\down\up}} & 
			\rule{0in}{6.3Ex}
			\atb{\sigma^x &= 2(\vec{S}_1\cdot\vec{S}_2)-S_1^zS_2^z) \\
				\sigma^y &=  2\hat{z}\cdot \vec{S}_2\times\vec{S}_1 \\
				\sigma^z &= S_1^z-S_2^z}
			\\ \hline
			& 	\lb{Singlet-Triplet \\ (ST$_+$)} 
			\rule{0Ex}{0.1Ex}
			& \qkets & \atb{&S_{12}=m=0\\
				&S_{12}=m=1} & \atb{&\ket{\text{S}}\\
				\ket{\text{T}_+}&=\ket{\up\up}} & 
			\rule{0in}{5.5Ex}
			\atb{\sigma^x&=\phantom{-}(S_2^x-S_1^x)/\sqrt{2}+\sqrt{2}(S_1^zS_2^x-S_1^xS_2^z)\\
			     \sigma^y&=\phantom{-}(S_1^y-S_2^y)/\sqrt{2}+\sqrt{2}(S_1^yS_2^z-S_1^zS_2^y)\\
				\sigma^z &= -(S_1^z+S_2^z)/2-\vec{S}_1\cdot\vec{S}_2-S_1^zS_2^z}
  \\[5pt] \hline\hline
 %
			\textbf{3} &
			\lb{Exchange-Only (DF Subsystem), RX, AEON, Hybrid} & \qkets & 
			\atb{\textcolor{gray}{S_{123}} &\textcolor{gray}{= 1/2}\\
				S_{12}&=0 \\
				S_{12}&=1 \\
				\phantom{+}&} &
			\atb{ &\phantom{+}\\\rule{0Ex}{3Ex}
				&\ket{\text{S}}\!\ket{m} \\
				(\sqrt{2}&\ket{\text{T}_{2m}}\!\ket{-m}\\
				-&\ket{\text{T}_0}\!\ket{m})/\sqrt{3}} &
			\rule{0Ex}{8.5Ex}
			\atb{
				\sigma^x &= 2(\vec{S}_2-\vec{S}_1)\cdot\vec{S}_3 / \sqrt{3}\\
				\sigma^y &= 4(\vec{S}_1\times\vec{S}_2\cdot\vec{S}_3)/\sqrt{3}\\
				\sigma^z &= 2[(\vec{S}_1+\vec{S}_2)\cdot\vec{S}_3-2\vec{S}_1\cdot\vec{S}_2]/3}
			\\[5pt] \hline\hline
		\textbf{4} &
			\lb{Exchange-Only (DF Subspace),\\ QUEX, \\ Singlet-Singlet} \rule{0Ex}{0.1Ex}& \qkets & 
			\atb{\textcolor{gray}{S_{1234} = m} &\textcolor{gray}{=0} \\ 
				S_{12}=S_{34}&=0 \\
				S_{12}=S_{34}&=1 \\
				\phantom{+}&} &
			\atb{
			&\phantom{+}\\
        \ket{\text{S}}\!&\ket{\text{S}}\\
    	(\ket{\text{T}_+}\!\ket{\text{T}_-}
				&+\ket{\text{T}_-}\!\ket{\text{T}_+}\\
        	-\ket{\text{T}_0}\!&\ket{\text{T}_0})/{\sqrt{3}}} &
			\atb{
				\sigma^x \! & =\! 2[\vec{S}_1\times\vec{S}_2\cdot
				                    \vec{S}_3\times\vec{S}_4
                     +(\vec{S}_2-\vec{S}_1)\cdot(\vec{S}_3-\vec{S}_4)/4]/\sqrt{3}
				\\
				\sigma^y \! &= \! [\vec{S}_1\times\vec{S}_2\cdot
				(\vec{S}_3-\vec{S}_4)	+\vec{S}_3\times\vec{S}_4\cdot
				(\vec{S}_1-\vec{S}_2)]/\sqrt{3}
				\\
				\sigma^z \! &= \! 2(
				\vec{S}_1\times\vec{S}_4\cdot\vec{S}_2\times\vec{S}_3
				+\vec{S}_1\times\vec{S}_3\cdot\vec{S}_2\times\vec{S}_4)/3
				\\			 &+[					
				(\vec{S}_1-\vec{S}_4)\cdot(\vec{S}_3-\vec{S}_2)
				+(\vec{S}_1-\vec{S}_3)\cdot(\vec{S}_4-\vec{S}_2)]/6}
     \rule{0Ex}{1Ex}
    \\
  \end{tabular}
  \end{center}
\end{table*}

\subsection{Singlet-triplet (ST$_0$ and ST$_\pm$) qubits} \label{Subsec:Basics:STintro}
Both the Loss-DiVincenzo~\cite{Loss1998} and Kane~\cite{Kane1998} proposals for quantum computing involve single-spin qubits manipulated with a combination of static and oscillating electric and magnetic fields. The oscillating fields can be difficult to localize in nanoscale devices, and the power dissipated by those fields can be problematic at cryogenic temperatures. In addition, the primary source of dephasing for single-spin qubits is the magnetic noise associated with the semiconductor environment, which can be large in materials such as GaAs, which have spinful nuclei (see \refsec{Sec:Decoherence}). In part to overcome these control and dephasing challenges, spin qubits can be realized through different sets of multi-spin states associated with groups of electrons (\reftable{table:layouts}). 

Conceptually, the simplest extension of the single-spin qubit is a qubit formed from two electrons in a double quantum dot (DQD), utilizing the controlled singlet-triplet splitting offered by the exchange interaction to define the singlet-triplet (ST$_0$) qubit~\cite{Levy2002,Petta2005}.  The $\ket{{\rm S}}$ and $\ket{{\rm T}_0}$ states are defined in \reftable{table:basis}.  Along with the basis states, the encoded qubit Pauli operators $\sigma^x, \sigma^y$ and $\sigma^z$ are defined such that the $\pm 1$ eigenstates of $\sigma^z$ are the encoded states and the $0$-eigenstates are leakage states (polarized triplet states $T_\pm$ in this case). Additionally, all of the encoded Pauli-operators have the correct commutation relations.  

To understand how physical interactions map to encoded qubit operations, any spin operator $X$ can be decomposed into encoded Pauli operators as $X=\sum_j c^j \sigma^j$ with $c^j=\Tr{X\sigma^j}/2$. Thus in the encoded ST$_0$ qubit subspace, and ignoring an overall phase factor, the ST$_0$ qubit Hamiltonian in the presence of exchange and magnetic field gradients is 
\begin{equation}
H_{{\rm ST}_{0}}=J_{12} \frac{\sigma^z}{2} + \mu_B \Delta(g^* B^z) \frac{\sigma^x}{2}.
\label{eq:singlet-to}
\end{equation}
Here, the exchange coupling $J_{ij}$ can be experimentally controlled by adjusting QD gate voltages \cite{Petta2005} and $\Delta (g^* B^z)$ is the effective difference in magnetic field between the two dots along an applied global field direction ($z$-direction).

The ST$_0$ qubit exists in a decoherence-free subspace (DFS) with respect to global magnetic fields that couple to the spin of the electron since $m$=0 for both $\ket{{\rm S}}$ and $\ket{{\rm T}_0}$
~\cite{Lidar1998}.  As a result of the tunable exchange coupling $J$, ST$_0$ qubits feature  full electrical control with baseband voltage pulses \cite{Petta2005}. Although the ST$_0$ qubit is insensitive to global magnetic fields, it remains sensitive to local magnetic-field fluctuations as a result of the $\Delta B^z$ term in the Hamiltonian. The $\sigma^x$ term may result from quasi-static hyperfine fields~\cite{Taylor2007,Petta2008}, $g$-factor variations \cite{Jock2018,Liu2021}, or micromagnet field gradients.

Pauli spin blockade, a manifestation of exchange coupling  (\refsec{Subsec:Gates:ST}), enables straightforward, rapid, and high-fidelity measurement of joint spin states. A spin blockade measurement converts singlets and triplets to different spatial configurations of the two electrons in the DQD, which can easily be distinguished with a nearby charge sensor \cite{Petta2005,Barthel2010,Borjans_sensing_2021}. 

Since the initial demonstration~\cite{Petta2005}, ST$_0$ qubits and variants thereof have been the focus of intense research. Single-qubit gates have been studied in GaAs QDs~\cite{Bluhm2010a,Shulman2014} and in Si QDs~\cite{Jock2018,Maune2012,Wu2014,Fogarty2018}. Capacitive coupling of ST$_0$ qubits can yield an entangling operation \cite{Taylor2007,Shulman2012,Nichol2017}. Early results on ST$_0$ qubits coupled via a superconducting resonator or exchange coupling are also encouraging \cite{Bottcher_arxiv}.

In the presence of a global magnetic field, the Zeeman energy can compensate for exchange, and the polarized triplet ($\ket{{\rm T}_+}$ in GaAs or $\ket{{\rm T}_-}$ in Si) can become degenerate with the singlet state.  This degeneracy can be lifted via transverse magnetic field gradients~\cite{Taylor2007}, spin-orbit coupling, or spin-valley coupling, and an effective ST$_+$ qubit can be formed in GaAs (or ST$_-$ qubit in Si; we will loosely refer to both types as ST$_\pm$ qubits, with the understanding that the relevant triplet state is dependent on the sign of the $g$-factor). In the $\{\ket{{\rm S}},\ket{{\rm T}_+}\}$ basis the encoded Hamiltonian is $H_{{\rm ST}_{+}}=E_{{\rm ST}_{+}} \sigma^z/2+\Delta_{{\rm ST}}\sigma^x/2$, where the electrically-tunable qubit splitting $E_{{\rm ST}_{+}}$ = $E_Z-J$, for average Zeeman energy $E_Z$. The size of the coupling $\Delta_{{\rm ST}}$ depends on multiple factors, including transverse nuclear fields \cite{Petta2010} and spin-orbit coupling \cite{Nichol2015}. Various methods relating to Landau-Zener-St\"uckelberg interferometry enable full control over ST$_+$  qubits \cite{Petta2010,Gaudreau2012}. To date, ac-driven ST$_\pm$ Rabi oscillations have not been observed. Two-qubit gates based on capacitive coupling have been proposed~\cite{Ribeiro2010Harnessing}.

A qubit related to the ST$_0$ qubit is the flip-flop qubit for two spins, which take $\ket{\up\down}$ and $\ket{\down\up}$ as eigenstates. As can be seen by the spin-operators defining the encoded $\sigma^j$ operators in \reftable{table:basis}, this qubit is effectively a rotation of the ST$_0$ qubit about the $y$-axis of the Bloch sphere. The rotated ST$_0$ Bloch sphere provides a more natural description when the effective magnetic field gradient between the two spins is the dominant term in the Hamiltonian. The large field gradient regime is more commonly encountered with LD qubits in the presence of a micromagnet field gradient \cite{Watson2018,Zajac2017CNot} or with electron spins bound to spin-carrying donor nuclei \cite{tosi_silicon_2017}.

\subsection{Exchange-only (EO) and resonant-exchange (RX) qubits} \label{Subsec:Basics:EOintro}

Quantum computing using LD qubits requires two different types of interactions described by Eq.~\eqref{eq:Heisenberg}: (i) an entangling spin-spin coupling, typically the exchange interaction, that can be used to realize two-qubit gates \cite{Petta2005}, and (ii) an effective local magnetic field that splits the qubit spin-up and spin-down states in a chosen basis and thus enables the execution of single-qubit gates~\cite{Koppens2006}. In a circuit-based model with separable initial states, single-qubit gates alone are not sufficient for universal quantum computation. However, universal quantum computation is possible with the exchange interaction alone if employing qubits defined by an encoded subspace with constant total spin \cite{Bacon2000,DiVincenzo2000ExchangeQC,Kempe2001}. 

In Sec.~\ref{Subsec:Basics:STintro} we indicated that singlet-triplet qubits consisting of two spin-1/2 particles require effective magnetic field gradients to realize two-axis qubit operations; however the Hilbert space of three or more spin-1/2 particles contains subspaces of dimension two or greater with identical spin quantum numbers, on which exchange may allow universal control. Mathematically, the Hilbert space of two spins may be combined by angular momentum rules as $\mathcal{H}_{1/2}\otimes\mathcal{H}_{1/2}=\mathcal{H}_{0}\oplus\mathcal{H}_{1}$
where $\mathcal{H}_{S}$ denotes the $2S+1$ dimensional representation space of  the rotation group for a spin-$S$ system.  Since exchange conserves total spin $S$ as well as all spin projections $m$, exchange can at most provide a phase difference between the $S=0$ (singlet) and $S=1$ (triplet) representations. In the case of three spin-1/2 particles, angular momentum rules decompose the total spin Hilbert space into a direct sum of two total-spin-1/2 subspaces and one total spin-3/2, i.e. $\mathcal{H}_{1/2}\otimes\mathcal{H}_{1/2}\otimes\mathcal{H}_{1/2}=\mathcal{H}_{1/2}\oplus\mathcal{H}_{1/2}\oplus\mathcal{H}_{3/2}$.   Exchange gives full control within the two copies of spin-1/2 subspaces, which provide the qubit.

The 2$S$+1 states in subspace $\mathcal{H}_S$ with total spin $S$ are characterized by the angular momentum projection, or $m$, quantum number.  Since exchange conserves $m$, this degree of freedom is not accessed by exchange-only control.  For ST$_0$ and for four-spin qubits in the $S=0$ subspace, $m=0$.  However, for the three spin-case there are two copies of the $S=1/2$ qubit corresponding to $m=\pm 1/2$.  Any exchange operation within a single, three-spin qubit behaves the same regardless of $m$. One possibility is to operate the three-spin qubit at high magnetic fields, where the $m$ value of the polarized ground state provides the ``decoherence-free subspace" qubit.  However, since the two subspaces perform equivalently, the second possibility is to leave $m$ unpolarized, and ignore this degree of freedom; doing so results in a ``decoherence-free subsystem" qubit.  This is straight-forward for single-qubit gates, but puts additional constraints on exchange-based two-qubit gates \cite{DiVincenzo2000ExchangeQC}. The states of the $S=1/2$ decoherence free subsystem qubit are shown in \reftable{table:basis}, for arbitrary $m$.  We note in this table that for 3-spin and 4-spin DFS qubits, unlike the single-spin or two-spin cases,  the decomposition of encoded Pauli-operators into spin operators feature no notion of direction; the qubit is controlled via the controlled fractional permutations of spins, rather than physical rotations about any preferred axis.

From \reftable{table:basis}, we see that for three spins in the $S_{123}=1/2$ subsystem, exchange coupling between spins 1 and 2, as for singlet-triplet qubits, appear as a $\sigma^z$.  Exchange coupling between spins 2 and 3 has weight both as $\sigma^x$ and $\sigma^z$, combining to the $\hat{n}$ axis shown in \reffig{fig:blochspheres}.  Composite gates enabling arbitrary single-qubit operations may be composed of combinations of these exchange operations.

Quantum gate operation for the EO qubit proceeds by sequentially pulsing on and off the exchange coupling $J_{ij}(t)$ for pairs of spins $i$ and $j$ (parallel pulsing of disjoint pairs is possible) while the magnetic field is held at a constant value identical for all qubits (e.g. zero field for all qubits). Since the pulse duration is chosen sufficiently long such that the pulse bandwidth in frequency space is smaller than $J_{ij}$, this type of operation is referred to as dc operation.  In the idle state without quantum gates being executed, the exchange coupling is set to zero everywhere ($J_{ij}=0$); all qubit states are degenerate, and ideally there is no phase evolution between superposed states in the laboratory frame. As we will discuss further in \refsec{Subsec:Gates:TQD}, exchange between pairs of qubits from distinct qubits allows for the implementation of a universal two-qubit gate \cite{DiVincenzo2000ExchangeQC}.

An alternative mode of operation for EO qubits is termed the resonant-exchange (RX) qubit. The RX qubit differs from the dc-mode EO qubit in that the nearest-neighbor exchange couplings are constantly set to the same non-zero value $J=J_{12}=J_{23}$, opening an energy gap between the qubit states $|0\rangle$ and $|1\rangle$.  Single-qubit gates can then be executed with ac exchange pulses $\Delta J(t) = J_{12}-J_{23} \propto \cos(\omega t)$ where $\hbar \omega=J$ \cite{Medford2013,Medford2013b,Taylor2013}. Two-qubit gates can be obtained using dc pulses for the exchange coupling between pairs of spins belonging to different qubits \cite{Doherty2013}, or presumably via capacitive couplings, as demonstrated for the case of ST$_0$ qubit~\cite{Shulman2012}.

While allowing for narrow-band ac operation, always-on exchange coupling also---to some extent---exposes the qubit to electric noise. The discussion of possible ways to protect RX qubits from electric noise at suitable operating points where the qubit is insensitive to noise (sweet spots) has led to the asymmetric resonant-exchange (ARX) qubit \cite{russ_asymmetric_2015} and always-on exchange-only (AEON) \cite{Shim2016} qubit concepts. The AEON qubit allows for one-qubit and two-qubit operations while always remaining at a sweet spot.  Magnetic field gradients are also a source of unwanted noise for exchange-only qubits.  For any of these three-spin encodings, matrix elements due to local gradients will, in general, cause leakage from the total $S$ subspace in which the qubit is encoded into another $S$ subspace.  

\subsection{Spin qubits with additional charge degrees of freedom}\label{Subsec:Basics:Otros}

The spin qubits discussed above operate in the regime of half-filling, with one particle per site, as represented by the diagonal entries in \reftable{table:layouts}.  The half-filled  charge configuration restricts the degrees of freedom to the spin of the particles, while particle hopping only occurs virtually (with small quantum amplitude) to produce the exchange interaction between spins.  The exchange interaction as so far described is a weak and temporary charge hybridization; always-on exchange qubits (RX, ARX, AEON) as described in the prior section weakly connect dots into larger structures to hybridize spin and charge.  In this section, we describe qubit variants that take this to the extreme of putting multiple spins into common sites, or correlating sites to spin, to more strongly exploit spin-charge hybridization for qubit initialization and readout, electric-field control, and electric-dipole coupling to other qubits or cavity electric fields.

An instructive example is the flopping-mode qubit, which consists of a single electron that can occupy either the left or right site of a DQD \cite{Croot_flopping-mode_2020,Benito2019b,Mutter2021}. The charge can be coupled to the spin by spin-orbit coupling or an external magnetic field gradient, and delocalization of the charge across the DQD near zero level detuning enhances the electric dipole moment compared to a single QD \cite{cottet_2010,Hu2012}.  Judicious control of the energy level detuning and tunneling strength between the two sites permits a tunability of the electric dipole. Therefore strong coupling to the electric field or other qubits can be obtained when needed, while there is a small susceptibility to  charge noise at small coupling or sweet spots when the qubit is idle. Increasing the number of sites available to a single particle to three allows for the formation of a charge quadrupole qubit \cite{Friesen2017,Koski2020}.

Rather than extending the number of sites for a single particle, one can also decrease the number of sites for the three-particle EO qubit.  Reduction from three to two sites leads to the QD hybrid qubit \cite{Shi2012,Koh2012,Kim2014,Shi2014}. While this design essentially fixes the intra-site exchange coupling to a non-zero value, it still allows for fast electrical control of a qubit via the energy detuning and tunnel coupling.  Although charge noise is a concern for the hybrid qubit, its impact is reduced due to the similarity of the orbital wavefunctions of the intra-site singlet and triplet states. Reducing the number of sites further to a single site, one obtains the spin-charge qubit~\cite{Kyriakidis2007}, see Table~\ref{table:layouts}.

Four spins in four dots can define a pulsed EO qubit via the decoherence-free-subspace with total spin $S=0$. This qubit is initialized into its ground state via two spin singlets~\cite{Bacon2000}. RX-like operation is possible using at least three always-on exchange interactions between four dots~\cite{Sala2017}. Alternatively, a hybrid quadrupolar exchange-only (QUEX) mode of operation is possible with four spins in three dots, using a valley or orbital splitting in the central dot as an effective always-on exchange coupling~\cite{Russ2018}.

\section{Mesoscopic physics of dots and donors} \label{Sec:Principles}
In this section, we review the basic principles behind the operation of QDs and donors, which form the basis for semiconductor spin qubits.
In subsection~\ref{Subsec:Principles:Hets} we discuss how electrons, which exist in bulk semiconductors as delocalized Bloch states, can be confined in QDs by the heterostructure and externally applied potentials. The essential role of Coulomb interactions in defining QD states and the exchange interaction is covered in subsection~\ref{Subsec:Principles:Coulomb}. Subsection~\ref{Subsec:Principles:Dots_and_sensors} summarizes the development of QD device designs and charge sensing technology. We conclude by covering interactions with other microscopic degrees of freedom in semiconductor QDs, such as spin-orbit coupling (SOC) and its relation to the Zeeman Hamiltonian (\ref{Subsec:Principles:SOC}), valley states in silicon (\ref{Subsec:Principles:Valley}) and lattice nuclei (\ref{Subsec:Principles:Hyperfine}). Several of these topics have also been reviewed elsewhere, e.g., \onlinecite{Wiel2002,Hanson2007,Zwanenburg2013}, and we will emphasize recent developments where applicable.

\begin{figure*}[t]
	\includegraphics[width=\textwidth]{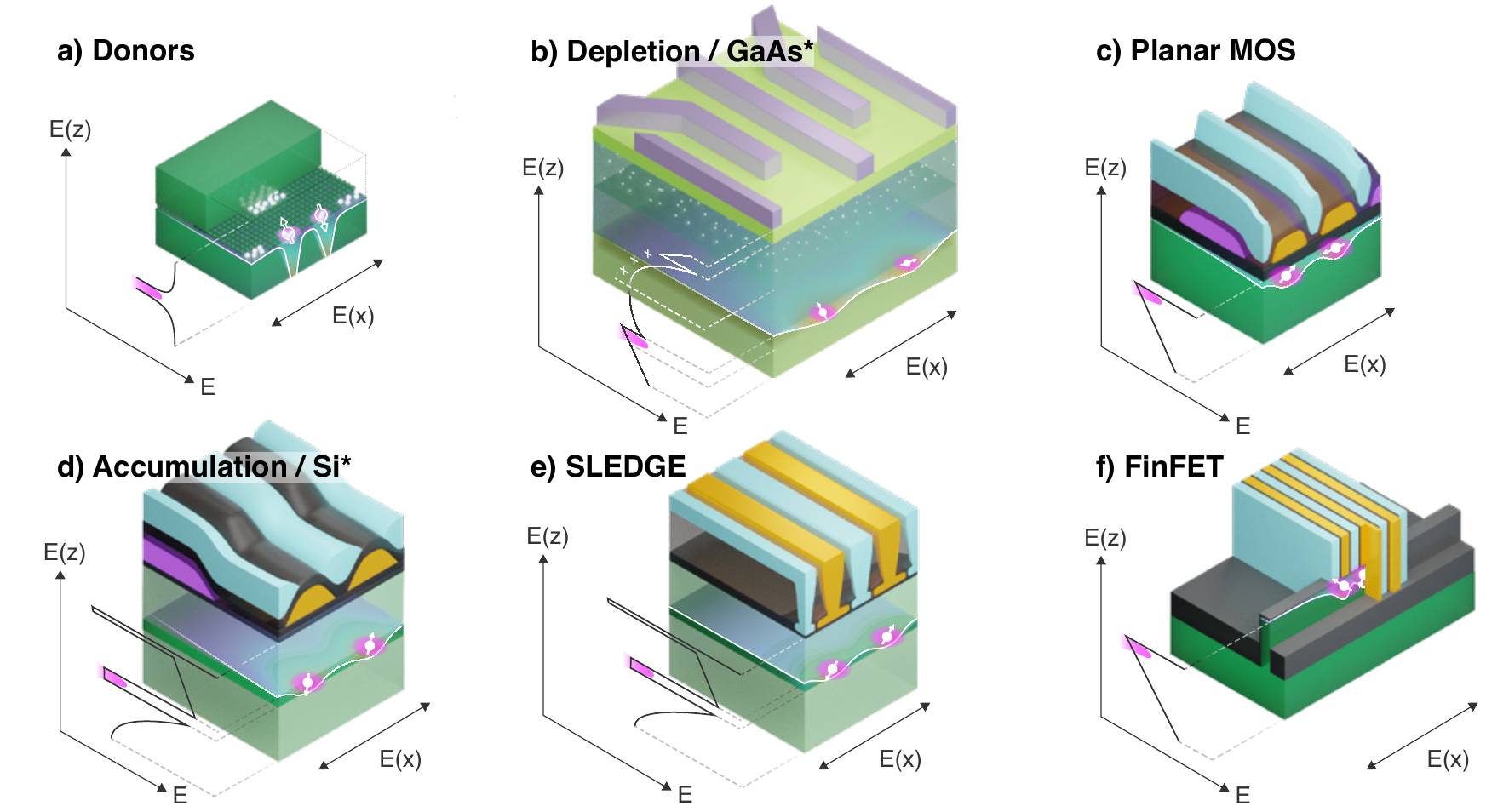}
	\caption{Device designs commonly used to confine electron spins. Vertical confinement is illustrated in the plots of $E(z)$ and lateral confinement is illustrated in the $xy$-plane. (a) Donor electrons are confined by the positive potential of the donor atom and manipulated with gates defined through conventional or STM lithography. (b) Depletion mode device design commonly used in early GaAs experiments.  (c,d) Modern SiMOS and Si/SiGe devices utilize overlapping gate architectures to achieve tight control of QD electrons. In SiMOS, electrons are localized at the SiO$_2$/Si interface (c). In Si/SiGe (d), the electrons reside in a buried quantum well. (e) SLEDGE (single layer etch-defined gate electrodes) devices utilize a single layer of gates patterned on the top surface of a Si/SiGe heterostructure. The gates are contacted from above using vias, which allows gate wiring to fan out away from the active area of the device in multiple planes. (f) FinFETs use a combination of dry etching and electrostatic gating to confine QD electrons.}
	\label{fig:confinement}
\end{figure*}

\subsection{Quantum confinement} \label{Subsec:Principles:Hets}

Semiconductor spin qubits rely on the full three-dimensional (3D) confinement of electrons. Figure \ref{fig:confinement} illustrates some of the most commonly employed spin qubit designs and the resulting electronic confinement potentials. In most planar QD systems, a layered semiconductor heterostructure generates confinement in the $z$-direction (generally the growth direction), while electrostatic gates confine electrons in the $xy$-plane [see Figs.\ \ref{fig:confinement} (a,c,d,e)]. In the case of donor spin qubits [see Fig.\ \ref{fig:confinement} (b)], 3D confinement is generated by the Coulomb potential of the dopant atom in the semiconductor. FinFET approaches [Fig.\ \ref{fig:confinement}(f)] use a combination of etching and electrostatic gating to define QDs. We begin our discussion of confinement by considering the bulk bandstructure of the most common materials used to fabricate spin qubits, namely GaAs and Si.

\subsubsection{Bulk bandstructure} 

\begin{figure}[t]
    \centering
	\includegraphics[width=\columnwidth]{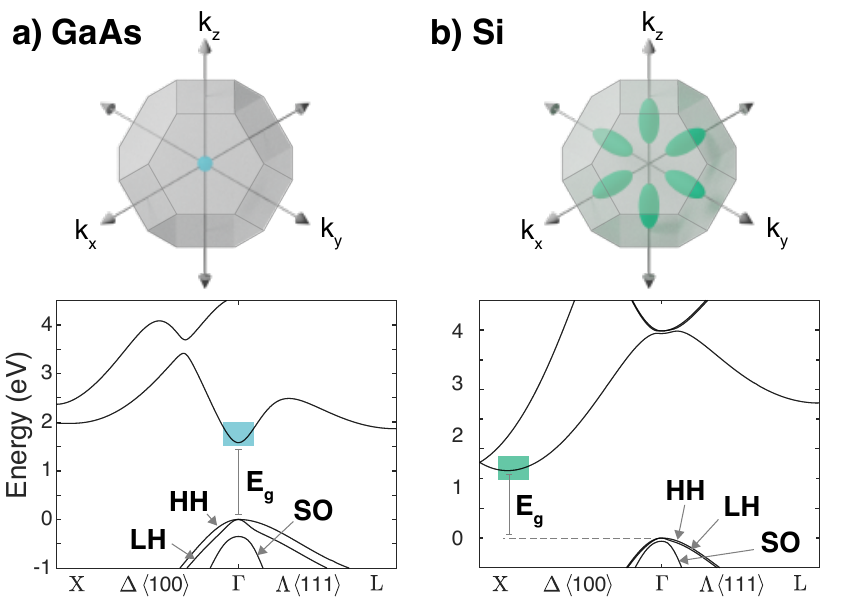}
	\caption{Bulk Brillouin zone (upper panels) and bandstructure (lower panels) as a a function of $k$ along the $\langle$100$\rangle$ and $\langle$111$\rangle$ directions for (a) GaAs and (b) Si. The nondegenerate conduction band minimum in GaAs is centered at the $\Gamma$ point ($k=0$), while Si has six equivalent conduction band minima along the high-symmetry $\braket{100}$ ($\Delta$) directions and an anisotropic effective mass. The heavy-hole (HH) and light-hole (LH) valence bands for both materials are separated in energy from the split-off (SO) band by the spin-orbit splitting.}
	\label{fig:bandstructure}
\end{figure}

Figure~\ref{fig:bandstructure} shows the first Brillouin zone and electronic bandstructure of GaAs and Si \cite{yu_fundamentals_2010}, which arise due to the crystalline potential of each material. While the full bandstructure is quite complex, much of its practical impact on the properties of QDs is captured by the effective mass approximation (EMA) describing the conduction band minima and valence band maxima. In this approach, the crystal potential effects are encapsulated by a renormalized kinetic energy operator in the Schr\"odinger equation, yielding the single-particle Hamiltonian \cite{yu_fundamentals_2010}

\begin{equation}
H_{\text{EMA}} = \sum_{i=x,y,z}\frac{-\hbar^2}{2 m^i} \frac{\partial^2}{\partial (r^i)^2} 
+U(\mathbf{r})
+\mu_B \vec{S} \cdot  \hat{g} \cdot \vec{B},
 \label{eq:effmass}
\end{equation}
with effective masses $m^i$ and the position vector $\mathbf{r}=(r^x,r^y,r^z)=(x,y,z)$. In this equation we have also included the slowly varying potential $U(\mathbf{r})$ which includes, e.g., the electrostatic potential generated by the gate electrodes, as well as the Zeeman term with the effective g-tensor $\hat{g}$ which is discussed further in \refsec{Subsec:Principles:SOC}.

The effective mass may be isotropic or anisotropic depending on the material; in the former case, we can define a single effective mass $m^* = m^{x,y,z}$. For instance, free electrons in GaAs [Fig. \ref{fig:bandstructure}(a)], occupy the isotropic $\Gamma$ ($k=0$) point conduction band minimum and are described by $m^* = 0.067m$, where $m$ is the bare electron mass. Bulk silicon [Fig. \ref{fig:bandstructure}(b)], by contrast, has a six-fold degenerate conduction band minimum along the $\braket{100}$ ($\Delta$) directions in $k$-space; each valley has an anisotropic effective mass of 0.92$m$ and 0.19$m$ in its longitudinal and transverse directions, respectively, where $m$ is the free electron mass. The six-fold valley degeneracy is broken in Si devices by heterostructure and electrostatic confinement which induces a valley splitting, which is discussed in more detail in \ref{Subsec:Principles:Valley}.

The effective mass approximation is sufficient for understanding many QD properties. However, microscopic details of important phenomena such as spin-orbit and valley splitting are sensitive to band mixing and atomistic effects beyond the effective mass approximation. Microscopic descriptions of such effects can be obtained from more complicated bandstructure models, for instance using $\vec{k} \cdot \vec{p}$ or tight-binding Hamiltonians \cite{yu_fundamentals_2010}. Such models are also useful in particular to describe valence band holes, where multiple bands are relevant due to $\Gamma$ point degeneracies and SOC. As pictured in Fig. \ref{fig:bandstructure}, this leads to heavy hole (HH) and light hole (LH) bands which are degenerate at $\Gamma$, as well as a split-off (SO) band which is lowered in energy by the bulk spin-orbit splitting.

\subsubsection{Bandstructure engineering}

To trap single spins, quantum confinement is necessary and is typically provided by a combination of material- and electrostatically-defined spatial barriers.
For donors in bulk silicon, 3D confinement is provided by the impurity potential itself as depicted in Fig.~\ref{fig:confinement}(a). This potential decays as $1/r$ away from the impurity, but has localized corrections in the immediate vicinity of the donor site; the latter short-range effects are called ``central-cell'' corrections \cite{pantelides_electronic_1978}. In epitaxial Si/SiGe and GaAs/AlGaAs heterostructures, by contrast, electrons are confined in the out-of-plane (growth) direction by the conduction band offsets occurring at semiconductor interfaces \cite{ando_electronic_1982,Abram_text,Bastard_text}. 

\newcommand{\AlGaAs}{Al$_x$Ga$_{1-x}$As}
\newcommand{\SiGe}{Si$_{x}$Ge$_{1-x}$}
For instance, many seminal results in mesoscopic physics were obtained with two-dimensional electron gas (2DEG) devices fabricated on Schottky-gated GaAs/AlGaAs heterostructures [Fig.~\ref{fig:confinement}(b)]. Sandwiching a thin GaAs layer between two \AlGaAs\ layers creates a 2DEG in the GaAs layer due to its lower conduction band edge. A 2DEG can also be formed at a single heterointerface, e.g., GaAs/AlGaAs, which confines electrons inside GaAs in a nearly triangular confinement potential. In most cases, the electrons are provided by doping the adjacent \AlGaAs\ layer with Si atoms \cite{manfra_molecular_2014}. Undoped enhancement-mode devices, where electrons are electrostatically forced into the quantum well with a top gate, are also being investigated \cite{Mak_undoped,Tracy_undoped}.

In Si metal-oxide-semiconductor (MOS) devices, the 2DEG is formed at the Si-oxide interface. The large band gaps of most oxides allow for very large band offsets, in turn enabling very high out-of-plane electric fields to be applied by metal gates without inducing leakage. As a result, MOS electrons are confined in an approximately triangular potential formed by the Si-oxide conduction band offset on one side and the gate-induced electric field on the other, illustrated by the potential cut in Fig.~\ref{fig:confinement}(c). 

2DEGs can be similarly formed in Si/SiGe heterostructures, where strain is appreciable due to the 4\% larger lattice constant of Ge compared to Si \cite{schaffler1997}. For spin qubit applications, a thin tensile-strained Si layer is typically sandwiched between lattice-relaxed \SiGe\ alloy layers, which induces a conduction band offset that traps electrons in the Si QW. Undoped heterostructures are now the norm for Si/SiGe QWs, as electron accumulation can be totally gate-modulated \cite{Deelman_MRS}. The induced out-of-plane electric fields in these structures are therefore comparatively modest, as shown in Figs.~\ref{fig:confinement}(d,e). Finally, FinFETs extend MOS architectures utilizing etching and electrostatic gating to confine QD electrons [Fig.~\ref{fig:confinement}(f)].

\subsubsection{Electrostatic gating}
\begin{figure}
	\includegraphics[width=\columnwidth]{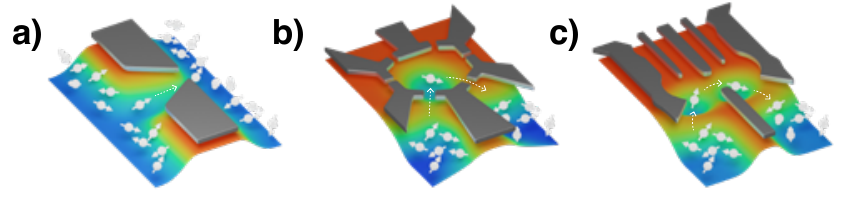}
	\caption{Electrostatic confinement. (a) 1D states can be formed in a QPC due to the potential constriction from a split gate. (b) Electrostatic confinement in both in-plane directions of a QW lead to 0D QD states. (c) Two QDs placed in series form a DQD. Depletion-mode gates are pictured here.}
	\label{fig:QPCandDot}
\end{figure}
Once a QW has been formed in a planar heterostructure, confinement in the in-plane dimensions can further reduce the effective dimensionality of the electronic states. In-plane confinement is achieved through the electrostatic potential $U(\vec{r})$ in Eq.~\eqref{eq:effmass}, which is typically induced by metal gate electrodes above the heterostructure. A confining potential along a single direction creates a quasi-1D channel, which can form a quantum point contact (QPC) [Fig.~\ref{fig:QPCandDot}(a)]. Finer-grained electrostatic confinement along both in-plane directions can form effectively 0D QDs. The potential minima define QD locations where electrons can be trapped [Fig. \ref{fig:QPCandDot}(b)]. 

Gate voltage changes alter both the QD electrochemical potential as well as the shape of the confining potential. QDs can be connected in series to make larger structures, such as the DQD depicted in Fig. \ref{fig:QPCandDot}(c). In a DQD, the interdot barrier height can be voltage-controlled to modulate the interdot tunnel coupling $t_c$ [Fig. \ref{fig:dotoccupation}(c)]. Typical devices use separate plunger and barrier gates to control the dot electrochemical potentials and interdot barriers, respectively. In practice, geometrical cross-capacitances influence the potential under neighboring gates \cite{Wiel2002}, and voltage compensation of multiple gates is required to independently control each dot potential, a procedure sometimes referred to as defining ``virtual gates'' \cite{Keller_Elecpump,Hensgens2017,VanDiepen2018,Mills2019}.

\subsection{Electron-electron interactions in QDs}
\label{Subsec:Principles:Coulomb}
Bandstructure and electrostatic confinement allow the formation of 0D QD states and trapping of individual electrons (and hence spins). As more electrons are added to a QD, the electron-electron Coulomb interaction becomes critical to the properties of the whole system. Trapped electrons in a QD electrostatically repulse any other electron attempting to join that dot. This classical effect defines the charging energy $E_C = e^2/C$, where $C$ is the total dot capacitance. Coulomb repulsion is drastically illustrated by the phenomenon of Coulomb blockade in electron transport through QDs. Biasing a QD in Coulomb blockade fixes its electron occupation, a prerequisite for defining any spin qubit  \cite{Kouwenhoven2001,Hanson2007}.

While Coulomb blockade can be understood conceptually by classical considerations, quantum effects further modify and enrich the physics. The full energy penalty for changing electron occupation is called the addition energy $\ts{E}{add}$, which can be qualitatively understood with a simple constant interaction model in which $\ts{E}{add} = E_C + E_{\rm orb}$. Here $E_{\rm orb}$ is the change in single-particle energy that appears when an extra electron must occupy a new orbital level to enter the QD, due to the Pauli exclusion principle prohibiting more than two electrons from occupying a single energy level.

\begin{figure}[t]
	\includegraphics[width=\columnwidth]{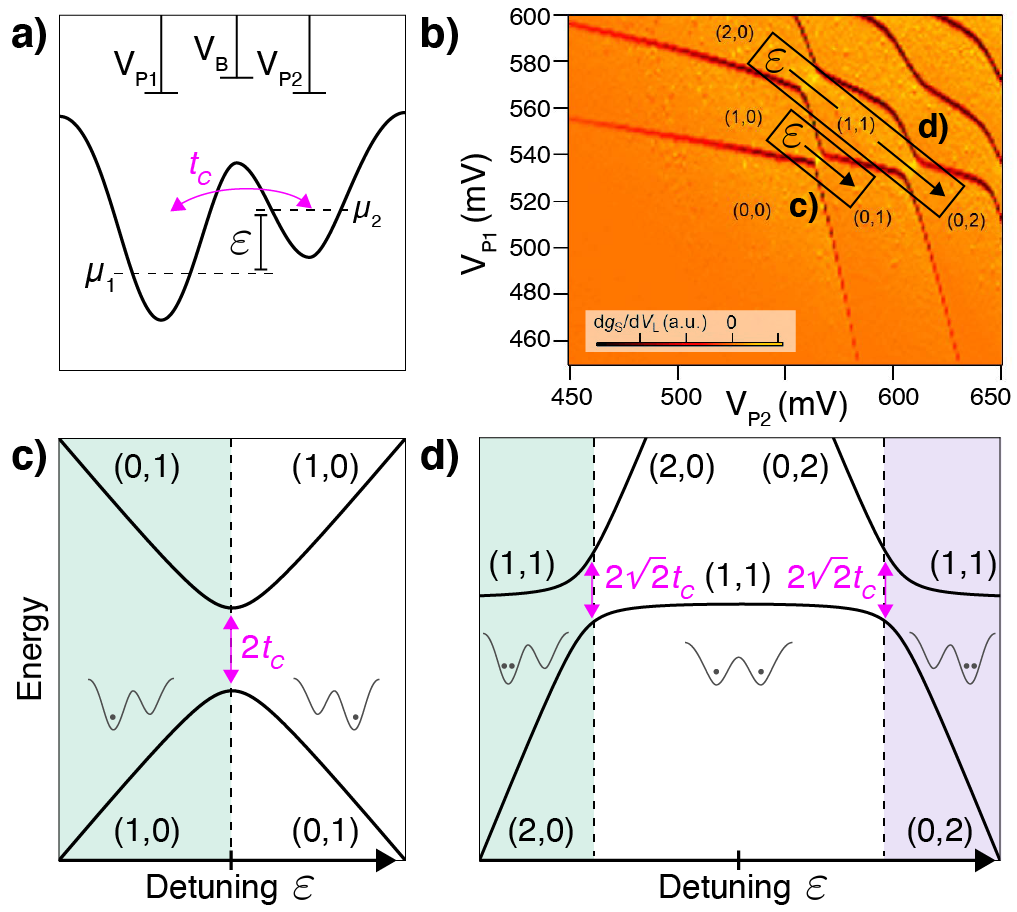}
	\caption{(a) DQD confinement potential. (b) DQD charge stability diagram fron \onlinecite{Zajac2017CNot}. (c) DQD energy levels near the (1,0)-(0,1) interdot charge transition. (d) DQD energy levels in the two-electron regime, showing the cross-over from the (2,0) $\rightarrow$ (1,1) $\rightarrow$ (0,2) charge state.}
	\label{fig:dotoccupation}
\end{figure}

Transport through multiple QDs connected in series proceeds when the electrochemical potentials of the individual QDs lie within the source-drain bias window and tunneling from one dot to the next is downhill in energy \cite{Wiel2002}. We consider the level structure of a DQD in detail [Fig. \ref{fig:dotoccupation}(a)], as it illustrates several key QD control principles. Figure~\ref{fig:dotoccupation}(b) shows a DQD charge stability diagram, with charge states denoted $(N_1,N_2)$, where $N_i$ is the number of electrons in dot $i$. For a single-electron DQD ($N_1 + N_2 = 1$), there are two relevant charge states, (1,0) and (0,1), and we can approximate the DQD in that basis as a two-level system with Hamiltonian
\begin{equation}\label{eq:dqd2level}
H_c = \begin{pmatrix}
\varepsilon/2 & t_c \\
t_c & -\varepsilon/2
\end{pmatrix},
\end{equation}
where the detuning $\varepsilon = \mu_1 - \mu_2$ is the difference in electrochemical potentials of the two dots. Hopping between different charge states is described by the tunnel coupling $t_c$, which is generally an exponential function of the interdot barrier height. As illustrated in Fig.~\ref{fig:dotoccupation}(c), the ground state charge occupancy changes from (1,0) to (0,1) as $\varepsilon$ changes sign, while around zero detuning the eigenstates are hybridized by $t_c$ into antibonding and bonding combinations of the charge states.

For a two-electron DQD (where $N_1$ + $N_2$ = 2), the (2,0), (1,1), and (0,2) charge states are possible. However, the DQD detuning must be highly biased for the doubly occupied (2,0) or (0,2) charge state to become the ground state due to Coulomb repulsion. As a result, the DQD ground state changes from (2,0) to (1,1) to (0,2) as $\varepsilon$ increases, as illustrated in Fig. \ref{fig:dotoccupation}(d). In practice, voltage modulation of detuning and tunnel coupling is critical for nearly all spin qubit control modalities.

\begin{figure}
	\includegraphics[width=\columnwidth]{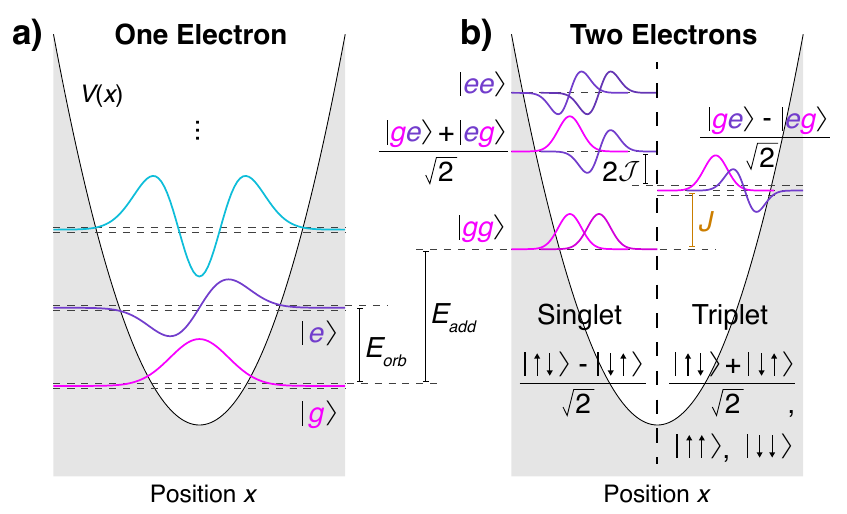}
	\caption{
	Low-energy orbital spectrum of a one- and two-electron QD. (a) A one-electron QD with a parabolic potential has excited states equally spaced by $E_{\rm orb}$ (only excitations along one dimension are shown for simplicity, and a small Zeeman splitting illustrates the spin degeneracy). (b) For two electrons, the total energy is increased by $E_{\rm add}$ and the lowest singlet and triplet eigenstates are shown with the combinations of the orbital wavefunctions that dominate each state. The singlet-triplet splitting $J$ is due to the triplet occupation of the excited orbital, though the energy of the latter is lowered from the one-electron orbital splitting by direct exchange $2\mathcal{J}$.}
	\label{fig:single_dot_ST}
\end{figure}

\begin{figure*}[t]
	\includegraphics[width=2\columnwidth]{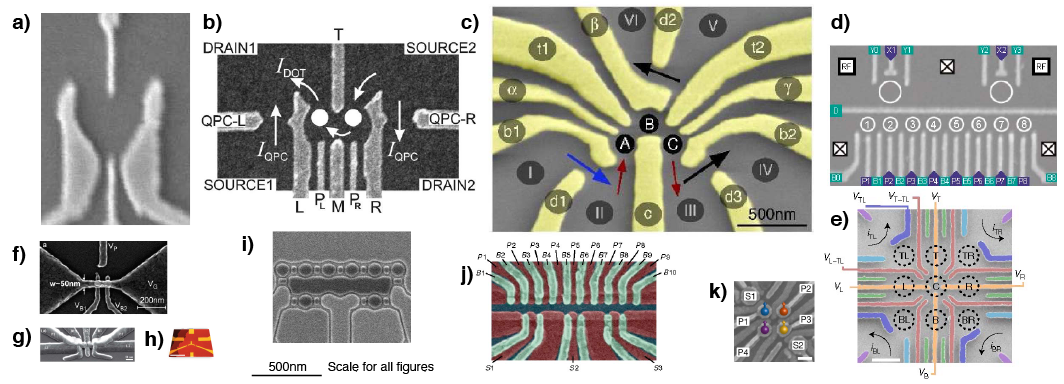}
	\caption{(a--c) Few electron single, double, and triple QDs 
	\cite{Ciorga2000}, \cite{Elzerman2003}, \cite{Schroeer2007}. (d) 8-site 1D QD array  \cite{Volk2019VirtualGates}. (e) 3 $\times$ 3 QD array \cite{mortemousque_coherent_2021}. (f,g) SiMOS single and DQD devices \cite{Angus2007},\cite{Lai2011SpinBlockade}. (h) Donor device fabricated using STM lithography \cite{He_2donorexchange_2019}. (i) Single-layer etch defined 1 $\times$ 6 QD array in Si/SiGe \cite{Ha_archive}. (j) 1 $\times$ 9 QD array fabricated using overlapping Al gates on Si/SiGe \cite{Zajac2016}. (k) Enhancement mode Ge/GeSi structure \cite{Hendrickx2021}. Holes are confined in (k), while the remaining devices isolate electrons. Images are sized to share common dimensional scales.}
	\label{fig:sec3devices}
\end{figure*}

Spin-spin Heisenberg exchange interactions are a key resource for spin qubits. Microscopically these interactions arise from the interplay of the Pauli exclusion principle, the external potential, and Coulomb interactions; given its complexity and importance, we refer the reader to \refsec{Sec:2spins} for a detailed discussion of this topic. Here we illustrate these principles by discussing the energy spectrum of two electrons in a single QD, which is also practically important for spin manipulation and measurement.

As illustrated in Fig. \ref{fig:single_dot_ST}, the one-electron states of a single QD include an orbital ground  and first excited state, separated in energy by $E_{\rm orb}$. When a second electron is added to the dot, the spatial wavefunctions must be either symmetric or antisymmetric under particle exchange, corresponding to spin singlet and triplet states, respectively. Singlets can have both electrons occupy the same or different (spin-degenerate) orbitals, while spatial antisymmetry requires that triplets must have electrons in separate orbitals. Restricting ourselves to the two lowest orbital states for simplicity, the ground state spin singlet comes from double occupation of the ground orbital, while the triplet is higher in energy as it must place one electron each in the ground and first excited orbitals, as shown in Fig.~\ref{fig:single_dot_ST}. Hence the singlet-triplet splitting $J = E_T - E_S$ is positive. This example illustrates the general principle that any two-electron system (even spanning multiple QDs) has a singlet ground state in the absence of magnetic fields \cite{Lieb1962}.

Note that in general for a two-electron QD, $J~<~E_{\rm orb}$, the single-particle orbital splitting, because the triplet state is lowered in energy by the direct Coulomb exchange interaction $2\mathcal{J}$\footnote{Literature on atomic and chemical systems may refer to the Coulomb exchange integral $\mathcal{J}$ as the ``exchange energy,'' which is the interaction between singlets and triplets occupying the same set of orbitals. In spin qubits, we define $J$ as the singlet-triplet splitting of the lowest two states, regardless of orbital content, as that is what gives an effective Heisenberg exchange interaction within the qubit Hilbert space.}. In practice, contributions from other orbitals are also quantitatively important, but they do not substantially change the qualitative physical picture. These arguments can also be extended to include excited valley states, which are often the lowest energy excitations in Si QDs; in such cases, the lowest excited triplet may occupy the excited valley rather than orbital state, giving rise to a even richer two-electron spectrum \cite{Hada2003,ercan_strong_2021}.

\subsection{Isolating and detecting single charges} \label{Subsec:Principles:Dots_and_sensors}

In this section we more closely examine spin qubit designs and the various approaches for detecting the number of charges trapped in a QD. Figure \ref{fig:sec3devices} gives an overview of the various single electron QD designs that have been utilized by the spin qubit community. Common ``stadium-style'' depletion-mode GaAs gate electrode designs are shown in Figs. \ref{fig:sec3devices}(a-e). The use of undoped Si/SiGe wafers, and overlapping gate stacks that gate the dots from the top, has been a paradigm shift for the community; one that has arguably propelled the field of Si spin qubits forward in recent years. Top gates allow for tighter confinement, yield larger capacitive coupling to QD electrons, and can be fabricated in multiple layers. Figures \ref{fig:sec3devices}(f,g) show examples of SiMOS single QD and DQD designs \cite{Angus2007,Lai2011SpinBlockade}. Figures \ref{fig:sec3devices}(h,i) illustrate dual-rail designs, where linear QD arrays are partnered with a parallel channel of charge detectors. The device in Fig. \ref{fig:sec3devices}(h) is a Si/SiGe TQD with an opposing charge sensor \cite{Reed2016}. A linear 9 dot array with 3 charge sensors is shown in Fig. \ref{fig:sec3devices}(i) \cite{Zajac2016}. These overlapping gate designs have been successfully extended to small 2D arrays in other material systems, as illustrated by the 2 $\times$ 2 Ge QD array in Fig. \ref{fig:sec3devices}(j) \cite{Hendrickx2021}. QD fabrication methods are also transitioning from academic-scale liftoff processes to industry-compatible subtractive processes that are more amenable to the development of multilayer devices \cite{Geyer_APL,Ha_archive}.  SiMOS CMOS nanowire devices fabricated in industrial-grade research foundries are similar to FinFETS, show single-electron, single-qubit operation, and have highlighted the promise of pathways to qubits which may scale in comparable fashion as silicon transistor technologies~\cite{ansaloni_single-electron_2020,zwerver_qubits_2021}.

\begin{figure}
	\includegraphics[width=\columnwidth]{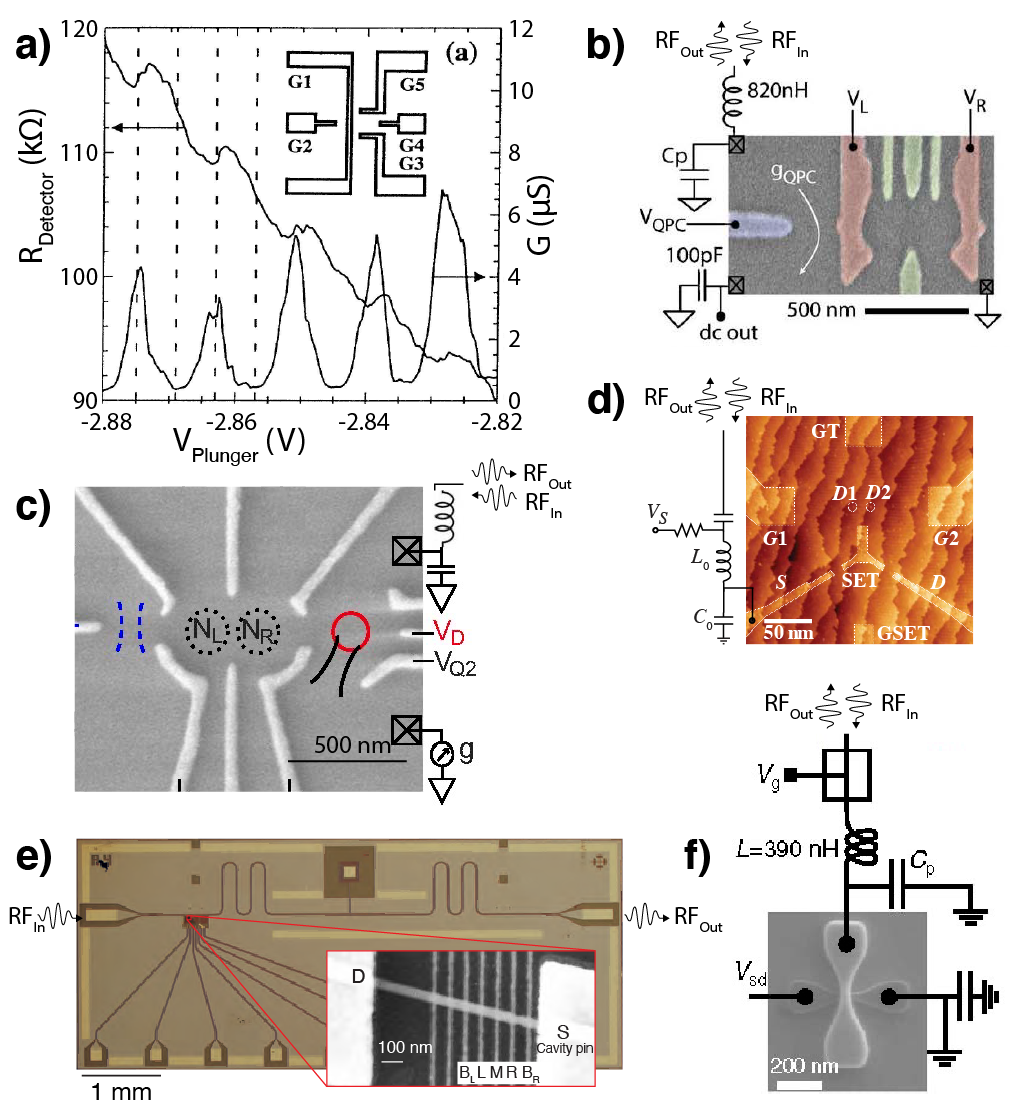}
	\caption{(a) QPC charge detector to probe the charge occupation of a single QD \cite{Field1993}. (b) RF-QPC for fast sensing of a DQD \cite{Reilly2007FastRF}. (c) Fast charge sensing of a DQD using a RF-QD charge sensor \cite{Barthel2010}. (d) Donor device fabricated using STM lithography and probed using RF-reflectometry \cite{Keith2019}.  (e) cQED device for detecting charge and spin states in a cavity-coupled InAs nanowire DQD \cite{Petersson2012}. (f) Dispersive gate sensing of charge states in a fin-FET device \cite{Gonzalez2014}. }
	\label{fig:sec3sensors}
\end{figure}

Charge sensing techniques can be adapted for highly sensitive single-shot spin readout by utilizing Pauli spin blockade, as will be presented in detail for different types of spin qubits in the following sections \cite{Elzerman2004,Barthel2009,pakkiam_single-shot_2018,West2019}. We now describe how the QPC charge sensors in the devices shown in Figs. \ref{fig:sec3devices}(b,c,e)] and QD charge sensors shown in Figs. \ref{fig:sec3devices}(d,h-j)] are used to measure changes in the charge occupation of QD devices \cite{Field1993,DiCarlo2004}.

The absolute number of electrons confined in a QD can be determined through charge detection using a QPC or a QD as a charge detector \cite{Field1993}. The measurement bandwidth can be greatly increased using radio frequency (RF)-reflectometry \cite{Schoelkopf_rf}, as later demonstrated with RF-QPCs [Fig.\ref{fig:sec3sensors}(b)] and RF sensor dots [Fig.\ref{fig:sec3sensors}(c)] \cite{Reilly2007FastRF,Barthel2010}. A recent development is dispersive gate sensing, where microwave reflection off of a QD gate is used to infer the QD charge occupation \cite{colless_dispersive_2013,West2019,urdampilleta_gate-based_2019,zheng_rapid_2019}. Dispersive sensing has the potential to scale to larger system sizes, as additional QD or QPC sensors are not needed. Finally, as will be discussed in detail in Sec. \ref{Sec:Hybrid}, dispersive charge and spin state readout can be achieved in the circuit quantum electrodynamics (cQED) architecture [Fig. \ref{fig:sec3sensors}(e)]. Baseband and microwave charge detection approaches have greatly benefited from the development of cryogenic amplifiers \cite{Vink2007,Macklin_TWPA}.

\subsection{Zeeman interactions and spin-orbit coupling}  \label{Subsec:Principles:SOC}

Direct magnetic manipulation of the electron spin $\vec{S}$ in solids is generally described by the Zeeman Hamiltonian
\begin{equation}
\label{eq:ESR}
H(t) = \mu_B \vec{S} \cdot  \hat{g}(t) \cdot \vec{B}_{\rm eff}(t),
\end{equation}
where $\mu_B$ is the Bohr magneton (= 58 $\mu$eV/T). In contrast to free electrons where the coupling is described by a scalar g-factor $g \approx 2$, the crystal field in solids can lead to an anisotropic magnetic response captured by an effective g-tensor $\hat{g}$ \cite{Slichter2010}. The effective magnetic field $\vec{B}_{\rm eff}$ can include externally applied fields as well as internal fields due to hyperfine or spin-orbit effects. Time-dependent modulation of this Hamiltonian enables coherent single-spin rotations, as detailed in \refsec{Subsubsec:Gates:LD:1QB}. As SOC is a crucial ingredient to both $\hat{g}$ and $\vec{B}_{\rm eff}$, we discuss it further here along with the ways it can be utilized to manipulate individual spins.

SOC arises from the relativistic coupling of spin to electric fields and is described by the Hamiltonian $H_{\rm SO} = \dfrac{g \mu_B}{\hbar m c^2} (\grad V \times \vec{p}) \cdot \vec{S}$, where $V$ is the electric potential and $\vec{p}$ is the electron momentum \cite{ZuticRMP}. In essence, an electron spin moving in a potential experiences an effective momentum-dependent magnetic field $\vec{B}_{\rm eff,SO}$. For spherically symmetric potentials, such as the hydrogen atom, this coupling takes the commonly cited isotropic form $\vec{L} \cdot \vec{S}$. In semiconductor heterostructures, the $\grad V$ term arises from internal crystal fields and potential discontinuities at material interfaces \cite{Hanson2007,ZuticRMP}.

The spin-orbit interaction in bulk solids increases with atomic number; thus, the spin-orbit splitting (equal to the valence band splitting in Fig. \ref{fig:bandstructure}) is 44 meV in Si but about 300 and 340 meV in Ge and GaAs, respectively. In bulk semiconductors, the $p$-like valence bands are particularly strongly coupled by SOC, while the effects on $s$-like conduction band electrons, such as in GaAs, are weaker but significant for spin qubit control, for example by altering the $g$-factor. In bulk silicon, the electron $g$-factor remains close to 2 and is only weakly anisotropic \cite{roth_g_1960}, while electrons in bulk GaAs have an isotropic g-factor of $-0.44$, which can be further (and anisotropically) modified in QWs \cite{Beveren2005,kogan_measurements_2004,yugova_universal_2007}.
\begin{figure}
	\includegraphics[width=\columnwidth]{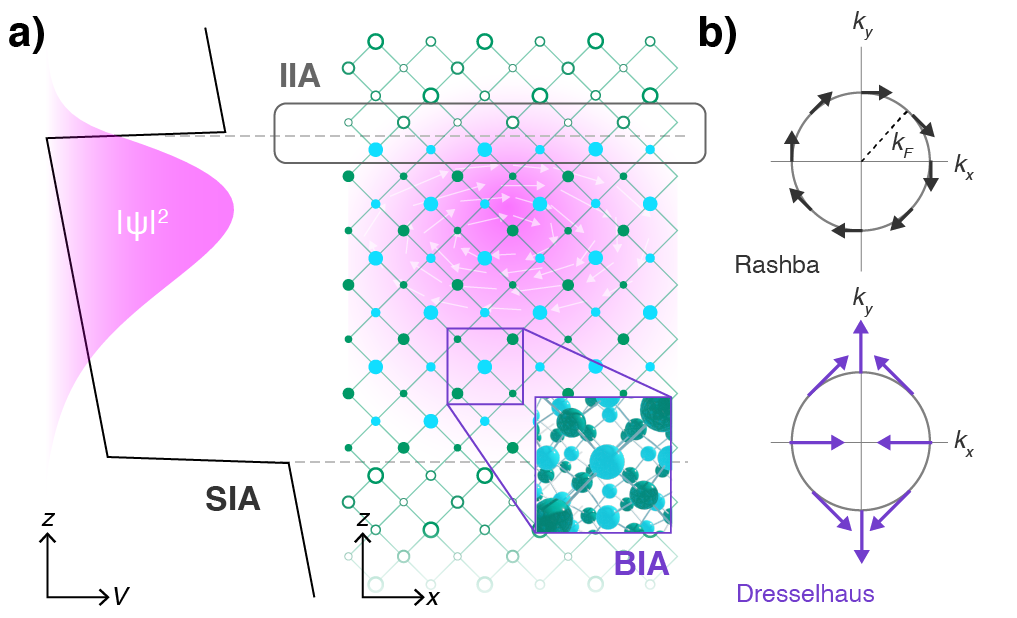}
	\caption{(a) Spin-orbit interactions in QDs arise microscopically from the inversion asymmetries due to the bulk crystal structure (BIA), structural effects (SIA) like external fields, and interfaces (IIA). Under an applied magnetic field into the page, the local momentum of the electron wave function rotates (as depicted by arrows), causing local couplings to the atomic-scale gradients induced by these asymmetries which sum to the effective couplings in Eq.~\eqref{Eq:SOC2D}. (b) Effective spin-orbit field direction for Dresselhaus- and Rashba-type interactions as a function of in-plane momentum at the Fermi surface momentum $k_F$ [see Eq. \eqref{Eq:SOI_B}].
	}
	\label{fig:spin_orbit_microscopics}
\end{figure}

Additional SOC effects arise in 2D QWs due to confinement and lowered symmetries, which for electrons are largely described by the effective Hamiltonian\footnote{The factor of 2 in this equation is due to our explicit use of spin rather than Pauli operators}
\begin{equation}\label{Eq:SOC2D}
	\ts{H}{\rm SO} = 2\gamma_R (p^y S^x - p^x S^y) + 2\gamma_D (p^x S^x - p^y S^y).
\end{equation}
where $\gamma_R$ and $\gamma_D$ are the so-called Rashba and Dresselhaus SOC coefficients. These interactions fundamentally arise from inversion symmetry breaking at different scales. Structural inversion asymmetries (SIA) due to confining electric fields lead to Rashba couplings, while Dresselhaus interactions relate to the bulk inversion asymmetry (BIA) of the zincblende lattice in GaAs and to heterostructure interface inversion asymmetry (IIA) in Si QWs \cite{golub_spin_2004,nestoklon_electric_2008,prada_spinorbit_2011}. Figure~\ref{fig:spin_orbit_microscopics}(a) illustrates these different sources of microscopic asymmetries and their connection to spin-orbit coupling. Intuitively, a QD electron undergoes cyclotron motion due to an applied magnetic field, leading to SOC effects as its local momentum samples these asymmetries \cite{Jock2018}. Additional spin-orbit couplings beyond the linear terms in Eq.~\eqref{Eq:SOC2D}, such as terms cubic in momentum $p$, can also be relevant, for instance for quantum-confined holes. 

The Hamiltonian of Eq. \eqref{Eq:SOC2D} introduces additional $g$-tensor modulations by coupling the vector potential $\vec{A}$ of an external magnetic field to spin via the momentum, $\vec{p}\rightarrow \vec{p}-e\vec{A}$. For example, choosing the Coulomb gauge for an in-plane magnetic field $\vec{B}=B^x\hat{x}$ one obtains $B$-dependent terms $eB^x z (\gamma_R S^x - \gamma_D S^y)$ in $H_{\rm SO}$. If the SO couplings $\gamma_{R,D}$ contain interfacial contributions, this introduces spin-dependent level shifts which contribute both diagonal and off-diagonal $g$-tensor terms $g_{xx}$ and $g_{xy}$.
Further $g$-tensor corrections arise from the admixture of excited orbital states \cite{de_sousa_gate_2003,stano_spin-orbit_2005} or valley states in Si QDs \cite{ruskov_electron_2018,nestoklon_electric_2008,prada_spinorbit_2011,Veldhorst2015b,harvey-collard_spin-orbit_2019}. These couplings can be sensitive to local device disorder, causing interdot $g$-factor gradients in Si/SiGe \cite{ferdous_valley_2018} and in MOS dots for electrons and holes \cite{Voisin2016,Jock2018,tanttu_controlling_2019}.
The effects of SOC on electronic $g$-factors have also been investigated in metallic nanoparticles \cite{Petta_SO_2001,Petta_SO_2002}, InAs \cite{Schroer2011} and InSb nanowire DQDs \cite{Nadj-Perge2012}, and self-assembled QDs \cite{Nakaoka2007}, among other systems.

The Hamiltonian in Eq. \eqref{Eq:SOC2D} can also be interpreted as the action of a momentum-dependent spin-orbit field
\be\label{Eq:SOI_B}
\ts{\vec{B}}{eff,SO} = (\gamma_D + \gamma_R) \sin\theta \hat{e}_{[110]} + (\gamma_D - \gamma_R) \cos\theta \hat{e}_{[1\bar{1}0]}
\ee
where $\theta$ denotes the angle between $\vec{p}$ and the [110] direction \cite{Kavokin2001}.
Figure \ref{fig:spin_orbit_microscopics}(b) shows the different orientation dependencies of Rashba and Dresselhaus SOC fields.
This effective field imparts a directional dependence to matrix elements involving momentum, including interdot tunneling and intradot orbital spin-flip transitions \cite{Stepanenko2012,hofmann_anisotropy_2017}.
This also enables controlling electron spins with orbital motion, or electric dipole spin resonance (EDSR), as first described in \onlinecite{RashbaPRL2003}.
For example, if we apply a static magnetic field $\vec{B}_0 = B_0 \hat{z}$ and take $\hat{g}(t)$ = $g \mathbb{1}$, the orbital motion of the electron with $p_y(t) = p_0 \cos(\omega t)$ yields $H_R$ = $2 \gamma_R p_0 \cos(\omega t) S^x$, which can be used to drive Rabi oscillations in a rotating frame.

\onlinecite{Golovach2006} developed the theory for EDSR in 2DEG-based QD systems, while \onlinecite{Flindt2006} considered EDSR in nanowire devices with strong SOC. 
Golovach and Loss find that a harmonic QD subject to an oscillating electric field can be described by an effective Hamiltonian
$H_{\rm eff} = \frac{1}{\hbar}(g \mu_B \vec{B} \cdot \vec{S} + \vec{h}(t) \cdot \vec{S})$, with $\vec{h}(t) = 2 g \mu_B \vec{B} \times \vec{\Omega}(t)$, where $\vec{\Omega}(t)$ is a dimensionless driving field. The coupling strength (and hence effective Rabi frequency) scales linearly with the amplitude of the oscillating electric field, and the drive is maximal when $\vec{\Omega}(t)$ and $\vec{B}$ are orthogonal. The driving strength is $\Omega(t) \sim r_0(t)/\lambda_{\text{SO}}$, where $\lambda_{\text{SO}}$ $\sim$ $\lambda_{\pm}=\hbar/m^*(\gamma_D \pm \gamma_R)$ is the spin-orbit length and $\vec{r}_0(t) = - {e \vec{E}(t)}/(m^* \omega_0^2)$ denotes the shift of the QD due to the electric field where $\vec{E}(t)$ and $\hbar\omega_0$ are the electric field and confinement energy of the QD.

An alternative to the ac-driven \textit{displacement} of the entire electronic wave function in a spin-orbit field, driving ac electric fields can also \textit{distort} the confining potential and hence the wave function, which manifests as an effective time-modulation of the anisotropic $g$-tensor which can also cause spin rotations. In general, $\hat{g} = \hat{g}(V(t))$, where $V(t)$ is a time-dependent gate voltage on the device \cite{VenitucciPRB}. The first demonstration of spin control using $g$-tensor resonance was in a 2D GaAs/AlGaAs heterostructure, where the Al concentration was purposely graded to achieve a spatially varying $\hat{g}$ \cite{Kato_g-tensor_2003}. Driving the system with an electric field yielded spin rotations that were optically detected using time-resolved Kerr rotations. Recent progress utilizing $g$-tensor modulation has occurred in hole spin qubits, taking advantage of the natural anisotropies of the valence band, as discussed in \refsec{Subsubsec:Implementations:Alt:Holes}.

``Synthetic'' spin-orbit fields can also be induced by translating a spin along an extrinsic magnetic field gradient, typically generated in QDs by a nearby micromagnet. As proposed by \onlinecite{Tokura2006}, this enables ``slanting Zeeman field'' spin resonance or EDSR in a magnetic field gradient, as external driving electric fields $E_{\text{ac}}$ displace the electron within the QD, allowing it to experience the spatially varying transverse magnetic field. The effective ac magnetic field strength can be calculated from perturbation theory as 
\begin{equation}
B_{\text{ac}} = \frac{e E_{\text{ac}} \ell_{\text{orb}}^2}{E_{\text{orb}}}|b_{\text{SL}}|,
\end{equation}
where $E_{\text{orb}}$ is the QD orbital splitting, $\ell_{\text{orb}}$ is the orbital length scale, and $b_{\text{SL}}=\partial B^z/\partial x$ is the transverse magnetic field gradient. The resulting Rabi frequency $f_{\text{Rabi}} = {g \mu_B B_{\text{ac}}}/(2h)$ is linearly proportional to $E_{\text{ac}}$ and $b_{\text{SL}}$ \cite{Pioro2008}.

Finally, while typical EDSR operation displaces the spin within a single QD, which limits the interaction strength in tightly-confined QDs \cite{Hu2012}, low power electrical spin control can be achieved by increasing the displacement of the electron through the use of DQDs [Fig. \ref{fig:sec4single-spin}(d)]. \onlinecite{Benito2019b} considered this ``flopping-mode'' spin qubit consisting of a single electron confined in a semiconductor DQD in the presence of both a homogeneous external magnetic field $B \hat{z}$ and a transverse field gradient created with a micromagnet $\Delta B^x \approx b_{\text{SL}}\Delta z/2$, where $2\Delta B^x$ is the difference in the $x$-component of the magnetic field from the left to right side of the DQD separated by $\Delta z$ \cite{Benito2017}. When $\Delta B^x$ is appreciable, ac driving of the electron across the DQD can lead to low-power single spin rotations \cite{Croot_flopping-mode_2020}.

\subsection{Valleys}\label{Subsec:Principles:Valley}
\begin{figure}[t]
	\includegraphics[width=\columnwidth]{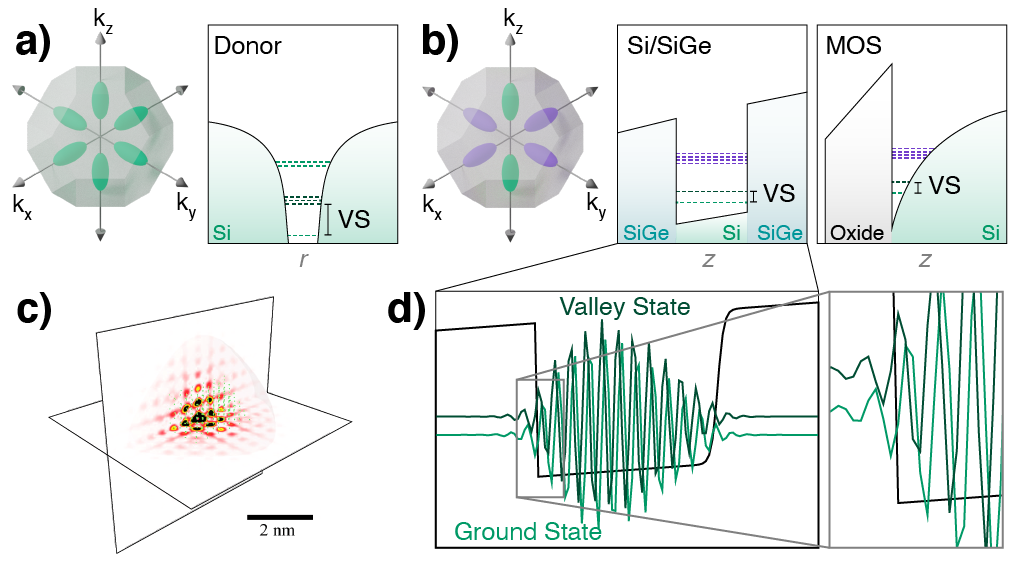}
	\caption{(a) The valley splitting of donors in bulk Si from the admixture of the six-fold degenerate valleys (depicted in the Brillouin zone) leads to three sets of states. (b) In Si QDs, the electric fields in MOS and strain in Si/SiGe raises four in-plane valley energies and the relevant valley splitting is between the two out-of-plane valleys. (c) The admixture of valley states leads to rapidly varying modulations in the donor ground state, pictured from an effective mass calculation presented in \cite{gamble_multivalley_2015}. (d) The full ground and excited state wave functions in Si QDs oscillate rapidly due to the intervalley phase. Interference of the valley Bloch functions minimizes the interface overlap for the ground state.}
	\label{fig:valley_dots_donors}
\end{figure}

A modification of the simple picture of electron confinement presented in Sec.~\ref{Subsec:Principles:Hets} occurs in Si, where the conduction band features six equivalent minima, referred to as ``valleys,'' as shown in Fig. \ref{fig:bandstructure}(b). The valley degree of freedom can complicate the level structure of quantum-confined states \cite{schaffler1997,Zwanenburg2013,Gyure2021}. For donors in bulk silicon, each valley contributes a degenerate state in the EMA. This degeneracy is lifted by valley-orbit coupling with the tetrahedral donor central-cell potential, leading to a nondegenerate ground state composed of a symmetric linear combination of the six valleys, as shown in Fig.~\ref{fig:valley_dots_donors}(a).
By contrast, the four in-plane ($x$, $y$) valleys are raised in energy by strain in Si/SiGe quantum wells \cite{schaffler1997} and higher subband quantization energy in MOS devices \cite{ando_electronic_1982}.
This leaves two longitudinal $k_z$ valleys whose degeneracy is lifted by the heterointerfaces, giving rise to the valley splitting, as illustrated in  Fig.~\ref{fig:valley_dots_donors}(b). Controlling and maximizing this splitting is critical for Si-based spin qubits, as it is typically the lowest energy excitation in a single-electron QD.

Valley splitting arises from atomic-scale interactions of the electron with the heterostructure potential, where the EMA is most questionable \cite{friesen_theory_2010,Saraiva2009} and numerical full-band calculations using tight-binding or pseudopotentials can offer atomistic insight \cite{boykin_valley_2004,zhang_genetic_2013}.
Nonetheless, many key features can be described within an augmented effective mass framework, where the full wave function is expanded in terms of envelope and Bloch functions for each relevant valley
\be\label{eq:valley_psi}
\psi(\vec{r}) = \sum\limits_{j=1}^{N_v=(2,6)} F_j(\vec{r}) e^{i\vec{k}_j \cdot \vec{r}} u_j(\vec{r}).
\ee
Here $\vec{k}_j$ and $u_j(\vec{r})$ are the wave vector and periodic part of the Bloch function, respectively, for the $j^{th}$ valley, and $F_j$ is the envelope function for that valley.
For donors in bulk silicon, the $N_v$ sum runs over all 6 valleys, whereas only the two $k_{\pm z}$ valleys matter for QDs.
Each valley envelope function is the solution of 
\be
\left(T_i + U(\vec{r}) \right)F_i(\vec{r}) + \sum\limits_{j \neq i}V_{ij}^{\text{VO}}(\vec{r})F_j(\vec{r})
=E F_i(\vec{r}),
\label{eq:valleyenvelope}
\ee
where $T_i$ is the effective mass kinetic operator for the $i^{th}$ valley, $U(\vec{r})$ is the external potential, and $V^{\text{VO}}_{ij}$ is the valley-orbit coupling matrix element, which can be fit to data or estimated from a model potential \cite{gamble_multivalley_2015}.
For donors, the magnitude of the valley-orbit splitting is mostly set by the central cell correction, though it is sensitive to local strain.
However, the superposition of valley states introduces a complicated interference pattern in the full donor wave function $\psi(\vec{r})$ [see in Fig.~\ref{fig:valley_dots_donors}(c)]. As a result, the inter-donor tunnel coupling and exchange coupling are very sensitive to placement of donors in the Si crystal lattice \cite{koiller_strain_2002,salfi_valley_2018,gamble_multivalley_2015}.

If we consider a QW with a sharp heterointerface at $z=z_i$, we can estimate the interfacial intervalley coupling as $V_{+z,-z}^{\text{VO}} = v_0\delta(z-z_i)$ \cite{Friesen2007,Saraiva2009}. Taking the valley-free envelope function $F(z)$ as the solution of the intravalley part of Eq.~\eqref{eq:valleyenvelope}, we can evaluate the intervalley matrix element of $V_{+z,-z}^{\text{VO}}$ to obtain the valley mixing $\ts\Delta{VO} = v_0 |F(z_i)|^2 e^{2 ik_{z} z_i}$. As this is a complex-valued matrix element, the valley splitting is equal to twice its norm ($\text{VS} = 2|\Delta_{\rm VO}|$). This simple example illustrates that the valley splitting is dependent on the electron overlap with the interface, which can be increased by using vertical electric fields or reducing the QW width.

Experiments show that the tunable out-of-plane electric fields in MOS structures allow for a wide range of valley splittings 50--500~$\mu$eV \cite{Yang2013,gamble_valley_2016,petit_spin_2018}.
Electric field tuning is weaker in Si/SiGe QWs due to the smaller conduction band offset and the valley splitting is most strongly influenced by interface quality and QW width, with values up to 200--300~$\mu$eV reported in high-quality interfaces and narrow wells \cite{Borselli2011,hollmann_large_2020,chen_detuning_2021}. Beyond improving the epitaxial quality, other methods have been proposed for achieving uniformly high valley splitting by modulating the Ge content of the barrier or QW regions \cite{zhang_genetic_2013,mcjunkin_valley_2021}.

The valley mixing phase $\phi_{V} = \arg (\ts\Delta{VO})$ is also significant as it characterizes the superposition of valleys in the ground state.
In general this phase minimizes the ground state overlap with the interface, lowering its energy, as shown in Fig.~\ref{fig:valley_dots_donors}(d).
Changes in this phase due to disorder modify the valley character of the ground and excited states of different QDs, enabling intervalley tunneling \cite{culcer2010interface,borjans_probing_2021,Burkard2016,mi_high-resolution_2017}.

\subsection{Hyperfine interactions}\label{Subsec:Principles:Hyperfine}

Nuclear spins in semiconductors act as both a nuisance and potential resource for spin qubits. For example, fluctuating hyperfine fields limit $T_2^*$ $\sim$ 10 ns \cite{Petta2005} in GaAs spin qubits, leading to strongly damped Rabi oscillations \cite{Koppens2006}. On the other hand, electric field control of the hyperfine coupling constant $A$ features prominently in Kane's proposal \cite{Kane1998}. The hyperfine interaction between one electron (carrying spin operator $\vec{S}$ and orbital angular momentum operator $\vec{L}$) with many nuclei at positions $\vec{R}_k$ carrying spin $\vec{I}_k$ is described by the Hamiltonian \cite{Abragam1961}:
\begin{multline}
	\ts{H}{hf}=\frac{\mu_0}{4\pi}g_0 \mu_B\hbar\gamma_n\int d^3\vec{r} \\
	\Psi^*(\vec{r}) \sum_k \biggl[\frac{\vec{L}-\vec{S}}{|\vec{r}-\vec{R}_k|^3}
	+3\frac{[(\vec{r}-\vec{R}_k)\cdot\vec{S}](\vec{r}-\vec{R}_k)}{|\vec{r}-\vec{R}_k|^5}
	\\+\frac{8\pi}{3}\delta(\vec{r}-\vec{R}_k)\vec{S}\biggr]\cdot\vec{I}_k \psi(\vec{r}).
	\label{eq:fundamental_hf}
\end{multline} 
Here, $g_0$ is the bare electron $g$-factor, $\gamma_n$ is the nuclear gyromagnetic ratio, and $\psi(\vec{r})$ is the full electron wave function (not the effective mass envelope function $F_j(\mathbf{r})$). The last term with $\delta(\vec{r})$ is the Fermi contact term and is dominant for conduction electrons in both GaAs and Si; it is isotropic and as such its effects are immune to the relative orientation of applied magnetic field with crystalline axes. The magnetic dipole-dipole terms are usually smaller, but they can contribute to the dephasing of electron spin resonance of donors and QDs in Si at low-magnetic field~\cite{Witzel2007,Zhao2019}.

For the Fermi contact hyperfine interaction, we obtain
\begin{equation}
\ts{H}{hf,contact} = \sum_k \hbar A_k\vec{S}\cdot\vec{I}_k,
\label{eq:contactHF}
\end{equation}
where
\begin{equation}
A_k = \frac{\mu_0}{4\pi}g\mu_B\gamma_n \eta |\psi(\vec{R}_k)|^2.
\end{equation}
Here $\psi(\vec{R}_k)$ is the effective mass envelope wavefunction at each nucleus location and $\eta$ is the \emph{bunching factor}, which captures the microscopic overlap of the Bloch wavefunction with the nucleus.  The  envelope wavefunction is normalized, $\sum_k |\psi(\vec{R}_k)|^2=1$, where the sum is over all nuclear sites in the crystal.  For \isotope{P}{31} in Si, as well as \Si\ in Si and all Ga and Al nuclei in GaAs, $\eta$ has been both measured \cite{Feher1959,Paget_1977} and calculated \cite{assali2011hyperfine,Philippopoulos2020}; however for some species such as \Ge\ in SiGe, only estimates are available, typically from spin-qubit experiments~\cite{Kerckhoff_quadrupolar_2020}. 

The dynamics of the nuclei themselves, in particular the magnetic nuclear dipole-dipole interactions, is also of critical importance in determining how the nuclear spin bath evolves. In the frequent case that one QD electron overlaps with many nuclear spins, the hyperfine interaction behaves as an effective ``Overhauser'' magnetic field that the electron spin experiences, which fluctuates in time due to nuclear dynamics \cite{Taylor2007}. These effects are central to spin qubit dephasing and decoherence, and are discussed extensively in Section~\ref{Sec:Decoherence}.

\section{Spin-spin interactions} \label{Sec:2spins}
The most important physical mechanism leading to interactions between spin qubits is the exchange interaction.
Exchange results from a combination of Fermi statistics, electron tunneling, and Coulomb repulsion; some common notation is required to combine these aspects.  We must first define a many-particle basis, which is generally done in terms of single-particle basis functions $\phi_{m}(\vec{r})\chi_\sigma$, for spatial orbitals enumerated by $m$, spin $\sigma=\up,\down$, and position $\vec{r}$.  The spinor obeys $\chi_\sigma^\dag\chi_{\sigma'}^\nodag=\delta_{\sigma\sigma'}$.  Exchange depends on the Pauli exclusion principle, which means that the multiparticle wavefunction $\Psi_{\sigma_1\sigma_2\ldots}(\vec{r}_1,\vec{r}_2,\ldots)$ must be fully antisymmetric for arbitrarily labeled electrons 1,2,\ldots.  This may be formally assured via the use of a Slater determinant, i.e.
\begin{multline}
\Psi_{m_1\sigma_1,m_2\sigma_2,\ldots m_N\sigma_N}(\vec{r}_1,\vec{r}_2,\ldots,\vec{r}_N) =
\\ \frac{1}{\sqrt{N}}\begin{vmatrix}
	\phi_{m_1}(\vec{r}_1)\chi_{\sigma_1} & \phi_{m_2}(\vec{r}_1)\chi_{m_2} & \dots & 
								\phi_{m_N}(\vec{r}_1)\chi_{\sigma_N} \\
	\phi_{m_1}(\vec{r}_2)\chi_{\sigma_1} & \phi_{m_2}(\vec{r}_2)\chi_{m_2} & \dots & 
								\phi_{m_N}(\vec{r}_2)\chi_{\sigma_N} \\
	\vdots & \vdots & \ddots & \vdots \\
	\phi_{m_1}(\vec{r}_N)\chi_{\sigma_1} & \phi_{m_2}(\vec{r}_N)\chi_{\sigma_2} & \dots & 
								\phi_{m_N}(\vec{r}_N)\chi_{\sigma_N}
\end{vmatrix}.
\end{multline}

Equivalently, this wavefunction may be described by anticommuting annihilation operators $c_{m\sigma}$. The operator $c_{m\sigma}^\dag$ creates a conduction electron in orbital state $\phi_{m}(\vec{r})$ and spin state $\sigma$ and we write
\begin{equation}
	\ket{\Psi_{m_1\sigma_1,m_2\sigma_2,\ldots m_N\sigma_N}}=
		c_{m_1\sigma_1}^\dag c_{m_2\sigma_2}^\dag \ldots c_{m_N\sigma_N}^\dag \vac,
\end{equation}
where $\vac$ is the vacuum containing no electrons.  
Using this notation, the general many-body Hamiltonian within the EMA approximation [\refeq{eq:effmass}] reduces to
\begin{multline}
	H = \\
		\sum_\sigma \sum_{mn} T_{mn} c_{m\sigma}^\dag c_{n\sigma}^\nodag
+
			\frac{1}{2}\sum_{\sigma_1\sigma_2}\sum_{mn\ell p} V_{mn\ell p}
			c_{m\sigma_1}^\dag c_{n\sigma_2}^\dag c_{\ell\sigma_2}^\nodag c_{p\sigma_1}^\nodag,
\label{eq:Ham_Nelec}
\end{multline}
with single-particle kinetic and potential energy integral
\begin{align}
	T_{mn} = \int d^3\vec{r} \ \phi^*_{m}(\vec{r}) \biggl[
	-\frac{\hbar^2}{2}\nabla \cdot (\boldsymbol{\beta}\cdot \nabla) + U(\vec{r}) 
	\biggr]\phi_{n}(\vec{r}).
\end{align}
Here $\boldsymbol{\beta} = (m_x^{-1},m_y^{-1},m_z^{-1})$ gives the inverse effective masses and $U(\vec{r})$ is the externally-applied potential due to gate biasing and built-in electric fields.  The general Coulomb integral is
\begin{multline}
	V_{mn\ell p}	=\\ 
	\int d^3\vec{r}_1 d^3\vec{r}_2 \phi_{m}^*(\vec{r}_1)\phi_{n}^*(\vec{r}_2) 
	\frac{e^2}{4\pi\epsilon_r\epsilon_0|\vec{r}_1-\vec{r}_2|}
	\phi_{\ell}(\vec{r}_2)\phi_{p}(\vec{r}_1),
	\label{eq:gen_Coulomb}
\end{multline}
where $\epsilon_r$ is the semiconductor relative permittivity (which may in general depend on position); any image effects due to metal gates are ignored for simplicity. 
Note that both of these integrals are independent of spin.

This notation allows us to distinguish two flavors of the exchange interaction, direct and kinetic.  Direct exchange is simply illustrated for two orbitals, perhaps labeled $1$ and $2$, with high spatial overlap, such as orbital states in a common dot or donor.  If we ask how the Coulomb interaction impacts the energy of a doubly-occupied orbital state, the dominant terms of the Coulomb integral in our single-particle basis can then be broken up into the direct Coulomb term $\mathcal{K}$, corresponding to the case $m\ne n$, $m=p$, and $n=\ell$, such as $V_{1221}$; and the direct exchange term $\mathcal{J}$, corresponding to $m=n$ and $\ell=p$, such as $V_{1122}$. These two terms separate a pair of two-electron energy levels by the energy $\mathcal{K}-\mathcal{J}/2$ for triplet spin states (spatially antisymmetric, spin symmetric), and by $\mathcal{K}+3\mathcal{J}/2$ for singlet spin states (spatially symmetric, spin antisymmetric).  Hence, the combination of Coulomb repulsion and Pauli exclusion \emph{raises} the energy of the singlet relative to the triplet state by the amount $2\mathcal{J}$.  Although this direct exchange term is important, leading in particular to Hund's rule when filling orbitals, spin qubit control mostly leverages the distinct and more highly-controllable kinetic exchange interaction, which is due to the effect of the Pauli exclusion principle on the (spin-independent) $T_{mn}$ and $\mathcal{K}$ terms.  We address this interaction in the next section.

\subsection{Kinetic exchange in the Fermi-Hubbard hopping model}  \label{Subsec:2spins:Exchange}

Kinetic exchange is most easily introduced using the simplified Fermi-Hubbard hopping model where we presume that electrons are rather tightly bound into their single-electron orbitals $\phi_{j}(\vec{r})$. Here, $\phi_{j}(\vec{r})$ describe ground-state occupation in dot $j$, with negligible dot-to-dot Coulomb interactions ($\mathcal{K}$) and dot-to-dot direct exchange interactions ($\mathcal{J}_{jk}$), as discussed above. In this approximation, the only relevant Coulomb interaction is the on-site Coulomb interaction with magnitude $U=V_{jjjj}$ and the kinetic energy transition matrix $T_{jk}$ is described in terms of a constant tunnel coupling $t_c=T_{12}$ between sites 1 and 2, and voltage-controlled chemical potentials $\mu_j$ for the diagonal elements $T_{jj}$.  Constraining the discussion to two electrically charged spin-1/2 particles (such as electrons) filling two sites, and neglecting any magnetic field at first for simplicity, the Fermi-Hubbard Hamiltonian is
\begin{multline}
	H_{\text{FH}}=
	\sum_{\sigma=\uparrow,\downarrow} \biggl[\sum_{j=1,2} \mu_j^\nodag c_{j\sigma}^\dag c_{j\sigma}^\nodag
	+t_c (c_{1\sigma}^\dag c_{2\sigma}^\nodag+c_{2\sigma}^\dag c_{1\sigma}^\nodag)\biggr]
	\\	+\sum_{j=1,2} U c_{j\uparrow}^\dag c_{j\uparrow}^\nodag 
	c_{j\downarrow}^\dag c_{j\downarrow}^\nodag.
	\label{eq:FH}
\end{multline}

\begin{figure}
    \centering
    \includegraphics[width=\columnwidth]{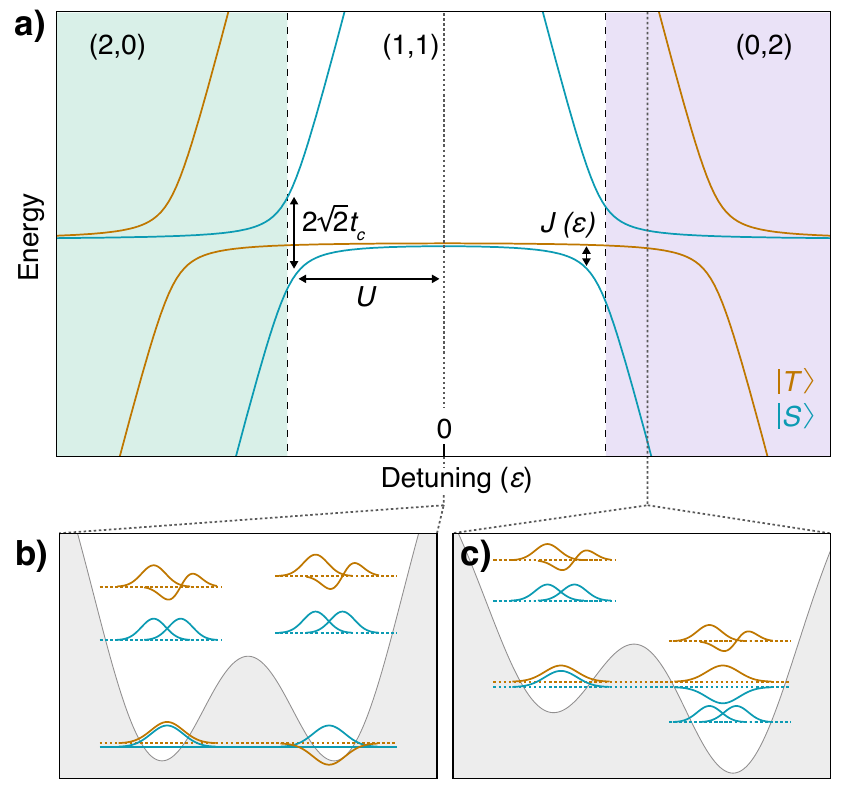}
    \caption{Energy levels, exchange coupling $J$, and wavefunctions in a DQD with two particles.  (a) Two-particle energy levels as a function of level detuning $\epsilon$. Tunnel coupling $t_c$ leads to level repulsion between the singlet states (blue) where the on-site Coulomb energy $U$ equals $\pm\epsilon$. $J$ is the energy difference between the low-energy spin singlet and the spin triplets (red). Wavefunctions for (b) the symmetric (i.e. $\epsilon$ = 0) and (c) detuned DQD.}
    \label{fig:exchange-levels}
\end{figure}

The possible (linearly independent) quantum states described by \refeq{eq:FH} can be characterized by their charge and spin configurations.  For two charges in two sites, the possible charge configurations are $(2,0)$, $(1,1)$, and $(0,2)$  where $(n_i,n_j)$ indicates the numbers of particles on sites $1$ and $2$.
The exclusion principle allows but one spin configuration for $(2,0)$ and $(0,2)$ with one spin up and one spin down particle, and hence total spin zero (spin singlet). For $(1,1)$ there are four possibilities, one spin singlet state and three spin triplet states. We may choose our energy-zero such that $\mu_1+\mu_2=0$ and define the detuning $\mu_1-\mu_2=\epsilon$. We therefore arrive at
\begin{align}
\label{eq:DQDcharge}
    H=&\left(U-\epsilon\right)|{\rm S}(0,2)\rangle\!\langle {\rm S}(0,2)|
      +\left(U+\epsilon\right)|{\rm S}(2,0)\rangle\!\langle {\rm S}(2,0)|\nonumber\\
     &+\sqrt{2}t_c\left(|{\rm S}(2,0)\rangle\!\langle {\rm S}(1,1)|
                       +|{\rm S}(0,2)\rangle\!\langle {\rm S}(1,1)|+\text{h.c.}\right),
\end{align}
where S indicates that all three states occurring in this Hamiltonian are spin singlets, while the three spin triplet states are at zero energy.
Diagonalizing this Hamiltonian for $|t_c|\ll U\pm \epsilon$ 
and $|\epsilon|<U$,
one finds a low-energy hybridized singlet state
\begin{align}
\label{eq:DQDsinglet}
|{\rm S}\rangle \simeq |{\rm S}(1,1)\rangle 
- \frac{\sqrt{2}t_c}{U-\epsilon}|{\rm S}(0,2)\rangle
- \frac{\sqrt{2}t_c}{U+\epsilon}|{\rm S}(2,0)\rangle ,
\end{align}
up to terms of order $t_c^2 / (U\pm \epsilon)^2$,
with energy $-J$ where
\begin{align}\label{eq:exchange-Hubbard}
    J = \frac{4Ut_c^2}{U^2-\epsilon^2} + O\left(\frac{t_c^3}{ (U\pm \epsilon)^3}\right)
\end{align}
represents the exchange coupling.
Virtual hopping between the two sites \emph{lowers} the energy of the lowest spin singlet by $J$ relative to the spin triplet energy [Fig.~\ref{fig:exchange-levels}]; this is the kinetic exchange interaction.

The other singlet states are at higher energies, separated by roughly $U\pm \epsilon$. Excited (2,0) and (0,2) triplets (discussed in 
\refsec{Subsec:Principles:Coulomb}) are at similarly high energies. Neglecting those higher states one finds as the effective Hamiltonian for the $(1,1)$ charge configuration
\begin{align}
\label{eq:exchange1}
H=-J|{\rm S}\rangle\!\langle {\rm S}|=\frac{J}{2}\left(S^2-2\right)=J\vec{S}_i\cdot\vec{S}_j + \textrm{const.},
\end{align}
where $\vec{S}=\vec{S}_i+\vec{S}_j$ denotes the total spin of sites $i$ and $j$, and the constant can be omitted to yield Eq.~\eqref{eq:Heisenberg}.

\subsection{Heitler-London and Hund-Mulliken models}  \label{Subsec:2spins:HLHM}

To gain a more microscopic understanding of the exchange $J$ in Eq.~\eqref{eq:Heisenberg} as well as the parameters of the Fermi-Hubbard model \eqref{eq:FH}, 
the localization of electrons to a single site realized by a QD in a 2D electron system can be modelled with high accuracy with a harmonic potential $V(\mathbf{r})$ = $m\omega_0^2 (x^2+y^2)/2$. Here, $\hbar\omega_0$ is the orbital level spacing of the QD and $\mathbf{r}=(x,y)$.  The exchange coupling between spins of electrons residing in two adjacent QDs $i$ and $j$ can then be modelled using a quartic potential $V(\mathbf{r})$ which is locally harmonic in its two minima, with $d$ the inter-dot spacing.  The exchange energy can be obtained as the energy difference of spin singlet and triplet states for the two-electron orbital Hamiltonian including the Coulomb interaction,
\begin{align}
    H =& \sum_{i=1,2}\left(\frac{1}{2m}\left(\mathbf{p}_i
    -e\mathbf{A}\left(\mathbf{r}_i\right)\right)^2
    +e \mathbf{r}_i\cdot \mathbf{E} + V(\mathbf{r}_i)
  \right)\nonumber\\
    &+\frac{e^2}{4\pi \epsilon_r\epsilon_0 |\mathbf{r}_1-\mathbf{r}_2|},
\end{align}
where $\mathbf{E}$,  $\mathbf{B}$, and $\mathbf{A}$ denote the electric and magnetic fields, and the vector potential.

The Heitler-London (HL) method evaluates the energies of the spin singlet (triplet) trial wavefunctions with antisymmetric (symmetric) spin state $|{\rm S}\rangle$ ($|{\rm T}_\alpha\rangle$) and corresponding symmetric (antisymmetric) orbital wavefunctions in the $(1,1)$ charge configuration,
\begin{align}
    |\Psi_\pm\rangle = \frac{1}{\sqrt{2(1\pm \Sigma^2)}}\left(|ij\rangle \pm|ji\rangle\right),
\end{align}
in order to guarantee an overall antisymmetric wavefunction under particle exchange as required for Fermions.
Here $\sigma=\langle i|j\rangle$ denotes the overlap between the single-particle ground-state wavefunctions of the electron localized on adjacent sites $i$ and $j\neq i$.
The exchange energy $J=\langle \Psi_-|H|\Psi_-\rangle - \langle \Psi_+|H|\Psi_+\rangle$ decays exponentially with increasing interdot spacing $d$ and magnetic field $B$ for large $B$.  The sign of $J$ can correspond to antiferromagnetic ($J>0$) or ferromagnetic ($J<0$) coupling.  While $J>0$ is obligatory for $B=0$ for a two-electron system, $J$ can display a sign change from positive to negative at finite $B>0$ \cite{Burkard1999,Zumbuhl2004}, or in multi-electron QDs \cite{Martins2017,Deng2018,Malinowski2019}.

The main shortcomings of the HL method are that it does not take into account doubly occupied sites and that, while it provides the exchange energy for the Heisenberg Hamiltonian \eqref{eq:Heisenberg}, it cannot deliver the parameters of the Hubbard model \eqref{eq:FH}. 

The Hund-Mulliken (HM) or molecular-orbital model extends the HL model to include doubly occupied sites by expanding the Hilbert space with two spin singlet states with orbital wavefunctions $|ii\rangle$ and $|jj\rangle$ corresponding to the $(2,0)$ and $(0,2)$ charge states \cite{Burkard1999}. The single-particle states $|i\rangle$ and $|j\rangle$ are first orthonormalized to form a convenient basis.  The exchange energy is found as
\begin{align}
    J = \frac{1}{2}\left(\sqrt{U^2+\frac{16t_c^2}{U^2}}-U\right)-2\mathcal{J}
    \approx \frac{4 t_c^2}{U} -2\mathcal{J},
\end{align}
where $U_i=U_j=U>0$ and $t_{ij}=t_c$ correspond to the effective on-site Coulomb and tunneling matrix elements in \refeq{eq:FH}, and $\mathcal{J}$ is the direct exchange contribution due to the long-range Coulomb interaction.
The approximation holds in the Hubbard limit $t_c\ll U$.  If direct exchange effects can be neglected we recover the result \refeq{eq:exchange-Hubbard} for $\epsilon=0$.

Extensions of the HL approach include the effect of an inhomogeneous field \cite{Sousa2001}, s-p hybridization of single-dot orbitals \cite{Burkard1999}, and a symmetry-breaking variational approach \cite{Yannouleas2002}.  The HM model has been extended to include on-site triplet states \cite{White2018}.  
Spin-orbit coupling, in the presence of a magnetic field, can render the exchange coupling anisotropic by contributing a Dzyaloshinskii-Moriya interaction $\mathbf{D}\cdot(\mathbf{S}_i\times\mathbf{S}_j)$ to the Hamiltonian \cite{Kavokin2001,Kavokin2004,Chutia2006,Baruffa2010a,Baruffa2010b,Liu2018}.   

\subsection{FCI calculations of exchange}  \label{Subsec:2spins:FCI}

The approximate analytic models described above give important insights into the exchange interaction, but do not completely capture the impact of band structure and electrostatic confinement. These can be fully accounted for by solving the complete Hamiltonian of \refeq{eq:Ham_Nelec}, which in general must be done numerically \cite{reimann_electronic_2002}. The full configuration interaction (FCI) method is an efficient and systematic way to solve multi-electron Hamiltonians and is thus an invaluable tool for understanding exchange interactions in realistic spin qubit devices.

In the FCI approach, first developed for quantum chemistry \cite{szabo_modern_1996}, a set of $2K$ single-particle spin orbital basis states $\{\phi_{m}(\vec{r})\chi_\sigma\}$ is chosen which are product states of real-space basis functions and spinors; the former may be convenient analytic functions or eigenstates of the single-particle operator $T$ in \refeq{eq:Ham_Nelec} \cite{Rontani_JCehmPhys_2006,Gyure2021,joecker_full_2020}.
Often, $K\approx 20-40$ orbitals are needed to obtain fully converged dot or donor states.
From this single-particle basis, the set of all possible $N$-particle Slater determinants is constructed, which is used as the multi-electron basis in which \refeq{eq:Ham_Nelec} is diagonalized. All matrix elements of the Hamiltonian in this basis can be expressed solely with single-electron terms and two-electron Coulomb integrals in Eq.~\eqref{eq:gen_Coulomb}, which can be computed using the single-particle states $\phi_{m}$. This ensures that all exchange and correlation effects are included, provided a large enough single-particle basis is used.

The resulting $N$-electron eigenstates are linear combinations of Slater determinants and (in the absence of spin-orbit or magnetic gradients) can be classified by their spin properties, including total spin $S^2$ and spin projection $S^z$.
For instance, exchange $J$ can be computed from the energy splitting between the lowest two-electron singlet and triplet eigenstates.
As the total number of Slater determinants scales as $\binom{2K}{N}$, FCI calculations become intractable for large $N$; however, realistic two- and three-electron systems are well within the capabilities of modern computers.

\subsection{Discussion of theoretical approaches for calculating exchange} 
\label{Subsec:2spins:Compare}

\begin{figure}[t]
	\centering
	\includegraphics[width=\columnwidth]{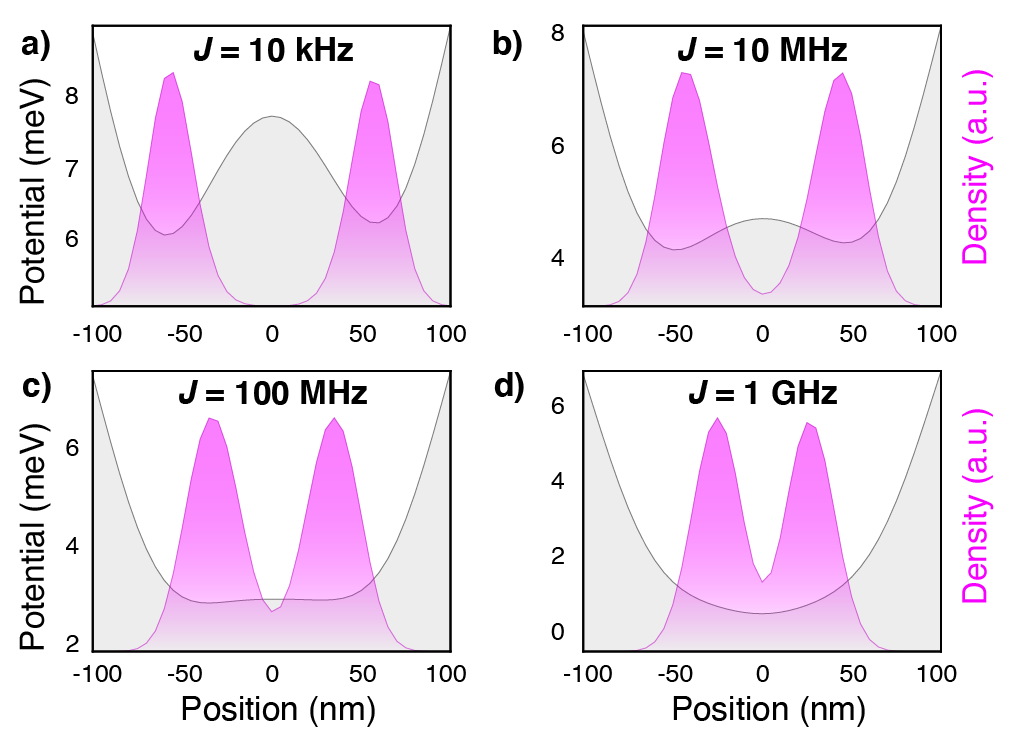}
	\caption{(a)-(d) Simulation of the change in the DQD potential (gray) and electron density (pink) as the interdot barrier is lowered to increase exchange from approximately 10 kHz to 1 GHz. The potential is generated by solving the Poisson equation for a representative Si/SiGe DQD, which is then used in an FCI calculation to obtain the wave functions and $J$ ($K=30$ single-particle eigenstates are used to construct the basis). At practically useful multi-MHz levels of exchange, the electrostatic barrier vanishes and the electrons shift closer together, separated primarily by their Coulomb repulsion.}
	\label{fig:fci-exchange}
\end{figure}

The most basic model for describing controlled exchange is the Fermi-Hubbard hopping model, \refeq{eq:FH}, with constant $U$, detuning $\epsilon$ taken as a linear function of gate voltage, and tunnel coupling $t_c$ taken as an exponential function of gate voltage.  The model makes predictions for exchange as a function of voltage that are not well replicated by experiments, with the largest deviations at high values of exchange~\cite{Reed2016}.  This is unsurprising, given the change in character of tunneling barriers as dots combine shown in Fig.~\ref{fig:fci-exchange}.  Nonetheless, this model is of high value for providing qualitative understanding in exchange-based experiment design.

The HL model is surely more quantitative, but has some limitations on its validity \cite{Calderon2006,Saraiva2007}; in the weak interdot coupling limit the HL results agree qualitatively with exact diagonalization results with some quantitative modifications \cite{Melnikov2006}. Experimental results in laterally coupled vertical DQD show that the Heitler–London model forms a good approximation of the two-electron wavefunction \cite{Wiel2006}.

Since the HM method takes into account double occupation of sites, its range of validity in charge configuration space is greater than that of the HL approach. The HM predictions have been experimentally verified in \onlinecite{Hatano2008}.
The validity of the single-particle description even for multi-electron QDs has been discussed in \onlinecite{Hu2001,Bakker2015}.  A comparison of the Hartree-Fock, HM, Heisenberg, and Hubbard models using a double-well potential consisting of a linear combination of Gaussians can be found in \onlinecite{Hu2000}. 

The determination of $J$ with high accuracy and predictive power is possible with FCI \cite{Hu2001}. Since the magnitude and sensitivity of $J$ depend on both material properties (such as the effective mass and permittivity) and device electrostatics, the accuracy depends in turn on accurate modeling of the device structure.
The sensitivity of FCI to material parameters reveals phenomena which may not be obvious from site-based methods, e.g., the specific charge configurations for ``sweet spots'' where a qubit is resilient against charge noise \cite{Vion2002}.

As an example, in Fig.~\ref{fig:fci-exchange} we compare the numerically computed electrostatic potential and electron density in a typical Si DQD as $J$ increases.
Qualitatively we expect to modulate exchange by lowering the tunnel barrier between well-separated electrons; however, in practice the reduced confinement displaces the electrons significantly towards each other as exchange is activated. 
Indeed, at large $J$ no external potential barrier between the electrons exists at all, and the Coulomb repulsion itself acts as the effective barrier; hence, the notion of a separable dot basis does not hold as the electron states transition smoothly between a double- and single-dot limit.
Such effects are particularly important when considering simultaneous exchange between multiple pairs of electrons~\cite{pan_resonant_2020,Qiao2020,vanDiepen2021}, which requires coordinated spatial displacements; describing such effects accurately within site-based approaches like the Fermi-Hubbard, Heitler-London and Hund-Milliken models discussed above requires major modifications.

More generally, numerical FCI calculations are important for describing the effects of electron-electron interactions on QD level structure, such as Wigner molecule behavior~\cite{ercan_strong_2021}.
Similarly, such calculations can capture the impact on $J$ of locally-sensitive parameters such as valley splitting and spin-orbit coupling. FCI calculations have revealed the complex dependency of exchange couplings in donors~\cite{gamble_multivalley_2015,tankasala_two-electron_2018} and QDs~\cite{Hu2001,Gyure2021,nielsen_many-electron_2012}, and been used to study charge noise sensitivity \cite{shim_barrier_2018} and mediated exchange in multi-electron dots~\cite{nielsen_six-electron_2013,deng_interplay_2020}.

\subsection{Pauli spin blockade} 
\label{Subsec:2spins:PSB}

\begin{figure}[t]
	\centering
	\includegraphics[width=\columnwidth]{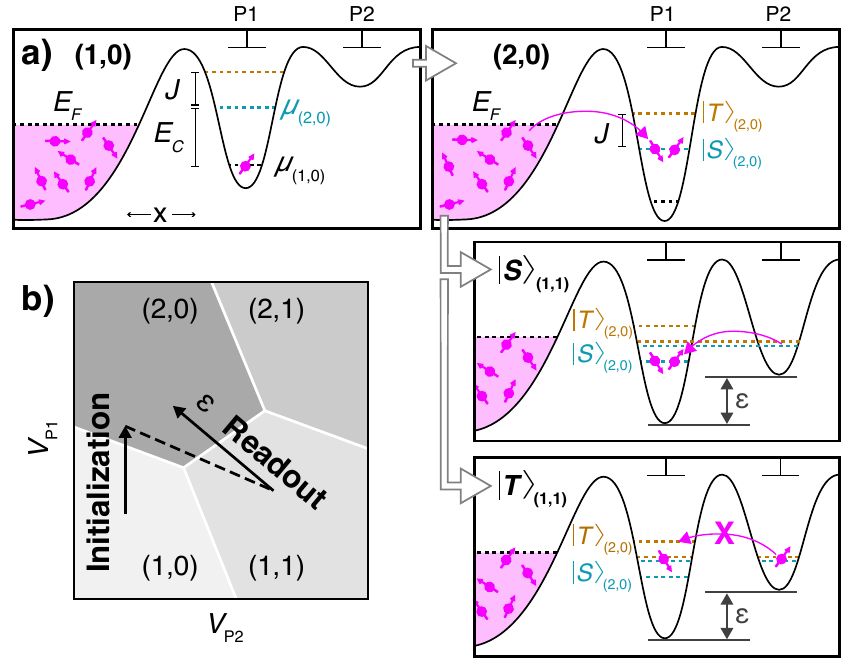}
	\caption{PSB in a DQD. (a) (2,0) singlet initialization occurs by biasing the left QD such that $\mu_{S_{(2,0)}}<E_F<\mu_{T_{(2,0)}}$. Qubit operations and readout are then performed by changing bias positions along the (2,0)-(1,1) detuning axis. Readout is implemented by detuning such that the singlet ground state is (2,0). Interdot tunneling is prohibited by PSB for the spin triplet state (lower right panel), allowing spin-to-charge conversion. (b) Charge stability diagram in the vicinity of the (2,0)-(1,1) anticrossing.}
	\label{fig:psb}
\end{figure}

An important manifestation of exchange, well understood from the Fermi-Hubbard model discussed in \refsec{Subsec:2spins:Exchange}, is Pauli spin blockade (PSB). As illustrated in \reffig{fig:psb}, the ground state of a two-electron DQD can be either the $(1,1)$ or $(2,0)$ charge configuration\footnote{We choose the (1,1)-(2,0) charge boundary for specificity, though the following applies also to dynamics at the (1,1)-(0,2) transition.} depending on the DQD level detuning $\epsilon=\mu_1-\mu_2$. As discussed in \refsec{Subsec:Principles:Coulomb}, the $(2,0)$ ground state is a spin singlet. Thus, when the detuning $\epsilon$ satisfies $-U-J^{\rm max}<\epsilon<-U$, singlets occupy the $(2,0)$ charge state, but the triplet spin states remain in the (1,1) configuration (Fig.~\ref{fig:exchange-levels}). The maximum value of the exchange coupling, $J^{\rm max}$,  depends on the energy separation between the ground and first excited states in the left QD. In GaAs QDs, this spacing typically depends on the orbital energy spacing, which can be of order meV. In Si QDs, this energy spacing can depend on the valley splitting, which can be tens to hundreds of $\mu$eV, or the orbital energy spacing, depending on the number of electrons. This phenomenon, wherein spin states map onto distinct charge configurations, constitutes PSB.

The experimental realization and confirmation of PSB first occurred in vertical GaAs DQDs, which are fabricated by etching semiconductor heterostructures~\cite{Kouwenhoven1998Dots}. Electrical transport measurements in the first experiments provided evidence of current rectification via PSB~\cite{ono_current_2002}. Even at this early stage, these experiments were motivated by the possibility of using electron spins as quantum bits. Following the initial demonstration of PSB, pulsed-gate measurements showed that the triplet-singlet relaxation time was much longer than charge relaxation times, confirming the suitability of singlet and triplet states for quantum information purposes~\cite{fujisawa_allowed_2002}. PSB was later observed in planar GaAs DQDs with higher electron occupations \cite{Johnson2005a} and used in pulsed-gate experiments to measure triplet-singlet relaxation as a function of magnetic field \cite{Johnson2005}.

\begin{figure*}[t]
	\centering
	\includegraphics[width=2\columnwidth]{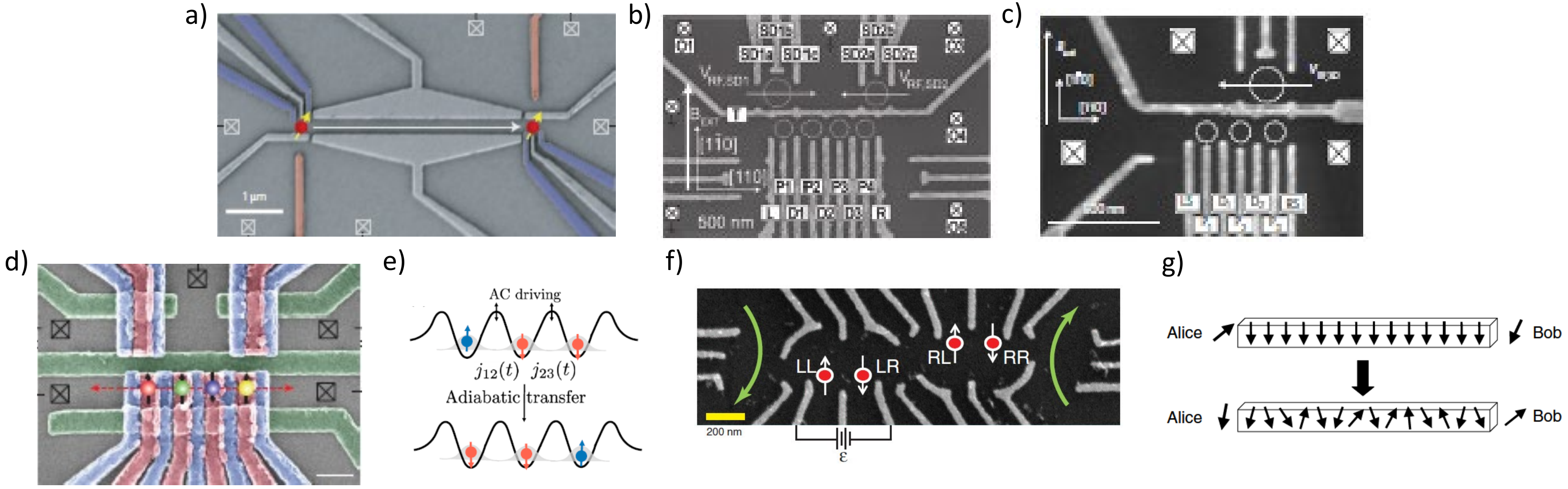}
	\caption{Various approaches for achieving long-range spin coupling: a) Surface acoustic waves \cite{Bertrand2016}, b) Charge transport \cite{Baart2016single}, c) Superexchange \cite{Baart2017}, d) Spin-SWAPs \cite{Kandel2019}, e) Spin-CTAP \cite{GullansCTAP2020}, f) Capacitive coupling \cite{Shulman2012}, and g) Coupling through a spin chain \cite{Bose2003}.}
	\label{fig:longrange-spin}
\end{figure*}

PSB is an essential tool for the initialization and readout of many types of spin qubits. Pairs of electrons in the same QD can easily be initialized as spin singlets by enabling electron tunneling between that dot and a nearby electron reservoir~\cite{Petta2005,Botzem2018,Maune2012}. After initialization, spin singlets can be separated via interdot tunneling into separate dots. If the two electrons are separated adiabatically in the presence of a magnetic gradient, the singlet transitions to a spin-zero product state, thus enabling the straightforward creation of product states~\cite{Petta2005,Foletti2009}.

Following evolution of the spin states, these steps can be reversed to project a pair of electrons onto the singlet-triplet basis. A simple readout method involves rapidly pulsing the detuning to $-U-J^{\rm max}<\epsilon<-U$ after manipulation. In this state, the singlet-triplet energy splitting is extremely sensitive to environmental charge noise. The joint spin state dephases rapidly, and an external charge detector, such as a QPC~\cite{Petta2005} or QD~\cite{Barthel2009}, can extract information about the charge state of the DQD using one of the techniques discussed in \refsec{Subsec:Principles:Dots_and_sensors}, thus projecting its spin state. If the detuning is pulsed adiabatically with respect to any magnetic gradients, one spin-zero product state maps to the singlet, and all other spin states map to triplet. Generally, PSB readout is straightforward to implement, and can enable rapid ($\mu$s-scale or shorter) and high-fidelity ($>98\%$) readout fidelity of different qubit types~\cite{Reilly2007FastRF,Barthel2009,Barthel2010,Elliot2020,Noiri2020RF,Borjans_sensing_2021}.

\subsection{Long-range couplers}  \label{Subsec:2spins:LR}

Despite its simplicity and speed, Heisenberg exchange only directly couples nearest-neighbor spins, as it relies on wavefunction overlap. The requirement for close proximity of the spins [see Fig. \ref{fig:fci-exchange}] poses challenges for the design, fabrication, and operation of large-scale spin-based quantum information processors. This section reviews the various approaches for creating an effective long-range coupling between distant spins. Many of these approaches are in the early stages of development. As such, the experimental characterization of quantum state transfer fidelities using protocols such as randomized benchmarking and gate set tomography is one important future avenue of research in this area.

\subsubsection{Spin transport, spin SWAPs, and spin-CTAP} \label{Subsubsec:2spins:LR:SWAP}

Perhaps one of the most conceptually straightforward ways to achieve long-range connectivity is to physically transport qubits across a device. The two main approaches that have been investigated include using a surface acoustic wave (SAW) as a conveyor belt for electrons and ``bucket brigade'' style single electron shuttling. SAWs are travelling acoustic waves that are typically generated in piezoelectric materials, such as GaAs, using interdigitated transducers \cite{Datta1986}. Early experiments in GaAs/AlGaAs heterostructures demonstrated single charge \cite{McNeil2011}  and spin \cite{Bertrand2016} transport between two QDs [Fig. \ref{fig:longrange-spin}(a)]. Spin state transport using SAWs has recently been demonstrated with high fidelity \cite{JadotShuttle2021}. SAW implementations of spin state transport may have long term limitations due to power dissipation, SAW directionality, and the relatively large size requirements of SAW transducers. Some of these scaling challenges may be alleviated using charge and spin shuttling.

Charge shuttling involves moving an electron through an array of QDs by periodically modulating the confinement potential. Early experimental implementations of charge shuttling in superconducting devices were motivated by the metrological desire to have a high-speed current standard \cite{Keller1999}. A theoretical proposal by Taylor \textit{et al.} suggested using a bucket brigade charge shuttle to transfer quantum information between semiconductor spin qubits \cite{Taylor2005}. To achieve charge transfer, the detuning between adjacent QDs is ramped across the interdot charge transition. Early experiments in GaAs demonstrated spin shuttling [Fig. \ref{fig:longrange-spin}(b)] \cite{Baart2016single,Fujita2017}. In Si, charge shuttling has been achieved in a linear array of 9 QDs \cite{Mills2019}, and spin shuttling has been quantitatively characterized in a SiMOS DQD \cite{Yoneda2020shuttle}.  Conveyor-mode charge shuttling through a 400~nm long open channel defined by a series of electrodes has been demonstrated in \onlinecite{seidler2021}.

Another approach for achieving spin state transfer without the physical transfer of charges is to use a sequence of pairwise spin SWAPs to couple spatially separated spin qubits [Fig. \ref{fig:longrange-spin}(d)]. Spin SWAPs can be achieved using exchange pulses, as proposed in the original Loss-DiVincenzo proposal \cite{Loss1998,Petta2005,kandel2021adiabatic}. Spin SWAPs can also be implemented in systems with a magnetic field gradient by periodically modulating the exchange coupling \cite{Nichol2017}. First demonstrations were achieved in GaAs, with more recent high fidelity demonstrations having been achieved in Si/SiGe QDs  \cite{Nichol2017,Sigillito2019SWAP}.

\onlinecite{Greentree2004} proposed using coherent transport via adiabatic passage (CTAP), in analogy to stimulated Raman adiabatic passage (STIRAP) commonly used in atomic physics\cite{Vitanov2017}, to achieve charge transfer in QD arrays. Theoretically, the idea has been extended to spin by \onlinecite{GullansCTAP2020}, where it was shown that time-varying exchange pulses can be used to transfer spin states with high fidelity [Fig. \ref{fig:longrange-spin}(e)]. Experimental results by \onlinecite{kandel2021adiabatic} in GaAs QD arrays give a proof of concept that such adiabatic protocols are viable.

\subsubsection{Superexchange} \label{Subsubsec:2spins:LR:SuperJ}

To create an effective long-range exchange coupling between distant spins, sometimes referred to as superexchange, an additional QD-based mediator (typically a single QD or a chain of occupied QDs) is physically interposed between the two spins of interest. Through a process involving a virtual occupation or excitation of the mediator, the spins coupled to the mediator experience an effective, indirect exchange interaction \cite{Bose2003,Friesen2007}.

When two electrons are coupled to a single QD mediator [Fig. \ref{fig:longrange-spin}(c)], they can experience an effective tunnel coupling, which depends on the electrochemical potential of the lowest unoccupied level of the mediator, through a virtual tunneling process ~\cite{Braakman2013,Loss1998}. This virtual tunneling process for electrons also creates a virtual exchange interaction for spin states. Although the occupation of the inner QD never physically changes, this scenario creates an indirect coupling between the outer QDs, which preserves the coherence of both charge~\cite{Braakman2013} and spin~\cite{Baart2017,Malinowski2019,Chan2021} states. Direct, coherent spin exchange with mediator electrons is also possible in a multiply-occupied QD mediator~\cite{Malinowski2019}.

Superexchange can also occur with a multi-QD mediator~\cite{Qiao2021Superexchange}. One of most commonly studied systems, which is predicted to exhibit superexchange, is an extended, strongly-coupled spin chain ~\cite{Wojcik2005,Campos2006,Oh2010}, to which two end spins are weakly coupled. The use of a spin chain as a long range coupler of spins, also referred to as the ``spin bus'', has been examined by \onlinecite{Bose2003}, \onlinecite{Bose2007}, and extensively by \onlinecite{Friesen2007}; these works show that a series chain of $N$ QDs with nearest neighbor exchange coupling $J$ may provide an effective end-to-end exchange coupling of $J$/$\sqrt{N}$ [Fig. \ref{fig:longrange-spin}(g)].

\subsubsection{Capacitive and electric dipole-dipole couplings}

Spin-qubit encodings with a charge-qubit character offer a natural coupling scheme with more reach than exchange: 
the electric field created by charge displacement in one qubit can be used to control the state by displacing the charge of another qubit [Fig. \ref{fig:longrange-spin}(f)]. At short range, this is effectively a quantum cross-capacitance effect; at larger distances, it has the character of an
electrically mediated effective dipole-dipole coupling.  It translates to a spin coupling due to exchange, field gradients or spin-orbit~\cite{Taylor2005,Stepanenko2007,Shulman2012,Cayao_PRB_2020}, or due to the hyperfine splitting between electrons and nuclei.  The latter effect may benefit the scaling of donor systems, since the electric dipole of a donor impurity may be “stretched” by the action of a gate above the device, enabling electric control of a long-distance dipole-dipole coupling.  Since this long-range coupling has a weak spatial dependence in comparison to exchange, it may allow donor devices to be fabricated through a controlled ion implantation process, with the inevitable placement straggle compensated for by gate calibration~\cite{tosi_silicon_2017}.  Coupling a donor to a dot may offer similar advantages~\cite{harvey-collard_coherent_2017}.  Such spin-relevant capacitive interactions are most effective when coupling to microwave excitations in a resonator, which we address in the next subsection and in Sec.~\ref{Sec:Hybrid}.

\subsubsection{Cavity QED} \label{Subsubsec:2spins:LR:CQED}

Three sets of experiments in 2004 demonstrated coherent coupling of solid-state qubits to photons, opening the door to long-range qubit coupling approaches employing photons in the microwave \cite{Wallraff2004} and optical regimes \cite{Yoshie2004,Reithmaier2004}. Long-range coupling of two superconducting qubits using a microwave cavity was achieved shortly thereafter \cite{Majer2007,Sillanpaa2007}. The concept of a cavity-bus for coupling superconducting qubits is now widespread~\cite{Blais2021}. Concepts for coupling spin qubits to cavities date as far back at 1999 \cite{Imamoglu1999}, with a resurgence of theoretical activity taking place again in 2004--2007 \cite{Childress2004,Jin2012,Trif2008cQED,Burkard2006}. Given the explosive growth of this area of quantum information science, we devote Sec. \ref{Sec:Hybrid} to a review of progress in QD cQED and its potential for providing long-range spin-spin couplings for qubits.  We also note for completeness various proposals and experiments demonstrating coupling of superconducting qubits to phonons, an area of which is ripe for exploration using QDs~\cite{Gustafsson2014}.

\section{Quantum gates and quantum circuits} \label{Sec:Gates}
Over the last two decades, there has been immense progress developing spin qubit technologies using the interactions and building blocks discussed in the previous sections.  In this section, we delve into the theoretical and experimental status of the qubit types introduced in \refsec{Sec:Basics}. For each qubit type, we discuss how initialization and readout have been physically implemented, strategies followed for performing single- and two-qubit gate operations, and the current status of gate fidelities.

For comparative fidelity in this review, we put particular emphasis on randomized benchmarking (RB).  The RB experiment consists of random sequences of quantum gates $C_RC_{N}\ldots C_{2}C_1$ applied to an initial state, where the $(N+1)^{\text{th}}$ ``recovery" gate $C_R$ is chosen so that each sequence would, in the absence of error, have the logical action of identity \cite{Magesan2011}.  The $C_j$s are drawn from the Clifford group, the group of gates which transform any multi-qubit Pauli-operator $P$ (as $C_j^\dag P C_j$) into another Pauli operator (i.e. the Clifford group is the normalizer of the $n$-qubit Pauli group.)  Besides forming a discrete group for computational ease of composing to identity, this choice of operations ``twirls" generic errors on the gates $C_j$ into a uniform, incoherent, depolarization-like error, enabling a potentially complex error structure to collapse into a single-exponential decay when averaging over the results of measuring the initial state probability after many random sequences.    The exponential decay constant resulting from simple least-squares fitting of repeated measurements over random circuits provides the single benchmark number, interpreted as an average gate infidelity.  The infidelity of a particular Clifford gate, such as the CZ or CNOT entangling gate, can be extracted by measuring the decay whilst interleaving this gate amongst all the Cliffords, and subtracting off the measured decay rate without interleaving.  For a review of RB and its variants, see \onlinecite{helsen2020}.  

Example randomized benchmarking data from a number of semiconductor spin qubits are shown in \reffig{fig:allRB}; these results will be discussed in more detail in following sections.  One-qubit RB (1Q RB) and two-qubit RB (2Q RB) are important accomplishments, in part because the ability to perform RB, which requires the application of many (preferably 1000s) of programmed, calibrated operations on a qubit, shows that the whole system, including cryogenics, control hardware, wiring, and qubits, are co-performing in a way necessary for operation as a future quantum computer. Quantum state, process, and gate-set tomography (GST)~\cite{mohseni2008} use repeated state estimation to identify specific errors, and may give complementary information to a qubit's computational utility, and hence these methods provide additional fidelity metrics in the sections that follow.

\subsection{Loss-DiVincenzo single spin qubits}  \label{Subsec:Gates:LD}

The control of a single LD qubit follows the same principles as the coherent control of large spin ensembles, a subject with a long history in electron spin resonance (ESR) and nuclear magnetic resonance (NMR) \cite{Abragam1961, Slichter2010}. However, single-spin control faces additional challenges that are absent in ensemble experiments. In bulk ESR/NMR, initialization is typically performed by waiting for the ensemble to thermalize; at typical magnetic fields and temperatures, the resulting polarization is quite small, but this is compensated for in the measurement signal-to-noise ratio by the large size of the spin ensemble. For single-spin qubits, an initialization routine giving nearly 100\% polarization is required, and waiting for thermalization is prohibitively time-consuming. Hence coherent single-spin control requires fast, high-fidelity initialization and measurement procedures, and this is where the review of LD qubits begins.

\subsubsection{Initialization and readout} \label{Subsubsec:Gates:LD:Init}

\begin{figure}[t]
	\centering
	\includegraphics[width=\columnwidth]{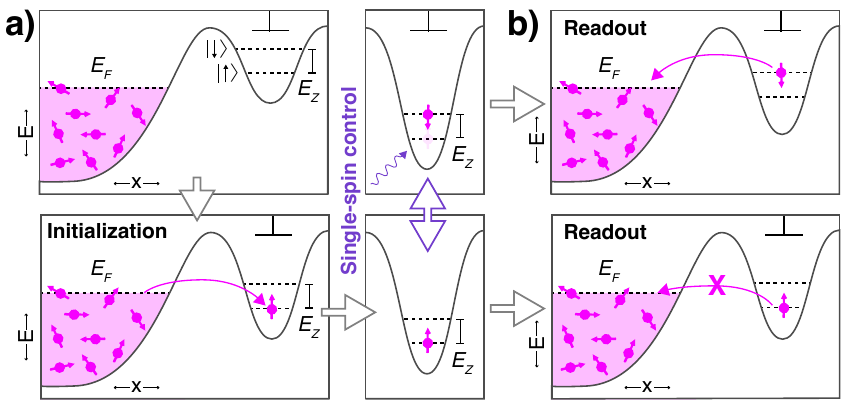}
	\caption{Energy-dependent tunneling for single spin initialization and readout of LD qubits. Note that the ground state in GaAs is $\ket{\uparrow}$ due to its negative electron $g$-factor. (a) $\ket{\uparrow}$ can be initialized by emptying the dot (top panel) and then applying a positive voltage pulse, such that $E_{\downarrow} > E_F > E_{\uparrow}$ (bottom panel). With $E_{\downarrow} > E_F > E_{\uparrow}$ an electron can only tunnel into the spin ground state. After spin manipulations, (b) spin readout is performed by pulsing back to the initialization bias condition. In this example, the presence (absence) of a tunneling event during the measurement period indicates $\ket{\downarrow}$ (or $\ket{\uparrow}$).}
	\label{fig:elzermanreadout}
\end{figure}

The first experimental demonstration of single spin readout was achieved by \onlinecite{Elzerman2004} in a GaAs QD.  In the same issue of Nature, electrical detection of single spin resonance in a Si transistor was also reported \cite{Xiao2004}. \onlinecite{Elzerman2004}, and many similar works since then, use energy-dependent tunneling, providing a high-enough magnetic field for the Zeeman splitting $E_Z$ = $g \mu_B B$ to greatly exceed the thermal energy $k_BT_e$ for electron temperature $T_e$\footnote{For context, if $g$ = 2 (as in Si) and $B$ = 1 T $E_z$ = 116 $\mu$eV, corresponding to frequency $f = E_Z/h$ $\approx$ 27.6 GHz, while $T_e$ ($k_BT_e$) is typically 50--300 mK (4-26 $\mu$eV).}.
Initialization and readout are then achieved through single-electron tunneling between the QD and an electron reservoir (see Fig. \ref{fig:elzermanreadout}). Tunneling is controlled by adjusting the QD energy level relative to the Fermi level of the reservoir $E_F$ using time-dependent gate voltage pulses $V_g(t)$. These gate voltage pulses can be very short $\sim$ 100 ps, as had been previously demonstrated in charge qubits~\cite{fujisawa_allowed_2002,Hayashi2003,Petta2004,Petersson2010}.

The modest $g$-factor in GaAs required \onlinecite{Elzerman2004} to operate with $B$~=~10~T.  The gate voltage pulse sequence for readout, illustrated in \reffig{fig:elzermanreadout},  first emptied the QD and then pulled the energy of both spin states below $E_F$ to randomly load the QD in $|\downarrow\rangle$ or $|\uparrow\rangle$. After waiting for a time $t_{\rm wait}$, the QD was biased to set $E_{\downarrow} > E_F > E_{\uparrow}$. Through the process of spin-to-charge conversion, an increase in the QPC current corresponds to a $\ket{\downarrow}$-spin measurement outcome, while no change in current is detected for an $\ket{\uparrow}$-spin.   Similarly, initialization is achieved by pulling only the spin ground state beneath $E_F$.  Single spin control is then generally implemented deep in Coulomb blockade (see Sec.~\ref{Subsec:Principles:Coulomb}) to prevent loss of the electron to the reservoir when microwave fields are applied to drive the spin.

Elzerman spin-dependent tunneling imposes several experimental constraints and must be carefully optimized to achieve high fidelity readout. First, by necessity, Elzerman readout is implemented on QDs that are adjacent to charge reservoirs. In contrast, readout of central dots in a large array would require transport of the spin to an end site of the array (see Sec.~\ref{Subsec:2spins:LR}). Second, there is a competition in time-scales. Since spin readout is achieved using charge detection, the electron must have sufficient time to tunnel off the QD during the readout pulse. If the tunnel rate is too fast compared with the measurement bandwidth, the charge signal can be missed, while if the rate is too slow, the spin can relax before measurement. Third, $E_z \gg k_BT_e$ is required to initialize into the ground state, which implies operation at high field and low temperature. In practice, the spin relaxation rate $\Gamma_1 = 1/T_1 \propto B^5$ in GaAs, which limits the practical field range (in addition to technical challenges associated with microwave control above 20 GHz). Finally, spin readout is destructive since the $\ket{\downarrow}$-spin is lost to the Fermi sea during tunneling. An overview of the conditions required to achieve a readout fidelity $F > 99 \%$ has been given by \onlinecite{Keith_2019_RO_theory}. \onlinecite{mills_2Q_2021} recently achieved $F > 99 \%$ in Si/SiGe quantum devices.

\subsubsection{Single-qubit gates} \label{Subsubsec:Gates:LD:1QB}

\begin{figure*}[t]
	\includegraphics[width=2\columnwidth]{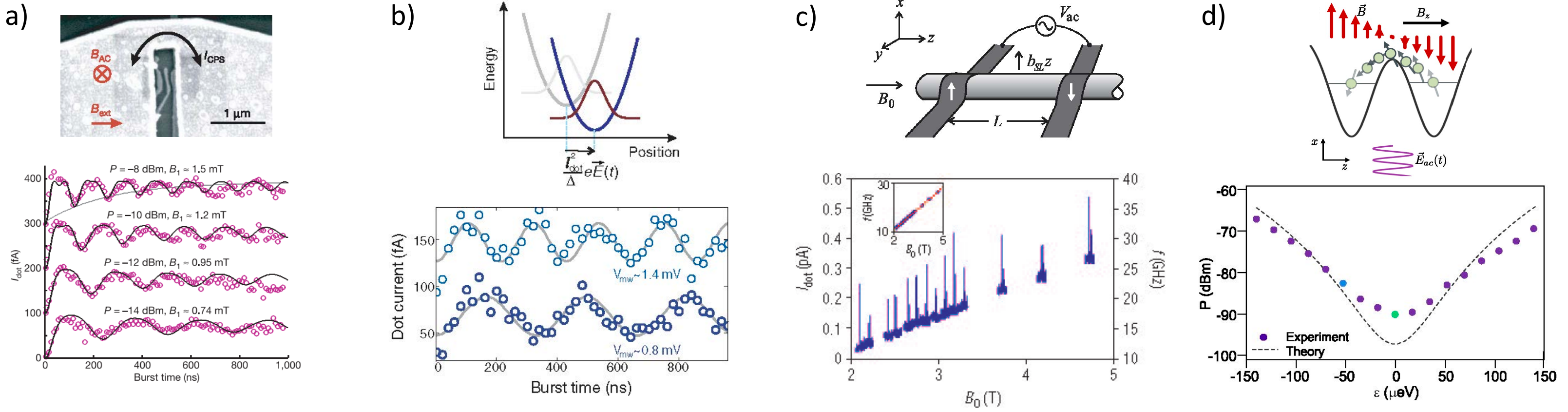}
	\caption{Single spin rotations driven with: (a) an ac magnetic field generated by a coplanar waveguide \cite{Koppens2006}, (b) an ac electric field in the presence of intrinsic SOC \cite{Nowack2007}, and (c) an ac electric field in the presence of synthetic SOC (a magnetic field gradient) \cite{Tokura2006,Pioro2008}. (d) Low power EDSR in a field gradient can be achieved in the flopping-mode regime of a DQD \cite{Croot_flopping-mode_2020,Benito2019b}.}
	\label{fig:sec4single-spin}
\end{figure*}

Coherent single spin control was first demonstrated by \onlinecite{Koppens2006} using ESR in a GaAs DQD. By applying a source-drain bias $V_{SD}$ across the DQD, a (1,1) polarized spin triplet state (T$_+$ or T$_-$) was initialized via transport in the PSB regime. Spin detection in this case occurred by measuring the DQD leakage current $I_{\rm dot}$ as a function of $B_0$ and the frequency $f_{\rm ac}=\omega/2\pi$ of an applied microwave magnetic field $B_{\rm ac}$ generated by driving an ac current through a stripline fabricated adjacent to the DQD. On resonance, when $B_0 = \pm h f_{\rm ac}/g \mu_B$ for one of the spins, single spin ESR drives transitions from the triplet to singlet, lifting PSB and increasing $I_{\rm dot}$. Measurements 
revealed a peak in $I_{\rm dot}$ around $B = 0$ due to hyperfine mixing of the spin states  \cite{Johnson2005,Koppens2006,Jouravlev2006}, as well as two satellite peaks following the resonance condition $B = \pm h f_{\rm ac}/g \mu_B$ [Fig. \ref{fig:sec4single-spin}(a)].

The physics of how applied transverse ac magnetic fields drive coherent spin rotations follows conventional ESR.  The transverse ac field may be assumed to point along $\hat{x}$, i.e. $B_1(t)\hat{x} = B_{\rm ac}\cos(\omega t + \phi) \hat{x}$, where $\phi$ is a phase relative to a local oscillator. The effective Hamiltonian in the rotating frame [see App.~\ref{app:gates}] is then $\tilde{H} = (g \mu_B B_0- \hbar \omega){S^z} + g \mu_B (B_{1}/2){S^x}$. The first term vanishes when the electron spin is driven on resonance (with $\hbar \omega$  = $g \mu_B B_0$) and the electron spin coherently rotates between $\ket{\uparrow}$ and $\ket{\downarrow}$ at the Rabi frequency $f_{\text{Rabi}}=g \mu_B B_{1}/(2 h)$. In the Bloch sphere representation of the LD qubit [see Fig.~\ref{fig:blochspheres}], the static $B_0$ field points along the $z$-axis and leads to Larmor precession of the spin, while the transverse field $B_1(t)$ points along the x-axis for $\phi$ = 0 and yields a $\sigma^x$ rotation. 

For \onlinecite{Koppens2006,Koppens2008}, Rabi oscillations at frequencies up to $\sim$10 MHz were achieved, but were highly damped in this first experiment due to hyperfine interactions [lower image in Fig. \ref{fig:sec4single-spin}(a)] which move the spin out of resonance and lead to imperfect rotations on the Bloch sphere. Hyperfine coupling is discussed in greater detail in \refsec{Sec:Decoherence}.  Later silicon-based ESR devices devices~\cite{Veldhorst2014,Pla2012} achieved comparable Rabi frequencies in a system with reduced hyperfine coupling.

Single-spin control based on ESR raises questions on how to selectively control one qubit in an array.  In some LD-based architectures, only global single-spin control  is possible~\cite{Jones2016}, but these require high dot-to-dot uniformity.
Tunable and selective single-qubit rotations require a unique Larmor resonance for each qubit, for example by engineering magnetic field gradients across the device \cite{Pioro2008} or through voltage-tunable $g$-factors~\cite{Veldhorst2014}.  A key concern of any ESR approach is power dissipation, as device heating often limits the maximum Rabi frequency that can be obtained, motivating new designs for resonators and approaches for local control with global fields~\cite{Vahapoglu2021}.

One year after ESR control of a single spin in a GaAs QD was shown, ~\onlinecite{Nowack2007} achieved electrically driven single spin rotations using EDSR with the intrinsic SOC of GaAs. An ac voltage excitation applied to a gate electrode shifted the orbital wave function, and coherent Rabi oscillations were again detected by measuring $I_{\rm dot}$ in the PSB regime [Fig.~\ref{fig:sec4single-spin}(b)]. The highest Rabi frequency achieved was 4.7 MHz; nevertheless, this important demonstration spurred the investigation of electrical control in strong spin-orbit systems (see Sec.\ \ref{Subsubsec:Implementations:Alt:SOQubit}) and added weight to the development of EDSR in the ``artificial SOC" created by magnetic field gradients~\cite{Tokura2006}. The transition from ESR to gradient-enabled EDSR not only affords more speed, but it also provides a clear mechanism for selectivity, since the ac driving field can be applied directly to a QD gate electrode. 

\onlinecite{Pioro2008} demonstrated the feasibility of electrically driving spin rotations using a magnetic field gradient resulting from a fabricated Co micromagnet. A time-dependent gate voltage $V_{\rm ac}$ periodically moved the electron in the inhomogeneous field of the micromagnet and spin rotations were detected in the PSB leakage current [Fig.~\ref{fig:sec4single-spin}(c)]. The  longitudinal magnetic field gradient from the magnet allowed the EDSR transitions of both spins to be spectrally resolved. \onlinecite{Yoneda2014fast} built upon these results by demonstrating $>$100 MHz Rabi frequencies, measuring Rabi chevrons in the time-domain, and achieving $Z$-gates in the field gradient.  

A larger displacement of the electron spin in the magnetic field gradient can be achieved in a DQD at $\epsilon$ = 0, which is known as the ``flopping-mode'' \cite{Croot_flopping-mode_2020}. As illustrated by the measurements in Fig. \ref{fig:sec4single-spin}(d), the power required to achieve an EDSR Rabi frequency $f_{\rm Rabi}$ = 6~MHz is reduced by a factor of $\sim$250 at $\epsilon~=~0$ compared to the far-detuned single dot regime. Flopping-mode operation may greatly reduce power requirements in larger QD device architectures. 

\subsubsection{Two-qubit gates} \label{Subsubsec:Gates:LD:2QB}

\begin{figure}[t]
	\centering
	\includegraphics[width=\columnwidth]{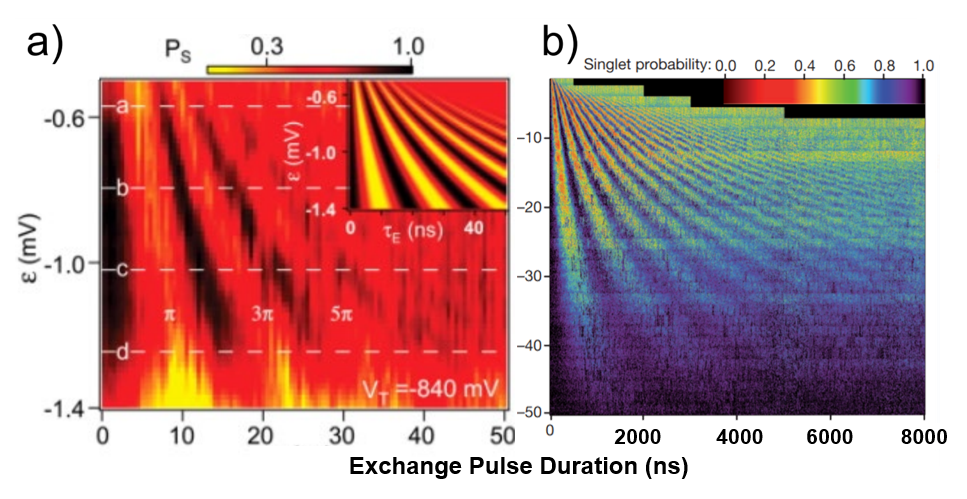}
	\caption{Coherent exchange oscillations as first observed in a DQD using PSB readout for (a) GaAs \cite{Petta2005} and (b) SiGe \cite{Maune2012}.}
	\label{fig:J-gate}
\end{figure}

LD qubits use voltage-controlled exchange for two-qubit gates (see \refsec{Sec:2spins}), which was first shown to coherently couple two single-spins by \onlinecite{Petta2005} (\reffig{fig:J-gate}a). In this experiment fast $\sim$~200 ps exchange oscillations were observed in a GaAs DQD. Time-domain control of $J(t)$ was also used to measure the inhomogeneous spin dephasing time $T_2^* \sim$ 10 ns and the spin-echo decay time $T_2 \sim 1 \mu$s. Many aspects of \onlinecite{Petta2005} were later repeated in Si/SiGe by \onlinecite{Maune2012}, \reffig{fig:J-gate}b, with longer coherence times and improved exchange coherence; the limiters of coherence for exchange oscillations will be discussed in \refsec{Sec:Decoherence}.  These early results only featured singlet-triplet readout by PSB; \onlinecite{Nowack2011} extended these results to a GaAs DQD that allowed for independent single-shot readout of each spin with a fidelity of 86\%.

True LD operation requires the ability to do both single-spin rotations for single-qubit gates and exchange operations for two-qubit gates, completing a universal control set. 
The exchange Hamiltonian of \refeq{eq:Heisenberg} 
couples $|\uparrow\downarrow\rangle$ to $|\downarrow\uparrow\rangle$; an exchange $\pi$ pulse (activating exchange for a time $\tau = \pi \hbar/J$) realizes a $\mathrm{SWAP}$ gate, while an exchange $\pi/2$ pulse generates the entangling square-root of swap gate $\sqrt{\mathrm{SWAP}}$. 
The effect of exchange can be seen by writing  \refeq{eq:Heisenberg} as the projection operator on the spin-singlet state, $H= -J|{\rm S}\rangle\langle {\rm S}|$, with the resulting unitary $U(\phi)=\exp(-i\phi|{\rm S}\rangle\langle {\rm S}|)=\openone +(e^{i\phi}-1)|{\rm S}\rangle\langle {\rm S}|$. For $\phi=J\tau/\hbar\pi$ we find $U(\pi)=1-2|{\rm S}\rangle\langle {\rm S}|=\mathrm{SWAP}$ while for $\phi=\pi/2$ we have $U(\pi/2)=(1+i)\openone/2+(1-i)\mathrm{SWAP}/2=\sqrt{\mathrm{SWAP}}$. Using this interaction and single-qubit rotations separately, the CNOT gate (up to a global phase) could then be obtained using the sequence
$\mathrm{CNOT}=e^{-i\pi S^y_2/2}
e^{i\pi S^z_1/2}
e^{-i\pi S^z_2/2}
\sqrt{\mathrm{SWAP}}e^{i\pi S^z_1}\sqrt{\mathrm{SWAP}}e^{i\pi S^y_2/2}$ \cite{Loss1998}.

In practice, however, exchange coupling and local magnetic fields typically act on a register of spin qubits simultaneously, e.g. in devices with magnetic field gradients or g-factor variations \cite{Brunner2011}.  Considering two exchange-coupled spins, we can investigate this situation with the Heisenberg Hamiltonian Eq.~\eqref{eq:Heisenberg} where $i,j=1,2$ such that 
$H=J\mathbf{S}_1\cdot\mathbf{S}_2 
    +g  \mu_B (\mathbf{B}_1\cdot \mathbf{S}_1 + \mathbf{B}_2\cdot \mathbf{S}_2)$,
where for simplicity we have assumed that the g-factor is the same for both sites, although similar principles may be applied with dot-varying $g$-factors~\cite{tanttu_controlling_2019,Jock2018}. Taking the magnetic field direction to be the same on both sites (i.e.~$\hat{z}$), $H=J\mathbf{S}_1\cdot\mathbf{S}_2 
    + B (S_{1}^z + S_{2}^z)
    + \Delta B (S_{1}^z-S_{2}^z)/2$,
with $B=g  \mu_B(B_1+B_2)/2=g  \mu_B B^z$ and $\Delta B=g  \mu_B(B_1-B_2)=g  \mu_B \Delta B^z$.
As this Hamiltonian includes two, potentially indepedently controllable non-commuting terms, a variety of adiabatic and diabatic control options exist for achieving entangling two-qubit gates.  For example, the direct time evolution of this Hamiltonian with all terms held constant generates the CZ (or CPHASE) gate, $U_{\textrm{CZ}}=\mathrm{diag}({1,1,1,-1})=i \exp(-i\tau H/\hbar)$, for a gate time $\tau=2\pi k/\Omega$ where $\hbar\Omega=\sqrt{J^2+\Delta B^2}$ with $k=1,2,\ldots$ and $J=(k-n-2m-1/2)\hbar\Omega/k$ with $n,m$ integers, and $B=(n+1/2)\hbar\Omega/2k$ \cite{Burkard:1999p1157}. A simple case is $k=1$ and $n=m=0$ where CZ can be realized for arbitrary $\Delta B\neq 0$, with $B=\Delta B/2\sqrt{3}$, $J=2\Delta B/\sqrt{3}$, and $\tau=\pi\hbar/J$.  When combined with single-qubit rotations, this gate lends itself to the implementation of the CNOT gate. An equivalent version of a CZ gate can also be derived from a two-site hopping model \cite{Meunier2011}.  

\onlinecite{Watson2018} utilized a dc exchange pulse to implement a CZ gate in the large magnetic field gradient regime. \onlinecite{Veldhorst2015} demonstrated full two-qubit control in SiMOS, achieving selective spin control by voltage-shifting the $g$-factors and therefore the ESR resonance frequencies of the two qubits. Fast CZ gates were implemented by pulsing on exchange. \onlinecite{Zajac2017CNot} demonstrated a resonantly driven CNOT gate by lowering the energy of antiparallel spin states ($\ket{\uparrow \downarrow}$, $\ket{\downarrow \uparrow}$) relative to the parallel spin states ($\ket{\uparrow \uparrow}$, $\ket{\downarrow \downarrow}$) with exchange while applying a single microwave pulse ~\cite{Russ_CNOT_2018}.  As each of these experiments also included site-selective single-spin initialization, control, and read-out, full-gate sets for LD qubits were demonstrated in all cases.

\subsubsection{Limits of fidelity - randomized benchmarking} \label{Subsubsec:Gates:LD:RB}

The transition to Si/SiGe spin qubits from GaAs has resulted in higher overall operation fidelities for LD qubit control. \onlinecite{Kawakami2014} demonstrated spin control in a Si/SiGe DQD with a Co micromagnet, observing $f_{\text{Rabi}}\sim$5~MHz and measuring $T_2$ and $T_{2}^*$ using spin-echo and Ramsey pulse sequences, and later single-qubit randomized benchmarking with 98.1\% fidelity~\cite{Kawakami2016}.
Using ESR for RB, \onlinecite{Veldhorst2014} showed a single-qubit control fidelity of 99.6$\%$ in a $^{28}$Si-MOS device, included in \reffig{fig:allRB}. 
Similarly, \onlinecite{Takeda2016} reported fidelities of 99.6$\%$ using EDSR in a field gradient in natural-Si/SiGe devices. 
\onlinecite{Veldhorst2015} extended these results to a SiMOS DQD, where selective ESR control of two spins was achieved. 
\onlinecite{Zajac2017CNot}  used RB to demonstrate single-qubit fidelities of 99.3\% and 99.7\% in a two-qubit Si/SiGe device. 
Isotopic enrichment has led to continued increases in the single-qubit gate fidelity, as discussed in Sec. \ref{Subsubsec:Gates:LD:RB}.
Using isotopically enriched Si/SiGe, \onlinecite{Yoneda2018} achieved single-qubit fidelities exceeding 99.9\%. Characterization of the electrical noise in this device indicates coherence is limited by charge motion in the presence of the micromagnet field gradient. \onlinecite{yang2019silicon} achieved single-qubit Clifford fidelities of 99.96\% in a SiMOS device using improved pulse engineering. Recently, \onlinecite{xue2021} reported single-gate fidelities of 99.69\% in a Si/SiGe QD notable for being operated by a cryogenic control chip.

Early attempts to characterize two-qubit gate fidelities employed quantum state tomography. \onlinecite{Zajac2017CNot} used the resonant CNOT gate to generate a Bell state with fidelity $F$ = 78\%. \onlinecite{Watson2018} achieved similar Bell state fidelities using decoupled CZ gates. Both of these experiments had to correct the tomography for significant SPAM errors. \onlinecite{Huang2019} more rigorously characterized two-qubit gate fidelities using RB in a SiMOS DQD, with  an average Clifford (CROT) gate fidelity of 94.7\% (98\%) achieved in a regime with always-on exchange. \onlinecite{xue_benchmarking_2019} implemented a variation on RB called character RB, enabling the interleaving of a two-qubit gate amongst single-qubit Cliffords, and obtained two-qubit gate fidelity estimates of 92\%. \onlinecite{xue_2Q_2021} recently achieved a two-qubit gate fidelity of 99.5\% using pulsed exchange. In the regime of always-on exchange, \onlinecite{noiri_2Q_2021} have also achieved RB with $>$99\% two-qubit gates. High fidelity overall operation of two qubits in a six QD device has been obtained by \onlinecite{mills_2Q_2021}, with sequential single spin rotation $F$ $>$99.9\%, simultaneous single spin rotation $F$ $>$99\%, and a two-qubit CZ $F$ $>$99.8\%. SPAM errors in this demonstration were $<$ 3\%. Fidelities are expected to further increase with reduced charge noise and higher levels of isotopic enrichment.

Efforts to control hole spins in Ge/GeSi heterostructures have advanced significantly in a short period of time. Due to strong SOC, hole spins can be manipulated electrically without the need for a separate ESR drive line or micromagnet. The smaller effective mass of holes in Ge also relaxes nanofabrication requirements, as the QDs are larger than in Si. \onlinecite{hendrickx_fast_2020} achieved short $\sim$20 ns single hole-spin rotations with $F$ $>$99.3\% and a two qubit exchange gate. Multi-qubit operations have been implemented in a 2 $\times$ 2 Ge QD array, culminating in the generation of a four-qubit Greenberger-Horne-Zeilinger state \cite{Hendrickx2021}. 

\subsection{Donor spin qubits} \label{Subsec:Gates:Donors}

When \onlinecite{Kane1998} was published, it was different from contemporary proposals based on QDs since basic GaAs QD devices had already been fabricated \cite{Kouwenhoven1998Dots}. While doped Si is common, the isolation of single donors in close proximity to gated nanostructures for single-electron control and measurement presented novel fabrication challenges. A number of groups have faced this challenge using bottom-up scanning-tunneling-microscopy (STM) lithography on hydrogen passivated silicon surfaces, enabling the placement of atoms nearly one-at-a-time into designated locations as both qubits and gates ~\cite{Lyding1994,Schofield2003STM,Bussmann2015}. 
Alternatively, \cite{Morello2010} has shown that the approach of detected ion-implantation of P into MOS-style devices allows single-donor-spin measurement and subsequent control. Electrostatically gated dot-donor devices are also being explored \cite{harvey-collard_coherent_2017}, and may provide unique opportunities for nuclear spin readout and coupling to microwave photons \cite{Mielke2021}. The ion implantation and STM lithography approaches have both shown steady progress in controlling single electron spin states, the nuclear spin of the donor, and the exchange coupling between donors, as we discuss in this section, concluding with a discussion of gate fidelities.

\subsubsection{Donor electron spin control and readout} \label{Subsubsec:Gates:Donors:1eQB}

\begin{figure}[t]
\centering
\includegraphics[width=\columnwidth]{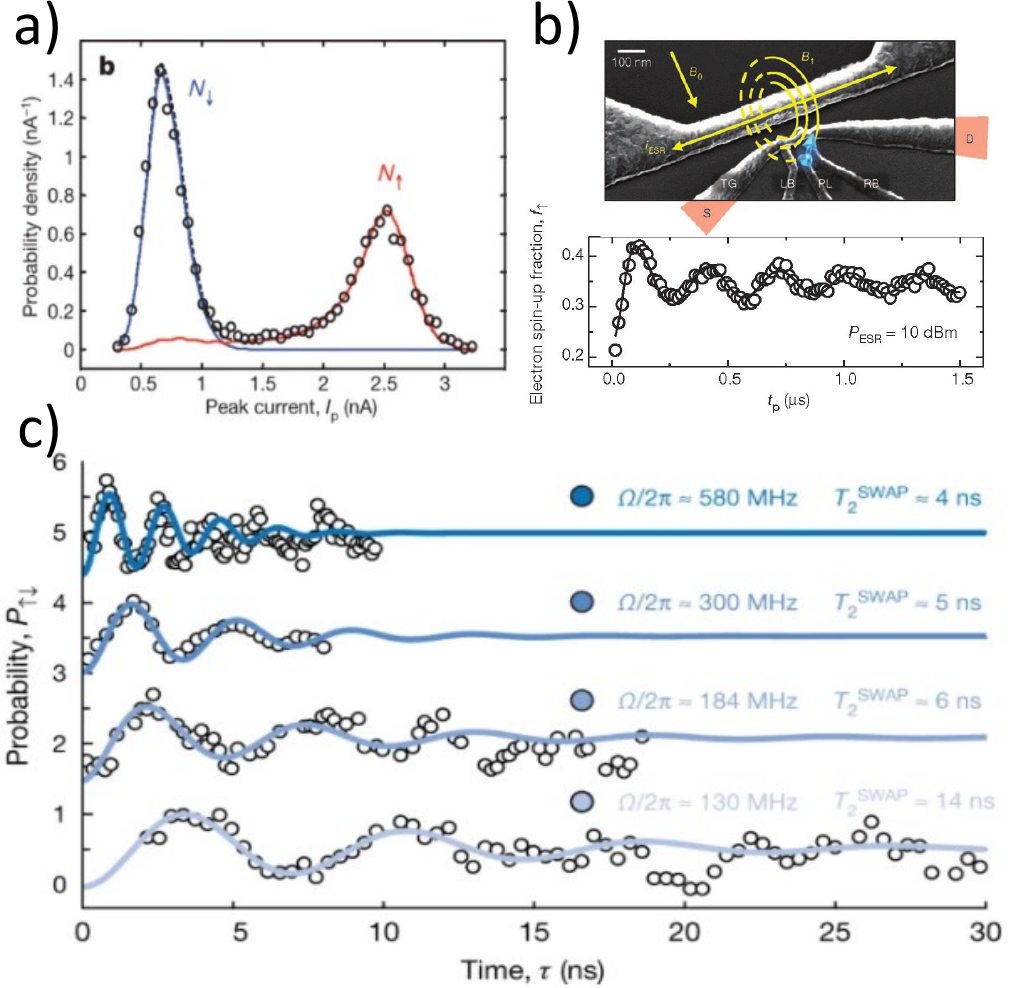}
\caption{a) Single shot readout of a donor-bound electron spin \cite{Morello2010}. b) Rabi oscillations of a donor-bound electron spin \cite{Pla2012}. c) Coherent control of exchange-coupled donor electron spins \cite{He_2donorexchange_2019}.}
\label{fig:gates_donor_electron}
\end{figure}

\onlinecite{Morello2010} used the Elzerman energy-dependent tunneling approach to spin initialization and readout discussed in \refsec{Subsubsec:Gates:LD:Init}, borrowing heavily from developments in QDs [Fig. \ref{fig:gates_donor_electron}(a)].  A single electron transistor (SET) was fabricated next to a 90 $\times$ 90 nm region that was implanted with P donors, and voltage control of a nearby plunger gate was used to control the electronic state of the donor. Single shot measurements allowed mapping of the electron spin lifetime as a function of magnetic field, with $T_1 = 6\, \mathrm{s}$ obtained at $B= 1.5 \,\mathrm{T}$, and the spin readout visibility was estimated to be around 92\%. Two years later, \onlinecite{Pla2012} showed coherent Rabi oscillations of a single donor electron spin in a natural-Si substrate [Fig. \ref{fig:gates_donor_electron}(b)]. These oscillations were highly damped due to hyperfine interactions, reminiscent of the first GaAs QD single-spin Rabi oscillations \cite{Koppens2006}. The use of a simple Hahn echo pulse sequence extended the coherence time out to 200 $\mu$s. 

The STM lithography approach achieved similar results: \onlinecite{Broome2017Readout} placed a small cluster of donor atoms next to a SET also defined using STM lithography, and demonstrated $F$ = 98.4\% single-shot readout of a donor singlet-triplet qubit. \onlinecite{Koch2019} later achieved an average measurement fidelity of $F$ = 97.9\% for single spin Elzerman readout using a SET, and \onlinecite{Keith2019} showed $F$ = 97\% measurement fidelity with a 1.5 $\mu$s SET measurement time. Dispersive gate-based sensing has also been explored, but as with QD systems, dispersive sensing yields lower fidelities and measurement bandwidths. \onlinecite{pakkiam_single-shot_2018} dispersively probed a donor singlet-triplet qubit with a moderate fidelity $F$ = 82.9\% and 3 kHz bandwidth. 

\onlinecite{Tettamanzi2017} took a first step towards donor quantum control by demonstrating pulse spectroscopy of a single P atom at frequencies up to 13~GHz. These experiments demonstrated that microwave signals could be transmitted down heavily doped P leads in silicon.  \onlinecite{Hile2018} later probed ESR spectra of a single P donor and 2P molecule, and \onlinecite{Koch2019} then extended these results to single shot measurements of a single P donor qubit using a SET.

\subsubsection{Donor nuclear spin control and readout} \label{Subsubsec:Gates:Donors:1nQB}

In \onlinecite{Kane1998}, the qubit is the \Pspin\ nuclear spin, not the electron; the electron is used for read-out and control leveraging the \Pspin\ hyperfine coupling of $A\approx 114$~MHz (\refsec{Subsec:Principles:Hyperfine}). \onlinecite{Pla2013} accessed the nuclear spin by using an ESR measurement time much less than the nuclear spin flip time.   This device was able to resolve ESR transition frequencies that jumped between $f_{\Uparrow} = g \mu_B B/h + A/2$ and $f_{\Downarrow} = g \mu_B B/h - A/2$. These jumps were interpreted as being due to flips of the nuclear spin state (denoted by $\Uparrow$ and $\Downarrow$). A broadband antenna on the device allowed for direct driving of the donor atom nuclear spin, with dephasing times $10^4$ times longer than for the donor electron spin. In a followup experiment in $^{28}$Si, nuclear spin control with a fidelity exceeding 99.99\% was demonstrated. \onlinecite{Muhonen2014} showed Carr-Purcell-Meiboom-Gill (CPMG) dynamic decoupling pulse sequences extended the nuclear spin coherence time beyond 30~sec.
\onlinecite{Laucht2015} showed that the Larmor resonances of each donor site could be selectively controlled by pushing the electron closer to its \Pspin\ using a gate, as proposed by Kane, enabling a global ESR field to selectively control one site at a time.

Recently, \onlinecite{asaad_coherent_2020} demonstrated coherent control of the \isotope{Sb}{123}\ donor. The \Pspin\ donor is a nuclear spin $I=1/2$ system, but nuclei with spin $I>1/2$ such as $I=7/2$ \isotope{Sb}{123}\ allow for richer and more complicated control possibilities. The uniform Zeeman splitting between adjacent states of different $m$ is shifted by the electric quadrupole interaction due to local strain, allowing individual addressability of all $2I+1=8$ nuclear spin transitions. Modulation of these quadrupole splittings by an ac electric field drives Rabi oscillations between transitions, and a dephasing time $T_2^*$ $\approx$ 92 ms was demonstrated.

\subsubsection{Two-qubit gates} 

\onlinecite{Kane1998} proposed coupling between donor nuclear spin qubits could be mediated via exchange between the electron spins on each donor, but it was soon noted that atomic-scale oscillations in exchange due to multi-valley interference would render this interaction highly sensitive to atomic placement~\cite{Koiller2001Exchange,Wellard2003Exchange,gamble_multivalley_2015,joecker_full_2020}, requiring either an architecture tolerant of such variation, extremely careful donor placement, or the use of asymmetric donor clusters with more than one phosphorous atom~\cite{Wang2016Exchange}. A variety of demonstrations of exchange on various donor devices have helped show a range of possibilities beyond Kane's original proposal.  \onlinecite{Weber2014SpinBlockade} used donor devices fabricated with STM-based lithography to show exchange and PSB of two electrons on the same donor site,  \onlinecite{Gorman2016}  demonstrated methods to calibrate tunnel couplings, and \onlinecite{Broome2017Readout} performed high-fidelity singlet-triplet (PSB) readout.   With sufficient control over the donor positions and of tunnel couplings, \onlinecite{Broome2018Correlations} was able to observe  two-electron correlations and \cite{He_2donorexchange_2019} showed fast coherent exchange oscillations between donor clusters [Fig. \ref{fig:gates_donor_electron}(c)]. As with the first exchange oscillations in GaAs and Si/SiGe DQDs, the oscillations were heavily damped due to charge noise \cite{Petta2005,Maune2012}.

An alternative coupling relevant to donors is the magnetic dipole-dipole coupling between electrons, as its long-range, magnetic nature avoids the atomic precision fabrication requirement for exchange.  Proposals to exploit this interaction through isotopic engineering and implanted donors employ a variety of methods to manage the interaction, including selective ionization and mechanical motion~\cite{PhysRevLett.89.017901,PhysRevA.70.052304,hill2015,ogorman_silicon-based_2016}, however execution of any such proposal will require devices with exquisite coherence.

\subsubsection{Limits of fidelity - randomized benchmarking} \label{Subsubsec:Gates:Donors:RB}

The demanding nanoscale fabrication requirements of donor devices have impeded their progress relative to gate-defined QDs. QCVV results are so far limited to ion-implanted devices, which are capable of supporting impressive quantum control fidelities. \onlinecite{Muhonen2015} performed comprehensive measurements of the electron and nuclear spin qubit gate fidelities using 1Q RB, included in \reffig{fig:allRB}. Average electron spin gate fidelities exceeded 99.95\%, while the nuclear spin fidelity was 99.99\%. The dependence of the fidelity on pulse power and shape in these early experiments suggests the overall fidelities are limited by quantum control hardware constraints, not the intrinsic performance of the qubit.

Recent characterization of two P ion-implanted donors coupled by a single electron using gate set tomography (GST) have demonstrated single-qubit fidelities of up to 99.93\% and two-qubit fidelities of 99.2\% \cite{madzik_arxiv}. GST allows for the distinction of coherent (stochastic) errors that transfer amplitude (probability) to erroneous states, as well as relational errors, where the errors incurred are dependent on the history of prior gate operations. \onlinecite{madzik_arxiv} found evidence for coherent $ZZ$ errors that were attributed to off-resonant leakage of microwave power near ESR frequencies. While an exchange gate has been demonstrated with an STM fabricated device \cite{He_2donorexchange_2019}, the fidelities are too low to support QCVV protocols. Quantitative characterization of the exchange gate through RB remains an important goal for the donor spin qubit platform.   

\subsection{Singlet-triplet qubits} \label{Subsec:Gates:ST}

The early demonstration of coherent exchange in a GaAs DQD \cite{Petta2005} showed not only the potential for two-qubit operations of LD qubits, but also basic single-axis control of the ST$_0$ qubit. The data in \reffig{fig:J-gate} show that the DQD level detuning $\epsilon$ enables control over the exchange coupling $J$, which is the energy separation between the S and T$_0$ qubit states, as discussed in \refsec{Sec:2spins}.  In these early demonstrations,  the longitudinal magnetic field gradient experienced by the two spins, $\Delta B^z$, which lifts the degeneracy between the flip-flop states $\ket{\uparrow \downarrow}=\frac{1}{\sqrt{2}}\left(\ket{{\rm S}} +\ket{{\rm T}_0}\right)$ and $\ket{\downarrow \uparrow}=\frac{1}{\sqrt{2}}\left(\ket{{\rm T}_0} -\ket{{\rm S}}\right)$, was provided by the random hyperfine fields of nuclear spins in the device. 

Figure~\ref{Fig:ST_basics} also shows that at a particular value of $\epsilon$, the $\ket{{\rm S}}$ and $\ket{{\rm T}_+}$ states become degenerate, where $J$ compensates the Zeeman splitting between triplet-states, $E_Z$. Near this detuning, the ST$_+$ qubit is formed. Here again we have the controllable qubit energy splitting $E_{ST_{+}}=E_z-J$ and the transverse coupling $\Delta_{\rm ST}$ can be introduced by various mechanisms such as microscopic hyperfine or spin-orbit interactions~\cite{Taylor2007,Petta2010,Stepanenko2012,Nichol2015}. For the ST$_{+}$ qubit we are assuming a device made with a negative g-factor material, such as GaAs, where $\ket{{\rm T}_+}$ is lower in energy than $\ket{{\rm T}_-}$; for a positive g-factor material (e.g. Si), the natural choice is a ST$_-$ qubit.

\subsubsection{Initialization and readout} \label{Subsubsec:Gates:ST:Init}

ST$_0$ and ST$_+$ qubit demonstrations \cite{Petta2005, Foletti2009, Maune2012, Botzem2018}  use PSB for initialization and readout (\refsec{Subsec:2spins:PSB}).  The high-fidelity of PSB initialization and readout in DQDs is enabled by the large exchange coupling in the (2,0) charge configuration. The energy splitting from the singlet ground state to the excited $(2,0)$ triplet states was shown to be meV or higher in energy in GaAs and tens to hundreds of $\mu$eV higher in energy in Si QDs, as discussed in \refsec{Subsec:2spins:Exchange}. These energy scales are larger than $k_BT_e$ at typical electron temperatures. Following initialization, the electrons are usually separated via tunneling to the (1,1) charge state. 

\begin{figure}
	\includegraphics{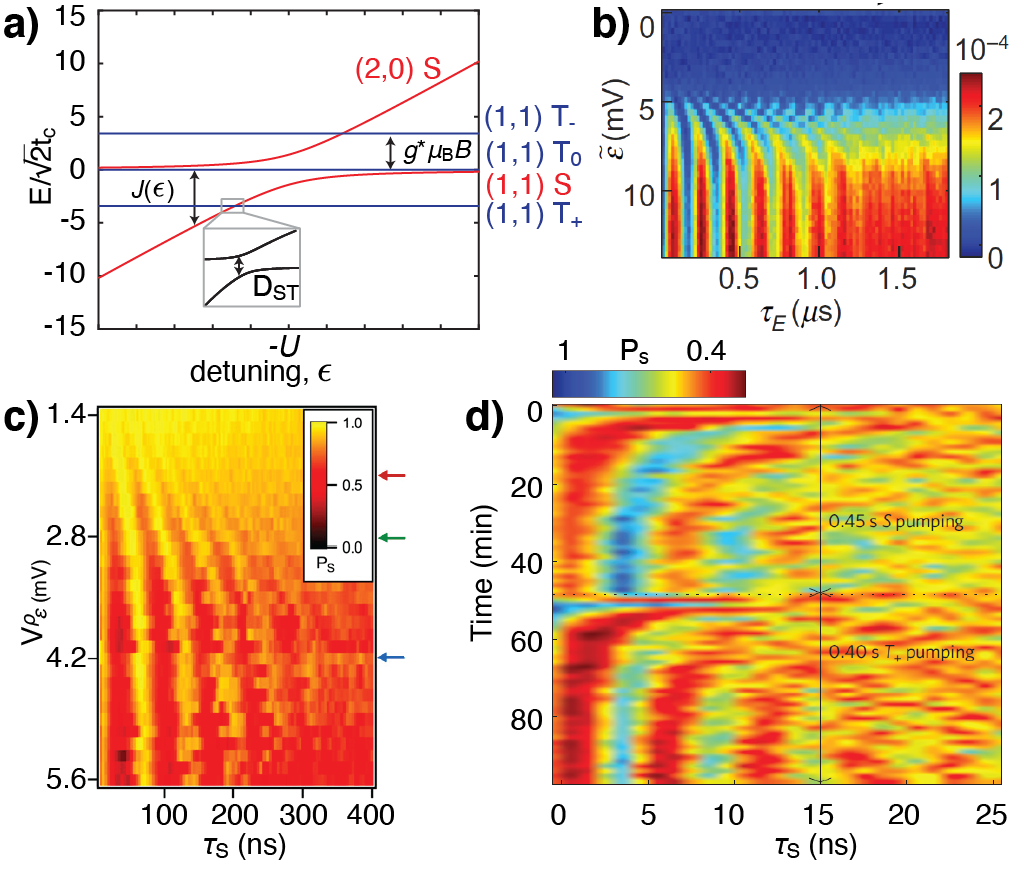}
	\caption{\label{Fig:ST_basics} a) Energy-level diagram for two electrons in a DQD. $\epsilon$ is the energy level detuning, and $(1,1)$ and $(2,0)$ indicate the DQD charge configurations. The Zeeman and exchange splittings are $g^*\mu_B B$ and $J(\epsilon)$ where $B$ denotes the magnetic field.
	The spin states are $\ket{{\rm S}}=\frac{1}{\sqrt{2}}\left(\ket{\up \down} - \ket{\down \up} \right)$, $\ket{{\rm T}_0}=\frac{1}{\sqrt{2}}\left(\ket{\up \down} + \ket{\down \up} \right)$, $\ket{{\rm T}_+}=\ket{\up \up}$, and $\ket{{\rm T}_-}=\ket{\down \down}$. Singlet-triplet oscillations driven by: b) $g$-factor differences between dots~\cite{Liu2021}, c) micromagnets~\cite{Wu2014}, and d) dynamic nuclear polarization~\cite{Foletti2009}.}
		\label{fig:ST_results}
\end{figure}

Experiments have leveraged adiabatic and nonadiabatic separation to complete qubit control [see the energy level diagram in Fig. \ref{fig:ST_results}(a)]. When electron separation occurs rapidly with respect to any magnetic gradients, tunneling preserves the spin state, so an initialized singlet remains a singlet \cite{Petta2005, Foletti2009,Maune2012, Botzem2018}.  If the separation occurs slowly with respect to magnetic gradients, the singlet state transitions to the lower-energy spin-zero product state~\cite{Petta2005,Foletti2009}.  Hence two orthogonal S-$T_0$-qubit basis initializations are available, and pulsing detuning $\epsilon$ or tunnel coupling $t_c$ enables characterization of the exchange coupling.  The spin-to-charge conversion offered by PSB reduces spin readout to dot-selective charge readout. A significant number of optimizations have been explored to increase readout speed and fidelity~\cite{Reilly2007FastRF,Barthel2009,Barthel2010,Elliot2020,Noiri2020RF,Borjans_sensing_2021}.  A key trade-off is that while larger gradient $B$-fields can drive faster single qubit operations, these persistent gradients reduce the fidelity of PSB readout due to enhanced spin relaxation \cite{Barthel2012}. Latched readout protocols first demonstrated with charge qubits \cite{Petersson2010} have been extended to singlet-triplet qubits and can overcome this limitation~\cite{Studenikin2012,Orona2018}. 

\subsubsection{Single-qubit gates} 
\label{Subsubsec:Gates:ST:Exchange}

As described in Sec. \ref{Subsec:Basics:STintro}, the Hamiltonian (Eq. \ref{eq:singlet-to}) governing the control of ST$_0$ qubits includes an exchange-driven $\sigma^z$ term and a $\sigma^x$ term that is set by an effective magnetic field gradient. Full two-axis control of the ST$_0$ qubit Bloch vector therefore requires control of exchange, which can be achieved by adjusting interdot barrier heights or DQD level detunings, and magnetic field gradients. Approaches to generate the required magnetic field gradients are varied and include dynamic nuclear polarization (DNP) \cite{Foletti2009,Bluhm2010}, the use of permanent micromagnets \cite{Wu2014,Fogarty2018}, $g$-factor differences \cite{Jock2018,Liu2021}, or spin-valley coupling \cite{Jock2021Valley}. Data acquired using some of these approaches are shown in Figs. \ref{fig:ST_results}(b--d). We elaborate on these approaches below.

For gate-defined spin qubits, typical exchange couplings are in the MHz to GHz range. Coherent exchange rotations are achieved by applying fast gate voltage pulses ($<$1 ns to 10's of ns). Voltage pulses of the opposite sign applied to the DQD plunger gates can rapidly change the detuning to configurations with large $J$, as first demonstrated by \onlinecite{Petta2005}. Such control at fixed tunnel coupling is capable of generating arbitrary single qubit gates~\cite{Hanson2007b}. However, detuning-controlled exchange oscillations are vulnerable to charge noise, and the number of coherent oscillations is typically around 10~\cite{Petta2005,Dial2013,Maune2012,Fogarty2018,He_2donorexchange_2019}. Exchange oscillations can also be observed with larger numbers of electrons in the QDs, in configurations where the inner electrons form a ``frozen core"~\cite{Barnes2011Screening, Higginbotham2014}.

\onlinecite{Bertrand2015} and \onlinecite{Martins2016}, working in GaAs DQDs, and \onlinecite{Reed2016}, working in isotopically enhanced Si TQDs, showed that improved qubit control results when the barrier height between electrons is pulsed to smaller values, as simulated in \reffig{fig:fci-exchange}.  The improvement occurs because the Coulomb-dominated exchange coupling is first-order insensitive to 
potential fluctuations in this ``symmetric" mode. As a result, the quality factor of exchange oscillations is higher than that for detuning-controlled oscillations, although the magnitude of the required voltage pulses is also significantly higher. Both of these methods of creating exchange coupling suffice to generate $\sigma^z$ rotations on the ST$_0$ Bloch sphere. In principle, both methods can also be used to control ST$_+$ qubits, though detuning sweeps have been more frequently used in these systems~\cite{Petta2010,Ribeiro2010Harnessing}.

Full control of the ST$_0$ and ST$_+$ qubit Bloch vectors also requires an effective magnetic field gradient for $\sigma^x$ rotations. The use of hyperfine field is particularly convenient for GaAs QDs, due to the many spinful nuclei.  A challenge with using hyperfine as a basis of control is that, as discussed in detail in \refsec{Subsec:Decoherence:HFT2}, the nuclear hyperfine field fluctuates randomly because the nuclear Zeeman energy is so small, typically less than 1 mK for fields of order 1 T, and magnetic dipole-dipole interactions lead to nuclear spin diffusion. 
However, various mechanisms can be employed to enhance and stabilize the nuclear polarization via the electron spins~\cite{Petta2008,Foletti2009,Bluhm2010,Shulman2012,Nichol2017}. These processes are collectively called dynamic nuclear polarization (DNP)~\cite{Abragam1978}.

In singlet-triplet qubits, DNP usually involves the degeneracy point between the $\ket{{\rm S}}$ and $\ket{{\rm T}_+}$ states. This degeneracy is lifted by a transverse gradient~\cite{Taylor2007,Petta2010,Stepanenko2012,Nichol2015}, which is typically generated via the hyperfine interaction between the electron and nuclear spins [Fig.~\ref{Fig:ST_basics}(a)]. As the DQD is adiabatically detuned across the ST$_+$ avoided crossing, the electrons transition from $\ket{{\rm S}}$ to $\ket{{\rm T}_+}$ via the transverse Overhauser field and a nuclear spin must change its state to conserve angular momentum in the electron-nuclear subsystem~\cite{Ribeiro2009,Brataas2011,Neder2014}. If repeated rapidly enough, this process can flip a large number of nuclear spins and can be used to ``pump" both the average $\frac{1}{2}(B_1^z+B_2^z)$~\cite{Petta2008} and difference $(B_2^z-B_1^z)$ longitudinal magnetic fields of the DQD~\cite{Foletti2009,Bluhm2010,Shulman2012,Nichol2015,Nichol2017}. It is not surprising that the average field should be affected, if one assumes that this process flips nuclear spins in both dots with approximately the same probability.
However, the underlying mechanism that builds up the difference field remains remains an active area of theoretical research~\cite{Gullans2010,Gullans2013}.

In addition to dynamic nuclear polarization, micromagnets can also be used to generate $\sigma^x$ rotations~\cite{Wu2014,Fogarty2018}. Although additional fabrication is required, micromagnets eliminate the requirement for DNP, which adds experimental overhead. In Si ST qubits, $g$-factor differences between dots can naturally lead to the existence of a $\sigma^x$ term, even in the presence of a uniform magnetic field~\cite{Kerckhoff_quadrupolar_2020,Liu2021}. Finally, when the Zeeman energy equals a valley splitting, the resonance that occurs between different valley states, together with spin-valley coupling, can also enable rapid $\sigma^x$ rotations in Si ST qubits~\cite{Jock2021Valley}.

Dynamical decoupling experiments illustrate the potential for using fluctuating hyperfine fields for full ST$_0$ control.
\onlinecite{Bluhm2010a} and \onlinecite{Malinowski2017b} have used exchange pulses to decouple ST$_0$ qubits from magnetic noise, resulting in nearly a 5 order of magnitude improvement in coherence. These experiments, in addition to later studies in SiGe~\cite{Kerckhoff_quadrupolar_2020}, also uncover the spectrum of the Overhauser field, revealing the significance of the Larmor precession of the individual nuclei~\cite{Neder2011Model}.  Stabilized magnetic gradients also enable decoupling ST$_0$ qubits from charge noise~\cite{Shulman2014,Dial2013} as well as charge noise spectroscopy~\cite{Dial2013,Jock2021Valley,Connors2021Spectroscopy}.

For ST$_+$ qubits, the $\sigma^x$ interaction typically comes from transverse magnetic gradients~\cite{Taylor2007,Petta2010,Stepanenko2012,Nichol2015} which can be created via hyperfine fields or micromagnets. However, unlike longitudinal gradients, transverse gradients are not amenable to DNP and are thus difficult to stabilize. Transverse gradients also contain spectral components at the Larmor precession frequencies of the individual nuclei~\cite{Nichol2015}; as a result, the naturally occurring hyperfine polarization is typically not stable enough to generate usable $x$-rotations. Spin-orbit coupling can also induce a ST$_+$ splitting~\cite{Stepanenko2012,Nichol2015}, but detuning charge noise in this case can create substantial decoherence. As an alternative to conventional qubit manipulation, repeated Landau-Zener sweeps through the avoided crossing have been proposed as a mechanism to achieve universal control of ST$_+$ qubits~\cite{Petta2010,Ribeiro2010Harnessing}. The axis of rotation on the Bloch sphere in this mode is controlled by the timing of two consecutive Landau-Zener sweeps.

In part to avoid issues associated with charge noise, a variant of the ST$_0$ qubit, the ``resonantly-driven ST$_0$ qubit," which is related to the ``flip-flop qubit,"~\cite{tosi_silicon_2017} has been developed~\cite{Klauser2006,Shulman2014,Nichol2017,Takeda2020}. This qubit's basis states $\ket{\up \down}$ and $\ket{\down \up}$ are equal superpositions of the original singlet and triplet states. In such a resonantly-driven ST$_0$ qubit, a large magnetic gradient, either from a micromagnet or hyperfine fields, generates the primary qubit energy splitting. An oscillating voltage applied to a plunger or barrier gate creates an oscillating exchange splitting. If driven at a frequency corresponding to the magnetic gradient, this oscillating exchange coupling can drive transitions. Because the qubit energy splitting does not depend on electric fields, decoherence due to charge noise can be suppressed.

\subsubsection{Two-qubit gates}  \label{Subsubsec:Gates:ST:2QB}

\begin{figure}
	\includegraphics[width=1.0\columnwidth]{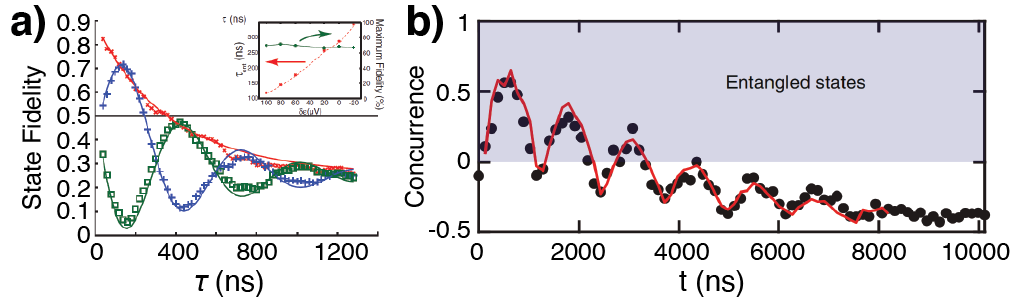}
	\caption{\label{Fig:ST_2q} Two-qubit operations in ST$_0$ qubits. a) Bell-state fidelity during a capacitive entangling operation between two ST$_0$ qubits, from~\cite{Shulman2012}. b) Concurrence during during a two-qubit operation between capacitively-coupled resonantly-driven ST$_0$ qubits, from~\cite{Nichol2017}.}
	\label{fig:ST_coupling}
\end{figure}

\onlinecite{vanWeperen2011Conditional} measured the shift in the exchange oscillation frequency of one ST$_0$ qubit due to changes in the charge configuration of another nearby ST$_0$ qubit, providing the capacitive interaction for a two qubit gate \cite{Taylor2005}. The electrostatic interaction translates to spin, as with spin initialization and readout, via PSB.  Consider two ST$_0$ qubits in close proximity. The first qubit will, depending on its state (singlet or triplet), have a slightly different charge configuration [(0,2) or (1,1)]. As a result, the second qubit experiences a different electrostatic potential and thus energy splitting $J$ depending on the state of the first qubit. This leads to an effective Ising interaction between the two ST$_0$ qubits of the form $H_{\rm int}\propto({dJ_1}/{d\mu_1})({dJ_2}/{d\mu_2}) (\sigma^z-I) \otimes (\sigma^z-I)$~\cite{Taylor2005,Stepanenko2007,Shulman2012}, which can be used to implement, for instance, a CZ gate [see Fig. \ref{fig:ST_coupling}].

Charge noise adversely impacts the performance of this capacitive coupling mechanism. Low-frequency charge noise may be refocused by applying spin-echo-like pulses to both qubits, using stabilized magnetic gradients~\cite{Dial2013,Shulman2012}. If refocusing pulses are applied to both qubits simultaneously, single-qubit dephasing is substantially reduced, while the two-qubit interaction is preserved. \onlinecite{Nichol2017} partially overcame charge-noise limitations this way using the resonantly-driven ST$_0$ qubit, where $\Delta B^z \gg J$. Although the qubit in this regime is sensitive to fluctuating nuclear fields, nuclear spin noise can be refocused much more effectively than charge noise~\cite{Bluhm2010a}. One complication with this approach, not present in the static ST$_0$ qubit case, is that the form and magnitude of the coupling depends on the frequencies of the two qubits~\cite{Calderon2018Ising}. By exploiting DNP, \onlinecite{Nichol2017} tuned the qubit energies to resonance, and performed a rotary echo to suppress low-frequency noise. Neighboring ST$_0$ qubits can also be coupled via the exchange interaction~\cite{Levy2002,Klinovaja2012,Li2012,Wardrop2014,Cerfontaine2020Exchange} and experimental investigations of this approach have recently been initiated \cite{Qiao2021Floquet}.

\subsubsection{Limits of fidelity - randomized benchmarking} \label{Subsubsec:Gates:ST:RB}

Single-qubit gate fidelities for conventional ST$_0$ qubits exceed 99.5\% in GaAs qubits, as measured via RB using stabilized hyperfine gradients~\cite{Cerfontaine2020Closed}. Based on simulations, the gate infidelities were attributed to charge noise.
For resonantly-driven ST$_0$ qubits in GaAs, single-qubit gate fidelities are $\sim$99\% as measured via RB, likely limited by both hyperfine and charge noise~\cite{Nichol2017}. 

Two-qubit operations for GaAs ST$_0$ qubits have so far only been assessed through state and process tomography. For conventional ST$_0$ qubits, the maximum Bell-state fidelity is about 70\%~\cite{Shulman2012}, limited by charge noise.
For resonantly-driven ST$_0$ qubits, the maximum entangling gate fidelity is about 90\%~\cite{Nichol2017}, as measured via process tomography, with a corresponding Bell-state fidelity above 90\%. A limitation associated with single- and two-qubit state tomography in ST$_0$ qubits is that the required tomographic rotations can be difficult to tune precisely~\cite{Shulman2012,Takahashi2013,Nichol2017}.

\subsection{Exchange-only qubits}  \label{Subsec:Gates:TQD}

A necessary first step in developing TQDs, identified early as the minimum system size for EO control~\cite{Bacon2000,DiVincenzo2000ExchangeQC}, was the determination of the voltage bias conditions for populating each dot with a single spin, and the identification of charge regimes enabling initialization, readout, and control~\cite{Gaudreau2006,Gaudreau2009,Granger2010,Pan2012,Schroeer2007}.  The familiar two-dimensional charge stability ``honeycomb" of the DQD becomes a three dimensional cell structure in gate voltage space.  For pairs of TQDs, six-dot arrays require calibration, necessitating even more complex, multidimensional bias tuning procedures to populate each QD with a single charge.  Recently, automation and machine learning have been brought to bear on this problem~\cite{Botzem2018,VanDiepen2018,Mills2019CA,Zwolak2020,Hsiao2020}. 

\subsubsection{Initialization and readout} \label{Subsubsec:Gates:TQD:Init}
For initialization and readout of TQD EO qubits, two of the QDs are used and subject to the same PSB procedure employed for ST$_0$ qubits~\cite{Petta2005,Maune2012,jones_spin-blockade_2019,DiVincenzo2000ExchangeQC}. In both cases, the initialization procedure creates a singlet state $\ket{{\rm S}}$ as described in Sec.~\ref{Subsubsec:Gates:ST:Init}; for the ST$_0$ qubit, this is exactly one of the qubit states, $\ket{0}$.  For a TQD, a third spin is present in a third dot, but this third spin need not be initialized.  As detailed in \refsec{Subsec:Basics:EOintro}, the encoded $\ket{1}$ state in the TQD case is a superposition of two of the triplet states; since Pauli blockade is based on spin parity, it distinguishes between singlet and triplet (but not triplet projections), which suffices for TQD qubit readout via PSB.  Importantly, however, a TQD qubit has a third leaked state, with total angular momentum $S=3/2$, which is also composed of a superposition of triplet states of the two dots undergoing Pauli blockade.  Therefore, a leaked state has the same PSB readout signature as the encoded $\ket{1}$ state.

TQDs present a convenient way to assess exchange for a single pair of QDs, even when full qubit control is not available.  By initializing a singlet on one pair of dots (1 and 2), and then pulsing exchange on a second overlapping pair (2 and 3), a ``triple-dot Rabi" experiment enables the measurement of coherent exchange oscillations without using magnetic field gradients.  \onlinecite{Laird2010} demonstrated such oscillations for early pulsed EO qubit experiments in GaAs, and \onlinecite{Reed2016} used it for the development of exchange sweet-spots in isotopically enhanced Si TQDs.  Unlike single-spin or singlet-triplet coherent oscillations, exchange oscillations decay due to a combination of charge noise and hyperfine dephasing, due to the ability of the encoded qubit to dephase into degenerate leakage states during exchange~\cite{Ladd2012}.

\subsubsection{Exchange-only single-qubit gates}  \label{Subsubsec:Gates:TQD:1QB}
Early coherent measurements of TQD states employed Landau-Zener transitions~\cite{Gaudreau2012,Poulin2015}, as utilized for ST$_+$ qubits (Sec. \ref{Subsec:Gates:ST}).  Such experiments validate energy level structure using tools familiar from DQD qubits, but they do not exploit true EO operation; indeed they explicitly rely on mechanisms other than exchange to traverse anticrossings.  

The EO modality takes its power from the ability to operate by idling qubits in a degenerate, non-evolving decoherence free subsystem or subspace, and then lifting selective singlet-triplet degeneracies with pulsed pairwise exchange~\cite{Bacon2000,DiVincenzo2000ExchangeQC,andrews_quantifying_2019}. In contrast to LD and resonant ST$_0$ qubits that use oscillating fields for quantum control, EO systems rely on the control of energy splittings which are dynamically increased and decreased by changing the trapping potential of electrons.

The TQD EO qubit is defined only by whether the first two spins are in a singlet state $\ket{{\rm S}}$ or any triplet state $\ket{{\rm T}}$.  Time-domain control of the exchange interaction $J_{12}(t)$ lowers the energy of the singlet state relative to any of the triplets, and therefore when pulsed on for a duration $T$ provides a phase such that $\alpha\ket{0}+\beta\ket{1}\rightarrow \alpha\ket{0}+\exp(-\frac{i}{\hbar}\int_0^T J(t) dt)\beta\ket{1}$.  This interaction may be taken as a rotation of the encoded qubit about $\hat{z}$.

Complete control of the EO qubit is accomplished by pulsing another overlapping pair, say dots 2 and 3.  To assess the geometric effect of exchange between these two dots, one may use angular momentum recoupling coefficients [Racah or Wigner 6$j$ coefficients~\cite{VMK1988}], i.e. the matrix elements $\braket{S_{12},S_{3},S_{123}|S_{1},S_{23},S_{123}}$, where $S_{jk\cdots}$ refers to the total angular momentum of spins $j,k,\ldots$ For $S_j=1/2$ and $S_{jk}$ being either 0 or 1 for singlet or triplet, these coefficients amount to a rotation of angle $2\pi/3$ about the $y$-axis from the singlet-triplet basis along the $z$-axis to an axis defined by unit vector $\hat{n}=\cos(2\pi/3) \hat{z}-\sin(2\pi/3)\hat{x}$.  The encoded qubit under exchange  $J_{23}$ between spins 2 and 3 therefore rotates about this $\hat{n}$ axis.  
At most 4 pulses are needed to perform an arbitrary Bloch sphere rotation under these geometric constraints~\cite{lowenthal1972}, generalized Euler angles for such constructions are known~\cite{chatzisavvas2009}, and a table of solutions for the 24 single-qubit Cliffords using 17 distinct angles and an average exchange-pulse count of 2.7 may be found in \onlinecite{andrews_quantifying_2019}. 

\begin{figure}
	\includegraphics[width=\columnwidth]{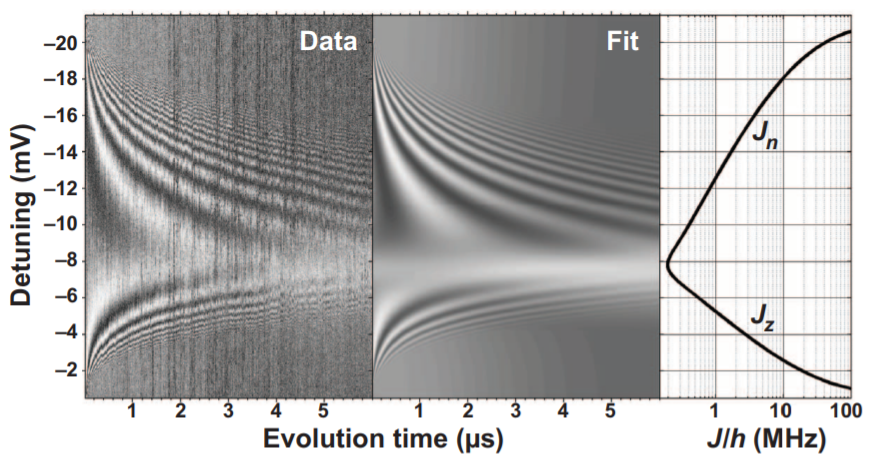}
	\caption{\label{fig:EO_oscillations} Gradient-free exchange oscillations from an isotopically enhanced Si/SiGe TQD \cite{Eng2015}. At very negative detunings, dots 2 and 3 are exchange-coupled and exchange oscillations are geometrically interpreted as a qubit rotation about $\hat{n}$ (see \reffig{fig:blochspheres}c); at less negative detunings, dots 1 and 2 are coupled, geometrically interpreted as rotation about $\hat{z}$.  Exchange increases exponentially with detuning. At $\epsilon = -7$~mV, both exchange couplings are active as would be required for operation in the RX reigime.
	}
\end{figure}

\onlinecite{Medford2013b} demonstrated complete EO qubit control in GaAs TQD.  Here $J_{12}(t)$ and $J_{23}(t)$ were controlled, sweeping the integrated phase during the exchange pulses. Singlet-triplet read-out via PSB was performed, and a self-consistent tomography technique showed the basic operation was consistent with theory.  Unfortunately, the decoherence free subsystem predicating EO control depends on homogeneous magnetic fields which maintain the total angular momentum of the spins, $S_{123}$, as a conserved quantum number.  Inhomogeneous magnetic fields, which are strong in GaAs due to hyperfine interactions (see \refsec{Subsec:Decoherence:HFT2}) prevent more than a few operations before leakage of the encoded qubit. A promising route to mitigate hyperfine effects is to implement EO systems in isotopically purified Si. \onlinecite{Eng2015}, \reffig{fig:EO_oscillations}, first demonstrated the longest composite single-qubit sequence (the 4-pulse $\pi$ rotation about the $\hat{y}$ axis) in a Si/SiGe QW structure with \Si\ content reduced to 800 ppm.  Calibrated operation of all composite gates for the 24 Clifford operations was later shown by \onlinecite{andrews_quantifying_2019} and will be further discussed in \refsec{Subsubsec:Implementations:EOQ:RB}.

\subsubsection{Resonant-exchange single-qubit gates} \label{Subsubsec:Implementations:EOQ:RX}

EO control in GaAs is more practical when multiple exchange interactions are constantly active, such as in the RX mode of operation (see~\onlinecite{Medford2013,Taylor2013} and \refsec{Subsec:Basics:EOintro}).  Such a qubit results from tuning a TQD to a regime where $J_{12}(t)$ and $J_{23}(t)$ are simultaneously active [see \reffig{fig:EO_oscillations}].  RX application is directly analogous to the rotating-frame Hamiltonian for single-spins (Appendix \ref{app:gates}), enabling the use of familiar rotating-frame RF sequences for decoupling and dynamic compensation.  As such, multipulse dynamical decoupling is a viable technique to mitigate hyperfine effects~\cite{Malinowski2017}.

In Si/SiGe, the valley degree of freedom has enabled a hybrid between RX and EO only qubits.  As discussed in \refsec{Subsec:Basics:Otros}, when two of the three electrons occupy a common dot whose valley splitting is within reach of microwave control, the resulting qubit has the same spin-encoding as an EO qubit, but the singlet and triplet states of the doubly-occupied QDs are perpetually split in energy by the valley splitting, analogous to the always-on exchange of the RX qubit.  A combination of microwave control, as in the RX qubit, and pulsed exchange, as in the EO qubit, similarly allow biasing to low charge-noise regions and complete qubit control, with demonstrations in isotopically natural Si showing fidelities in the mid-90\% range~\cite{Shi2012,Koh2012,Kim2014,Shi2014}.

\subsubsection{Two-qubit gates}  \label{Subsubsec:Gates:TQD:2QB}

There are three strategies for EO two-qubit gates.  One is to exploit the singlet-triplet character of the EO encoding and use capacitive charge-coupling in the high-detuning regime, as discussed in \refsec{Subsubsec:Gates:LD:2QB}. This would be possible both for EO and RX qubits, admit a wide variety of two-qubit gating modalities~\cite{Pal2014,Pal2015}, and be able to exploit long-distance transmission line couplers~\cite{Srinivasa2016}.   \onlinecite{Doherty2013} proposed a second strategy for the RX qubit modality, in which large exchange values are maintained within each TQD qubit, and a smaller exchange is activated to couple the two EO qubits.  The lowest order perturbative effect of the small inter-qubit exchange generates an entangling gate, with leakage effects occurring at higher order in the ratio of the inter- to intra-qubit exchange.  Both of these coupling mechanisms are susceptible to charge noise.  

The third method is to use true EO sequences between spins, in which charge-noise sensitivity during the two-qubit gate is no worse than that between spins during single-qubit operations. Schemes using a combination of single-pair and multi-pair exchange for the four-spin qubit were shown by \onlinecite{Bacon2000}, and pairwise entangling exchange sequences for the three-spin qubit were proposed by \onlinecite{DiVincenzo2000ExchangeQC} in the same year, although this latter sequence presents another subtle difficulty.  The decoherence-free subsystem of a TQD is insensitive at the single-qubit level to its total spin projection $m=m_1+m_2+m_3$, which may take values $\pm 1/2$ in the $S_{123}=1/2$ encoded subspace. This total spin projection is referred to as the ``gauge spin" and may be left unpolarized in single-qubit experiments. However, when two such qubits are combined, the two gauge spins may combine into a singlet or triplet states, and the action of intra-qubit exchange will behave differently in these two distinct subsystems.  The sequential gate from \onlinecite{DiVincenzo2000ExchangeQC} requires the gauge-spins to be in a triplet state, which would most likely be achieved via spin polarization.  Such polarization is generally not available in an EO system.

Fortunately, \onlinecite{Fong2011} derived a sequential gauge-independent CNOT sequence.  It has the same entangling action on the two-qubit encoded subsystem regardless of whether gauge spins are in singlet or triplet subspaces.  Such gauge invariance also means they function equivalently on four-spin EO qubits as three-spin EO qubits.  This sequence has a core gauge-invariant structure consisting of 12 $\pi/2$-pulses pairwise connecting five of the six spins spins (i.e. spin $\sqrt{\textsc{SWAP}}$ gates), some number of $\pi$-pulses to SWAP spins into place to achieve a particular connectivity of spins~\cite{Setiawan2014}, and some number of single-qubit pulses to convert to a desired operation. The CNOT gate implemented in a linear device architecture then summed to 22 pulses \cite{Fong2011}.  It was shown by \onlinecite{Zeuch2016} that the core entangling part of this gate may be decomposed into three uses of a primitive 5-spin sequence which swaps two spins depending on the encoded state of a single EO qubit; this decomposition and other constructions may lead to other two-qubit gate constructions beyond the Fong-Wandzura sequence family~\cite{Zeuch_2QBTQD_2020}. Other constructions based on decoupling concepts have also been proposed~\cite{vanMeter2019}.  Given the per-exchange error observed in \onlinecite{andrews_quantifying_2019}, a Fong-Wandzura CNOT sequence may have reasonable fidelity in existing device configurations, but it requires 6 dots with 5 well-calibrated exchange axes and sufficient valley splitting across the device.  New gate designs and Si/SiGe heterostructures with larger valley splittings may soon enable such a demonstration.

\subsubsection{Limits of fidelity - randomized benchmarking} \label{Subsubsec:Implementations:EOQ:RB}

\onlinecite{andrews_quantifying_2019} performed RB using a TQD in an isotopically enhanced Si/SiGe QW device using overlapping aluminum gates.     
The RB procedure was modified by randomly choosing whether a sequence of Cliffords composed to identity or $\sigma^x$.  Recalling that a measurement of a triplet state may correspond either to encoded $\ket{1}$, which responds to exchange, or to a $S_{123}=3/2$ leakage state, which does not, the presence of leakage could be deduced on average over many random sequences.  An error-rate per Clifford of 0.35\% was observed, with half of the error resulting from leakage.  
\onlinecite{Ha_archive} performed the same experiment using the SLEDGE architecture for similar Si/SiGe QW, and observed an error-rate per Clifford of 0.12\%.

The fidelity in this experiment depended on the details of the quantum control sequence. With substantial ``idle time" added between calibrated exchange pulses, error was limited by hyperfine dephasing which occurs due to leakage between degenerate $S_{123}=1/2,3/2$ states. If pulses are applied more quickly, the leakage per Clifford improves by simply outracing the leakage process, but another error limit is then reached due to the dynamic miscalibration of exchange pulses.  The limitations of such an error is a key outcome of RB, as it may be hard to observe in state or process tomography experiments, and it is ``contextual" (i.e. it depends on the control sequence employed).  Improved pulse delivery to the qubit as well as increased isotopic enhancement should further improve EO qubit operation fidelities.  The results however are very promising for exchange-based gates in silicon QDs in isotopically enhanced materials, as the noncontextual, non-hyperfine error from exchange pulses themselves (e.g. due to charge noise, see \refsec{Sec:Decoherence}), which occur an average of 2.7 times per Clifford gate, is substantially less than $10^{-3}$ in this experiment.   

\subsection{Alternative material platforms} \label{Subsec:Implementations:Alt}

\begin{figure}
	\includegraphics[width=1.0\columnwidth]{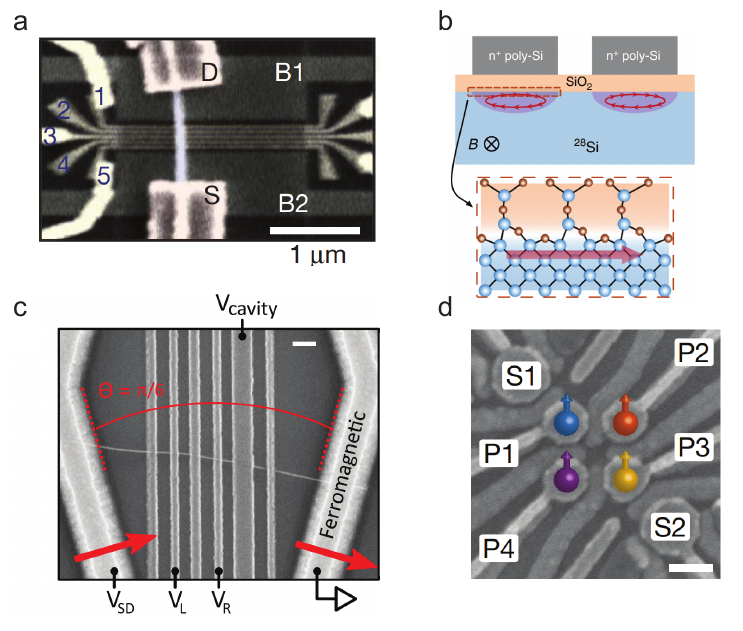}
	\caption{\label{Fig:others} a) Nanowire spin-orbit qubit, from~\cite{Nadj2010}. b) Spin-orbit qubit in SiMOS DQD, from~\cite{Jock2018}. c) Carbon nanotube qubit, from~\cite{Cubaynes2019}. d) Four-qubit quantum processor based on holes in Ge/SiGe~\cite{Hendrickx2021}.}
\end{figure}

Spin qubits have been realized predominantly using electrons in GaAs and Si, with recent encouraging results from holes in Ge as well. In this subsection we review results from several other materials systems, shown in Fig. \ref{Fig:others}, that have been investigated as suitable platforms for spin-based quantum information processing.

\subsubsection{Carbon nanotubes}  \label{Subsubsec:Implementations:Alt:CNT}

Carbon (C) is another group IV element that naturally occurs mostly in the form of a $I=0$ isotope (the natural abundance of $^{12}$C is 99\%).
One can therefore expect long electron spin decoherence times since the deleterious effects of the hyperfine coupling will be weak.  The fact that the valence electrons of C are in the atomic $p$-shell further reduces the hyperfine coupling (see Sec. \ref{Sec:Decoherence}). 

Carbon nanotubes (CNTs) are a one-dimensional form of carbon with an electronic bandstructure that can be either metallic or semiconducting \cite{Laird2015}.  The presence of a band gap in semiconducting CNTs allows for the formation of QDs using electrostatic gating \cite{Sapmaz2006}.  \onlinecite{Kuemmeth2008} measured the spin and valley degeneracies of single electrons in a QD formed in a clean CNT, as well as their coupling via spin-orbit interaction due to the CNT curvature.  PSB in the transport through a CNT DQD \cite{Palyi2010} enables measurement of the spin relaxation and dephasing times in $^{13}$C-enriched \cite{Churchill2009b} and natural \cite{Pei2012} CNTs.  \onlinecite{Pei2012} and \onlinecite{Laird2013} realized mixed spin-valley qubits in bent single-walled CNT devices, and \onlinecite{Cubaynes2019} observed the coupling of an electron spin localized in a CNT QD to a microwave cavity.

\subsubsection{Spin-orbit qubits}  \label{Subsubsec:Implementations:Alt:SOQubit}

As described in \refsec{Subsec:Principles:SOC} and \refsec{Subsubsec:Gates:LD:1QB}, electrical control of single spins can be achieved using the intrinsic SOC of a material and electrical driving. The theory for EDSR in a spin-orbit field predicts an effective ac magnetic field strength that is inversely proportional to $\lambda_{\rm SO}$, with a Rabi frequency that is proportional to the electronic $g$-factor \cite{Golovach2006}. While $\lambda_{\rm SO}~\sim$~8~$\mu$m in GaAs, heavier III/V compound semiconductors have a much shorter $\lambda_{\rm SO}$. For example $\lambda_{\rm SO}$ = 100 nm for InSb and 400 nm for InAs. In addition, the bulk electronic $g$-factor is 15 in InAs and 50 in InSb. These factors, combined with the small effective mass, resulted in the development of spin-orbit qubits beyond early demonstrations in GaAs \cite{Nowack2007}.

\onlinecite{Nadj2010} implemented EDSR in a bottom-gated InAs nanowire DQD. Due to the strong spin-orbit coupling present in InAs, the $g$-factors for the left and right dots were different, allowing for selective control of each spin. Fast Rabi frequencies were achieved $f_{R}$ = 58 MHz, but as in GaAs, the Rabi oscillations were strongly damped due to hyperfine coupling. Ramsey decay times $T_2^*$ = 8 ns and spin-echo coherence times $T_{2}$ = 50 ns were extracted. Possible reasons for the short $T_2$ relative to that observed in GaAs include the large, quadrupolar-split nuclear spin $I$ = 9/2 of indium and charge noise.

These experiments have been extended to different operating regimes and materials platforms. \onlinecite{Schroer2011} used EDSR to spectroscopically probe the strong anisotropy of the electronic $g$-factor in a InAs nanowire DQD. In a related experiment, \onlinecite{Nadj-Perge2012} performed spectroscopy of InSb spin qubits in the two-electron regime with highly anisotropic $g$-factors. Spectroscopy of the energy levels as a function of magnetic field allowed for a direct measurement of the spin-orbit gap $\Delta_{\rm SO}$, which was largest when the external magnetic field was parallel to the nanowire axis. Subsequently, it was shown that the EDSR driving mechanism strongly depends on the DQD energy level detuning \cite{Stehlik_harmonics}. While early experiments performed EDSR at high level detuning in effectively a single QD regime, EDSR when driven around $\epsilon=0$ exhibited a standard single photon resonance condition $hf = g\mu_B B$ as well as multiple harmonics $nhf = g\mu_B B$, with $n$ as high as 8. An even-odd dependence in the strength of the PSB leakage current was also observed. These observations were attributed to Landau-Zener physics, where near $\epsilon=0$ the DQD is repeatedly driven through avoided crossings in the energy level diagram \cite{Nadj2010,Schroer2011,Schroer2012,Petersson2012}. Similarly,  \onlinecite{Jock2018,Jock2021Valley} observed large spin-orbit and spin-valley couplings in SiMOS devices, leading to demonstrations of DQD spin-orbit singlet-triplet qubits. The stronger spin-orbit interaction of valence band states has also led to further experiments in Si and Ge hole qubits, described below.

\subsubsection{Holes in Si and Ge/GeSi}  \label{Subsubsec:Implementations:Alt:Holes}

Hole spin qubits have shown rapid progress in recent years, particularly in Si and Ge \cite{scappucci_germanium_2020}.
Holes have several attractive features: stronger SOI (and hence faster EDSR) as well as weaker nuclear hyperfine coupling, low in-plane effective mass, and the absence of degenerate valleys. However, the degenerate $p$-like states and SOI lead to strong band mixing. Strain and confinement further complicate the band mixing; the HH versus LH nature of the ground state differs for planar and nanowire devices, and strong structure- and tuneup-dependence of key parameters is expected.

Si holes can be confined in MOS QDs due to the large valence band offset \cite{ando_electronic_1982} and it is even possible to make ambipolar devices capable of confining electrons or holes \cite{betz_ambipolar_2014}.
Early demonstrations of PSB in planar \cite{Li2015} and SOI nanowire \cite{bohuslavskyi_pauli_2016} hole devices, followed by a qubit demonstration in the latter platform \cite{Maurand_holes_2016}, have occurred in the few-hole regime.
The relatively short $T_2^* = 60$ ns implies that decoherence is not hyperfine-limited at present.
Recent work showing shell filling \cite{liles_spin_2018} and single-hole $g$-tensor measurements in a planar MOS dot \cite{liles_electrical_2020} are promising for single-hole coherent operation.
In general, the observation of highly voltage-sensitive anisotropic $g$-tensors in MOS QDs \cite{crippa_electrical_2018,liles_electrical_2020} and Ge nanowires \cite{brauns_electric-field_2016} demonstrate the microscopic complexity of these devices. For few-hole Si nanowire MOSFET devices, it is predicted that $g$-tensor resonance can yield Rabi frequencies exceeding 600 MHz \cite{Voisin2016}. 400 MHz Rabi frequencies have been achieved with hole spin qubits fabricated in Ge/Si core/shell nanowires, with wide tunability of the SOC strength, Rabi frequency, and electronic $g$-factor \cite{Froning_hole}.

Holes in Ge have demonstrated promise on several fronts.
\onlinecite{higginbotham_hole_2014} showed extrinsic noise-dominated measurements of $T_2^* = 180$ ns in a Ge/Si core/shell nanowire and \onlinecite{watzinger_germanium_2018} demonstrated single-qubit control in the few-hole regime of Ge hut nanowire DQDs on Si. Recently, more emphasis has fallen on planar Ge/GeSi QWs; the compressive strain in such wells enforces a HH ground subband, with a HH-LH splitting of 10--50 meV and the in-plane effective mass is predicted to be about 0.06$m_0$ \cite{schaffler1997,terrazos_theory_2021}. The low disorder of this system and its ability to leverage design concepts and infrastructure from GaAs and Si/SiGe devices has enabled rapid experimental progress in the last few years.

\onlinecite{hendrickx_fast_2020} demonstrated
single- and two-qubit operation in the multi-hole regime with a single-qubit fidelity of 99.3\%. These results were quickly followed by reports of a single-hole qubit \cite{hendrickx_single-hole_2020}, singlet-triplet qubit \cite{jirovec_singlet_2021}, and hole manipulation in a 2x2 array \cite{lawrie_quantum_2020,van_riggelen_two-dimensional_2021,Hendrickx2021}.
Dephasing times out to 1 $\mu$s and $T_1$ $>$ 32 ms have been reported. As the theoretical hyperfine limits for Ge holes are not yet quantified, more work is needed to understand whether nuclear spins or transduced charge noise is the dominant dephasing mechanism.

\subsection{Discussion}
\begin{figure}[t]
	\centering
	\includegraphics[width=\columnwidth]{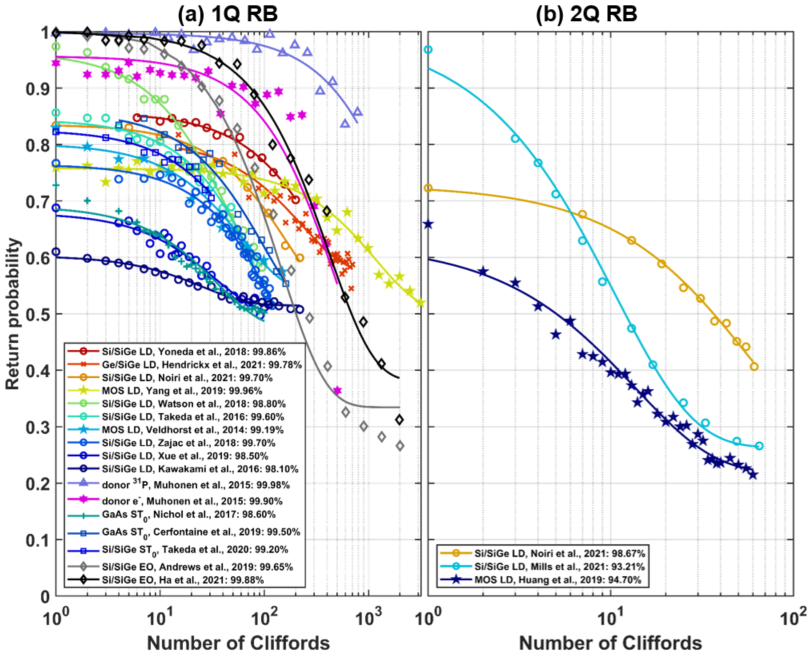}
	\caption{Fidelity of single qubit (a) and two qubit (b) gates in Si as evaluated by randomized benchmarking (RB).  In each experiment, an initial qubit state is prepared, random sequences of $N$ random Clifford gates $C_j$ are applied, a single Clifford recovery $C_R$ is applied to each random sequence which would, in absence of error, return the qubit or qubits to their initial state, and a read-out is performed.  The initial-state probability is plotted as a function of $N$.  Experimental data shown use different initialization, readout, and Clifford implementations, but in all cases a least-squares fit to an exponential decay with $N$ provides a fidelity benchmark.  A measure of state preparation and measurement (SPAM) fidelity is indicated by where each decay starts and saturates.  Ideally each $n$Q RB curve would saturate to return probability $1/2^n$ as $N\rightarrow\infty$, but leakage and SPAM errors generally lead to other saturation values. Note that two-qubit Clifford operations generally involve multiple two-qubit entangling, SWAP, and/or single-qubit gates.}
	\label{fig:allRB}
\end{figure}

Figure~\ref{fig:allRB} plots single- and two-qubit RB data drawn from many, but not all, recent publications on a common axis.  Return probability $P$ (that is, probability of returning to the $n$-qubit initialized state) is shown; some works report the difference $y$ between a measured return and a measured spin flip, which is converted here to return probability with the unbiased model $P\approx 1/2^n+(1-1/2^n)y$.  The $x$-axis counts the number of Clifford gates prior to a single (uncounted) recovery Clifford.  The fidelities indicated are per-Clifford-gate, which may include multiple primitive gates depending on the control modality. As can be seen, there is significant variance in the state preparation and measurement (SPAM) fidelity, approximately indicated by the intercept at 0 Cliffords, but recent spin-qubit fidelities, indicated by the decay rates of the exponential curves, are rather similar.  As randomized benchmarking requires a substantial amount of elements of a qubit apparatus to behave correctly, a key conclusion here is that all spin qubit technologies we have discussed have passed a critical test of showing the practical reality of performing quantum gates.  Fidelities still have room for improvement, but values greater than 99\% for basic gates are now firmly established across the semiconductor qubit community, and continue to advance.  

\section{Dephasing and decoherence}  
\label{Sec:Decoherence}
In the previous section, we assessed the operation and performancce of each major qubit type. For semiconductor qubits, the fidelity is limited by some dephasing or decoherence process. Consider the first exchange oscillations observed in GaAs and Si/SiGe (reproduced in \reffig{fig:J-gate}).  The first oscillation in each trace corresponds to a $\pi$-pulse, which may be considered a SWAP gate for an LD qubit, or a $Z$-gate for a ST$_0$ or EO qubit.  Critically, the visibility of this fringe is imperfect, and its reduction is a rough measure of the infidelity of the associated gate.  The loss of visibility is evident both as a function of time and as exchange is reduced. Why does the fringe visibility decay at the rate it decays?  What noise process is responsible for making these qubits imperfect, and if we identify that noise process, how may it be eliminated to improve fidelity? In this section we review the decoherence processes that are most relevant to semiconductor spin qubits.   

The processes leading to decoherence may be classified into a few important categories. In a \textit{relaxation} process, nondegenerate spin sublevels exchange magnetic energy with the environment (via phonons, photons, etc.).  In \textit{pure dephasing}, random energy-conserving elastic processes dynamically alter the phase of the qubit.  For \textit{inhomogeneous dephasing}, a single qubit's phase remains steady for long periods of time but is poorly synchronized with either a clock, another qubit, or with itself a significant period of time later.  
In the context of the Bloch equations, which describe NMR, the timescales corresponding to these effects are $T_1$ (relaxation), $T_2$ (decoherence), and $T_2^*$ (inhomogeneous dephasing) \cite{Abragam1961,Slichter2010,Vandersypen2005}.  ``Rotating frame" analogs of these timescales, which are relevant during coherent driving, include $T_{2,R}$ (the timescale for the decay of Rabi oscillations) and $T_{1\rho}$ (the timescale for decay when driving spins along a parallel rotating-frame axis).

\begin{figure*}
    \centering
    \includegraphics[width=1.8\columnwidth]{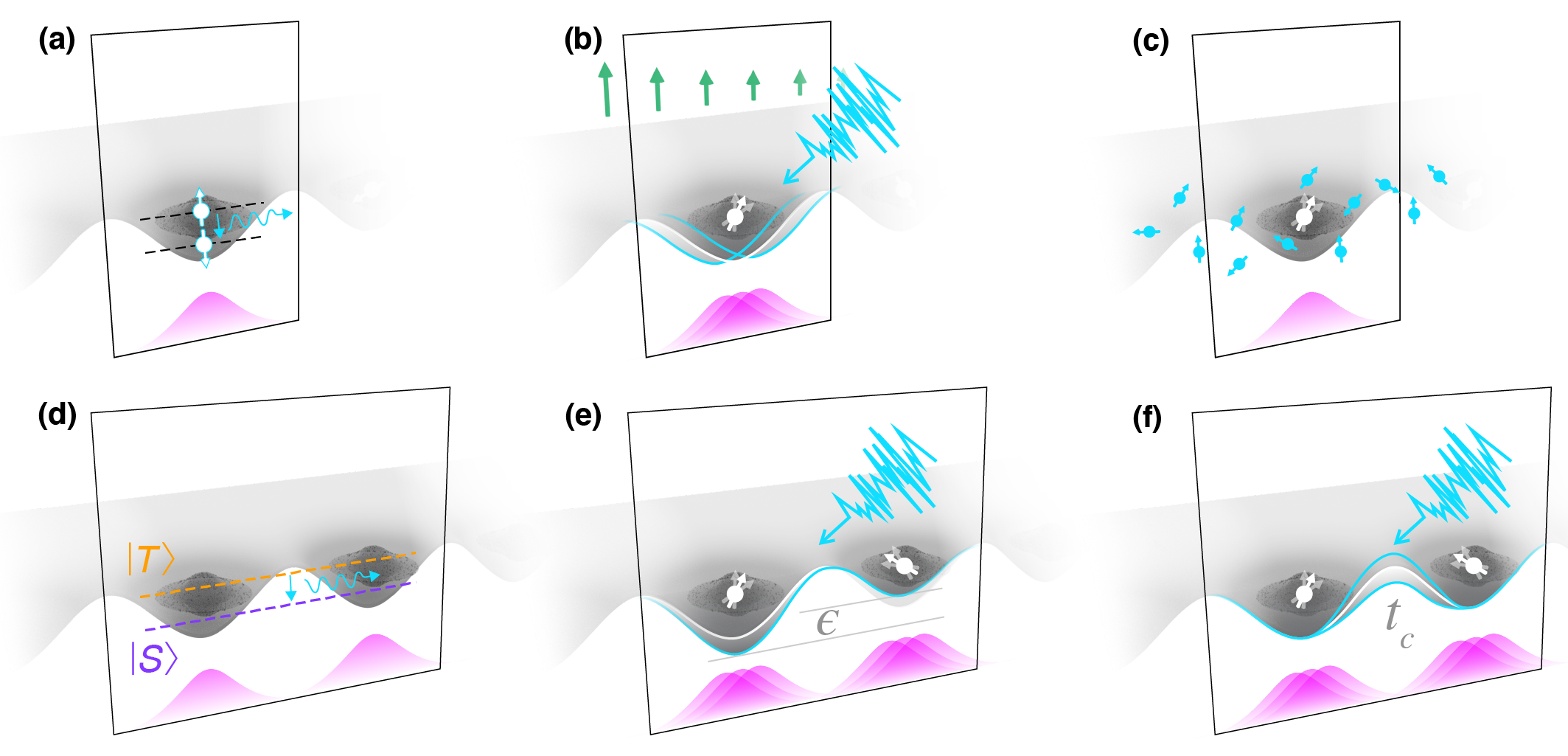}
    \caption{ Decoherence and relaxation mechanisms for spin qubits in semiconductor QDs or donors. The electronic spin state is indicated in blue, the electronic confinement potential (wavefunction) as a function of position in black (pink). Energy levels are shown as horizontal black bars. (a)--(c) Decoherence and relaxation mechanisms for a LD qubit, (d)--(f) for singlet-triplet qubits in DQDs. (a) Spin relaxation through emission (absorption) of energy quanta (blue, e.g. phonons) to or from the environment. (b) Charge noise (blue) leading to a fluctuating confinement potential and electronic wavefunction. When SOI is present, charge noise leads to spin dephasing. (c) Electron spin dephasing due to the hyperfine coupling to nuclear spins (blue). (d) Singlet-triplet spin relaxation. (e) Charge noise affecting detuning $\epsilon$. (f) Charge noise affecting interdot tunneling $t_c$.}
    \label{fig:decoherence}
 \end{figure*}

While the Bloch equations successfully describe the dynamics observed in many ensemble NMR and ESR experiments, the phenomenological exponential decay they describe is seldom observed for semiconductor spin qubits (e.g. the time decay in \reffig{fig:J-gate} is Gaussian).  An improved language for describing dephasing and decoherence phenomena in terms of the power spectral density (PSD) of the responsible environmental noise mechanism and the filter on that noise provided by the experiment which probes that decoherence mechanism, is the \textit{filter function formalism}~\cite{Ithier2005}.  We briefly review this formalism in \ref{SubSec:Filter_Function}, and then proceed to describe some of the most prominent physical noise sources that cause dephasing and decoherence in spin qubits (see Fig.~\ref{fig:decoherence}).  Many, more thorough reviews of the formalism are available; for example, see \onlinecite{Chirolli2008}.

\subsection{Filter function formalism}\label{SubSec:Filter_Function}

\subsubsection{$T_1$ via noise correlation function}\label{SubSubSec:T1}
A basic model for noise impacting spin qubits is captured by a spin's coupling to a noisy magnetic field, via Hamiltonian
$\ts{H}{noise} = -\hbar\vec{b}(t)\cdot \vec{S}$, where $\vec{b}$ is a vector angular frequency describing a stationary, zero-mean noise process.  For example, if the noise is a literal fluctuating magnetic field $\delta \vec{B}(t)$, $\vec{b}(t)=g \mu_B \delta \vec{B}(t)/\hbar$ and if $\vec{b}$ is parallel to a large applied magnetic field, $b(t)=|\vec{b}(t)|$ is the fluctuation of the spin's Larmor frequency.  

Starting from this Hamiltonian, if we presume an applied magnetic field along the $z$ axis providing spin Zeeman splitting $\hbar\omega_0$, and define $b^\pm(t)=b^x(t)\pm i b^y(t)$, then Bloch-Redfield-Wangsness (BRW) theory~\cite{Abragam1961} approximates that $T_1$ at temperatures $T\ll \hbar\omega_L/k_B$ is given by
\begin{multline}\label{eq:T1BRW}
	\frac{1}{T_1} = 
	\frac{1}{4}
	\int_{-\infty}^\infty \biggl(\langle b^-(0) b^+(t)\rangle +\langle b^+(0) b^-(t)\rangle \biggr) e^{-i\omega_L t}dt  
	\\
	= \frac{1}{2} S_{b}^\perp(\omega_L).
\end{multline}

Exponential relaxation at rate $1/T_1$ is due to the density of noise in transverse fluctuating magnetic fields at the Larmor frequency $\omega_L$, an intuitive result given that noise at $\omega_L$ is required to overcome the  Zeeman splitting $E_Z=\hbar\omega_L$ between opposite spin states. 

In the context of BRW theory, energy-conserving dephasing processes are described as exponential decay with rate $1/T_2$: 
\begin{equation}
	\frac{1}{T_2} = \frac{1}{2T_1}+
	\int_{-\infty}^\infty \langle b^z(0) b^z(t)\rangle dt.
\end{equation}
Here, the dephasing rate depends on the spectral density of longitudinal noise at zero frequency $S_{b}^{zz}(0)$. This expression suggests that only true dc noise contributes to dephasing. As discussed in the next section, however, noise at low frequencies also contributes to dephasing. The filter-function formalism provides a prescription for understanding how such noise contributes to dephasing.

\subsubsection{Filter function derivation}

The concept of a filter function has been formalized in a quantum information theory context for qubits by \onlinecite{Ithier2005}, \onlinecite{Uhrig2007}, \onlinecite{Cywinski2008}, and \onlinecite{Green2012}. Notable extensions and higher-order corrections, especially for the slow noise processes typical of spin qubits, are detailed in ~\onlinecite{Barnes2016}.  The filter function derivation utilizes an interaction picture, in which $\vec{S}$ acquires time dependence due to the action of the control of some experiment, 
\begin{equation}
	\tilde{\vec{S}}(t)=\ts{U}{control}^\dagger (t) \vec{S} \ts{U}{control}^{\phantom{\dagger}}(t).
\end{equation}
As a result of $\ts{\tilde{H}}{noise}(t)=\vec{b}(t)\cdot\tilde{\vec{S}}^z(t)$, a spin or qubit evolves according to a quantum process during a total time $T$ $\rho(T)=\Lambda[\rho(0)]$.   If $\Lambda$ is decomposed into a Choi matrix $\Lambda[\rho]=\sum_{jk} \lambda^{jk}\sigma^j\rho \sigma^k$, for Pauli matrices $\sigma^j$ and including the $\sigma^0$ as the identity matrix, then the infidelity of this noise process is taken as $1-\lambda^{00}$ which is cast into a decay function $\exp[-\chi(T)]$.  Using cumulant expansion considerations, $\chi(T)$ is expanded to second order in the noise field $\vec{b}(t)$ and filter functions $F^{\alpha\beta}(\omega T)$ depend on $U_{\rm control}$~\cite{Cywinski2008} 
and are defined to satisfy
\begin{equation}
	\chi(T) = \int_0^\infty \frac{d\omega}{2\pi\omega^2} \sum_{\alpha,\beta=x,y,z}F^{\alpha\beta}(\omega T) S_b^{\alpha\beta}(\omega), 
	\label{eq:chidef}
\end{equation}
where the single-sided noise spectral density corresponds to the noise-correlation function via 

\begin{equation}
	S_b^{\alpha\beta}(\omega)=2\int_{-\infty}^\infty \langle b^\alpha(0)b^\beta(t)\rangle \cos(\omega t) dt, \qquad \omega > 0.
\end{equation}

In the most commonly encountered situation where $\omega_0 \gg |\vec{b}|$, one appeals to a rotating reference frame in which the perpendicular terms $b^\pm(t)$ oscillate at frequency $\omega_0$, and therefore integrate to noise contributions of order $|\vec{b}/\omega_0|^2$, which we neglect. For dominant noise terms, which we discuss later in this section, this secular approximation is valid for applied magnetic fields above a few mT, and in some cases remains valid even in fields as low as the Earth's magnetic field.  However, transverse noise terms should not be forgotten, as they do play roles in multi-pulse experiments in regimes in which pulses occur at rates comparable to $\omega_0$, e.g., in fast-pulsing and low-magnetic field cases.  It's often assumed that the control pulses described in $\ts{U}{control}$ are instantaneous $\pi$-pulses about an axis orthogonal to the $z$-axis, from which it follows that $\tilde{S}^z(t)$ may be written as $\tilde{S}^z(t)=r(t)S^z$, where $r(t)$ takes only the values $\pm 1$, switching between the two for each $\pi$ pulse.  Under these simplifications, only $z$ components of $\vec{b}$ and only $S_b^{zz}$ are important, and we may therefore drop component superscripts.  Moreover, it may easily be derived that the filter function $F(\omega)=F^{zz}(\omega)$ is simply
\begin{equation}
	F(\omega T)=\biggl|\frac{\omega}{2} \int_{0}^T r(t) e^{i\omega t} dt\biggr|^2.  
\end{equation}

\subsubsection{Dephasing time $T_2^*$}
With the filter function derived, we are now in a position to calculate dephasing (decoherence) rates $T_2^*$ ($T_2$). $T_2^*$ is the rate of decay during a ``free evolution" experiment, analogous to a free-induction decay experiment in magnetic resonance. For single-spin qubits, in which measurements of $S^z$ are performed (see Sec.~\ref{Subsec:Basics:LDintro}), the relevant experiment is a time-ensemble of Ramsey experiments, in which the spin is prepared along an axis orthogonal to a large applied field using a single RF pulse, precession happens for a swept time duration $T$, and the spin then undergoes a second RF pulse of known phase, mapping the $xy$-plane precession onto the ensemble-measured observable $\langle S^z \rangle$.  In either case, $r(t)$ is constant for the duration $T$, and $F(\omega)$ is proportional to $\sin^2(\omega T/2)$.  For very slow noise phenomena (i.e., when $S_b(\omega)$ is strongly peaked near $\omega=0$, as it is for hyperfine noise to be discussed in \refsec{Subsec:Decoherence:HFT2}), the shape of the decay curve $\exp[-\chi(t)]$ then predicts decay of oscillations going as $\exp[-(t/T_2^*)^2]$, which defines $T_2^*$.  More generally, the structure of the low-frequency noise may lead to a power law decay, $\exp[-(t/T_2^*)^\alpha],$ including $\alpha=1$ for white noise; either way, $T_2^*$ is defined via $\chi(T_2^*)=1$.

The interpretation of $T_2^*$ defined above requires some care, as spin qubit systems often violate the assumption of ergodicity (i.e. that a series of Ramsey measurements made sequentially in time accurately reflects an ensemble average.)   For example, for $S_b(f) \propto 1/f^\alpha$, and without a low-frequency cutoff, $\chi(T)$ diverges. The usual resolution of this divergence is to introduce a low-frequency cut-off determined by the total amount of time used to average an experiment.  Formal treatments of such low-frequency cut-offs can be found in \onlinecite{Burkard2009},
\onlinecite{Barnes2016}, and \onlinecite{madzik_controllable_2020}. \onlinecite{Eng2015} and \onlinecite{Jock2018} in particular presented measured $T_2^*$ as a function of averaging time in Si/SiGe and SiMOS dots, and in both the logarithmic dependence on averaging time expected for $1/f^\alpha$ noise is observed. For a spin qubit in GaAs, the dephasing time was measured to be dependent on the acquisition time  \cite{Delbecq2016}. In short, a measurement of $T_2^*$ for a qubit does not by itself indicate an intrinsic property of the qubit, as it depends on experimental averaging details.

The relationship between $T_2^*$ and the overall performance of a qubit depends critically on control. Since  $T_2^*$ results from very slow drifts in the qubit frequency, it is well known that it can be compensated for via dynamical decoupling or noise compensation sequences. For GaAs spin qubits, $T_2^*\sim$10~ns (see \refsec{Subsec:Decoherence:HFT2}), meaning that noise compensation is critical for scaling into useful processors. For silicon, $T_2^*$ is generally much longer.

\subsubsection{Decoherence time $T_2$ and rotating frame timescales}

If compensation is employed, its efficacy will depend on how quickly $S_b(\omega)$ reduces with $\omega$ relative to the available speed of control.  This efficacy is somewhat captured by the parameter $T_2$, often taken as the $1/e$ point ($\chi(T_2)=1$) for decay in the Hahn spin-echo experiment, in which the Ramsey experiment described above is interrupted halfway by a single $\pi$-pulse applied orthogonal to the $z$-axis.  Then $r(t)$ has one switch from $+1$ to $-1$, and $F(\omega)=4\sin^4(\omega T/4)$. Since $F(\omega) \propto \omega^4$ as $\omega\rightarrow 0$, the Hahn echo cancels noise at $\omega=0$ and passes noise at higher frequency. Once again, $T_2$ is defined relative to the experiment used to measure it~\cite{Cywinski2008}.

Coherent driving of a spin will also extend pulse sequence times, as Rabi flopping at frequency $f_{\rm Rabi}$ acts as continuous dynamical decoupling. Two types of noise may be relevant.  First, there may be noise on the control field itself, e.g. $f_{\rm Rabi}=f_{\rm Rabi}(t)$ due to charge noise in EDSR, and second, noise from spurious magnetic fields such as Overhauser fields.  If the phase of the microwave signal causes rotations about an axis on the Bloch sphere equator and if the spin is initialized along $\hat{z}$, the decay time of Rabi oscillation is $T_{2,R}$, which is different in general from the $T_2^*$ of a freely evolving spin. If the spin is initialized along an axis on the Bloch sphere equator and then driven along that same axis, the associated decay timescale is $T_{1\rho}$.  If $f_{\rm Rabi}(t)\gg \int S_b(\omega)d\omega$, the decay exponent $\chi(T)$ for either experiment is likely to be limited by noise on $f_{\rm Rabi}(t)$, with a filter function comparable to that for $T_1$ above and relating to $f_{\rm Rabi}(t)$'s phase stability.  If $f_{\rm Rabi}$ is highly stable and only a transverse magnetic noise $b(t)$ is present, this experiment will correspond to shifting the filter function for noise on $b$ by $-f_{\rm Rabi}$, which leads to drastically slower decay for the same amount of noise $S_b(\omega)$.  

\subsubsection{Filters for multi-spin qubits}

Natural generalizations of the filter function formalism may be made for qubits composed of several spins (ST$_0$, EO, etc.).  In this case generalized spin-operators and magnetic fields are defined (see Table~\ref{table:basis}), 
which may be related back to physical spin operators usually with suitable sums over QDs~\cite{Kerckhoff_quadrupolar_2020}.  A key difference relative to LD qubits, however, arises from the fact that both ST$_0$ and EO qubits are degenerate when idle.  This means the $T_2^*$ and $T_2$ experiments track dephasing for two degenerate levels (a two-spin singlet state is prepared, allowed to mix with triplet due to noise, and compared again to a singlet).  As observed experimentally by \cite{Johnson2005,Koppens2005}, rapid hyperfine mixing can occur at very low magnetic fields, where any direction of $\vec{b}$ is important in each dot, leading to more complex filter functions~\cite{Kerckhoff_quadrupolar_2020}.  In these qubits, the analog of a Rabi experiment involves preparing two spins in a coherent superposition of degenerate singlet and triplet states, and then driving oscillations between them with a DC voltage bias that induces exchange.  Oscillations occur at frequency $J$, and dephasing occurs due to noise on $J(t)$, with decay envelope given by the filter function equations above, replacing $S_b(\omega)$ with $S_J(\omega)$.  A constant pulse still has filter function proportional to $\sin^2(\omega T/T)$, and modifications employing $\pi$-pulses or rotating-frame-type experiments are also possible~\cite{Dial2013,Eng2015}, enabling characterization of the charge noise $S_J(\omega)$.

\subsubsection{Non-Markovian and contextual noise}

The filter function formalism presented assumes an independent, stationary noise source.  However, dephasing, decoherence, and relaxation timescales have been observed to depend on the very control sequences used to measure them, due to such phenomena as RF heating, DNP (Sec. \ref{Subsec:Gates:ST}), and pulse-driven nuclear spin dynamics~\cite{madzik_controllable_2020,Kerckhoff_quadrupolar_2020}; such measurement-induced back-action is not easily accounted for in filter function theory.  Just as important, dephasing of qubits is made relative to a clock or a timed control sequence.  If that clock or control sequence dephases, it is equivalent to the qubit dephasing from a quantum control standpoint. Unfortunately, measurements of long dephasing and decoherence times are a necessary but not sufficient criterion for high fidelity qubit control.

\subsection{Spin dephasing due to hyperfine interactions} \label{Subsec:Decoherence:HFT2}

\onlinecite{Burkard1999,Coish2004} and \onlinecite{Coish2005} predicted that dephasing due to the hyperfine interaction between an electron spin and the spins of many lattice nuclei in the host crystal would be a significant challenge to spin qubits. Indeed, early experiments in GaAs DQDs extracted $T_2^* \sim 10$~ns and $T_2$ $\sim$ 1 $\mu$s ~\cite{Petta2005}.  Fortunately, hyperfine induced spin dephasing can be mitigated by a variety of methods, including isotopic purification, nuclear polarization, and dynamic decoupling.  In this section, we overview the physics of the hyperfine interaction in QDs and donors, including nuclear dynamics, followed by a summary of nuclear-limited measurements of $T_2^*$ and $T_2$.  Despite significant research in understanding hyperfine dynamics~\cite{Chekhovich2013}, questions remain about the ultimate limits of hyperfine coherence and the fundamental timescales for nuclear spin dynamics.

The dominant effect of nuclear spins is static dephasing mediated by the hyperfine interaction [\refeq{eq:contactHF}], impacting $T_2^*$. In this case, $T_{2,\infty}^*$ depends on $\sigma^2_b$, the variance of the magnetic field experienced by electron spins due to full randomization of the nuclear magnetization. 
The variance of the effective angular-frequency magnetic field $\vec{b}=A_k \vec{I}_k$ for an ensemble of independent nuclei, all with spin $I$, is summed:
\begin{equation}
\sigma^2_b = \frac{I(I+1)}{3}\sum_k A_k^2.
\label{eq:sigsq}
\end{equation}
The factor of 3 in the denominator is relevant at high field, where the $I^+_kS^-$ flip-flop terms average away at a timescale negligibly short relative to dephasing experiments. At zero field, all three nuclear spin directions are of relevance and $\sigma^2_b$ is three times higher.

Nuclear fluctuations are very slow in both GaAs and Si (of order  1 ms) relative to the $\mu$s timescales of qubit coherence measurements~\cite{Reilly2010,Ladd2005,madzik_controllable_2020}.  As such, $S_b(\omega)$ is strongly peaked at $\omega=0$ and the Ramsey decay is Gaussian.  For a  LD qubit, the envelope decay for an experiment lasting time $T$, following Eq.~(\ref{eq:chidef}), goes as $\exp{-\sigma_b^2 T^2/2}$, and $T_{2,\infty}^*=\sqrt{2}/\sigma_b$.  For a ST$_0$ qubit, the assumed independent, identical distributions of static noisy fields from two dots have adding variances, and ST$_0$ FID (in which a singlet is prepared, allowed to evolve for time $T$, and then measured) decays as $\exp{-\sigma_b^2 T^2}$, with $T_{2,\infty}^*=1/\sigma_b$.

All Ga and As isotopes have nuclear spin $I=3/2$, leading to $T_{2,\infty}^*$ $\sim$ 10 ns \cite{Petta2005}. In Si, however, only 4.7\% of naturally occurring Si isotopes feature non-zero nuclear spin (\Si, $I$=1/2), and in Ge only 7.8\% of naturally occurring isotopes (\Ge, $I$=9/2) have non-zero spin. The reduced number of spin carrying nuclei in natural Si (no enrichment) leads to a significant improvement in $T_{2,\infty}^*$. Further increases are feasible using isotopic enrichment, which was demonstrated as far back as 1958, when a $^{31}$P-doped sample with under 1200 ppm \Si\ was observed to have longer $T_2^*$ than a natural sample using ensemble ESR~\cite{Feher1958,Gordon1958}.

In addition to isotopic content, the overall size of the electronic wavefunction also impacts $\sigma_b^2$.  With the envelope wavefunction overlapping many nuclear sites, Eq.~(\ref{eq:sigsq}) leads to
\begin{equation}
T_{2,\infty}^* \propto \sigma_b^{-1} \propto \sqrt{\frac{N}{p_I}},
\end{equation}
where $N$ is the total number of nuclei for which $|\psi(\vec{r}_k)|^2$ is larger than some threshold and $p_I$ is the probability that a given lattice nucleus has spin.  Therefore, in the many-nuclei limit, electronic wavefunctions enveloping a larger \emph{number} of spin-carrying nuclei have a longer dephasing time, due to averaging over more nuclear spins. This occurs because the individual $A_k$ diminish as the electron wavefunction spreads out over more nuclear spins. However, this scaling cannot extend to very small wavefunctions, since $T_2^*$ cannot go to zero.  In fact, for small dots with low $p_I$, the value of $T_{2,\infty}^*$ varies widely from device to device [the standard deviation of $1/(T_{2,\infty}^*)^2$ scales as $p_I(1-p_I)/N^3$]. Whether dephasing occurs quickly or slowly will depend randomly on how often spinful nuclei are located in regions of the electron wavefunction in which $|\psi(\vec{r}_k)|^2$ is high. In a device such as a \isotope{P}{31}\ donor, it is plausible to have only one spinful nucleus, the \isotope{P}{31}\ nucleus itself, which may be coherently controlled and rarely undergoes randomization. Under these circumstances our approximations for the ergodic $T_{2,\infty}^*$ do not apply~\cite{madzik_controllable_2020}.

In order to observe a pure dephasing effect on a single qubit due to nuclear spins, the nuclear spins cannot be frozen; they must fluctuate on the timescale of the measurement. Moreover, if nuclear hyperfine effects limit $T_2$, and if we wish to compensate for nuclear dephasing, some notion of how quickly nuclei change their polarization state is required. In a dense, homogeneous crystal of nuclear spins, flip-flops driven by the dipole-dipole interaction happen frequently, causing Brownian spin diffusion with a noise spectrum $S_b(f)$ scaling as $1/f^2$~\cite{Abragam1961}.  In sparse spin systems and in the presence of field gradients and highly localized dot or donor electrons, the strength of this coupling varies drastically between nuclear spin-pairs, as it depends on the inverse cube of the distance between randomly placed nuclei and any changes in their local magnetic field due to field gradients or the hyperfine field of electron spins.  Hence pairs will have varying flip-flop rates, and the noise spectrum $S_b(f)$ might be expected to be closer more $1/f$, as anticipated from a broad range of two-level fluctuators.  Such a spectrum is observed in silicon systems~\cite{Eng2015,madzik_controllable_2020}.  Solving with more rigor the problem of how coupled nuclear spins impact the coherence of a central electron spin, the ``central spin problem," depends on the rich and efficacious use of many-body-physics approximations.  Theoretical headway on this problem occurred in the spin-qubit context employing coupled-cluster expansion techniques, which were able to theoretically predict Hahn $T_2$ values in silicon donor and other systems~\cite{PhysRevLett.105.187602}, but still do not capture all relevant effects, especially the very slow dynamics governing $T_2^*$.  

Experimental measurements of spin coherence have been performed in a variety of systems. In GaAs QDs, $T_2$ $\sim$ 1 $\mu$s for the ST$_0$ qubit \cite{Petta2005}. \onlinecite{Koppens2008} measured $T_2$ = 500 ns in GaAs using ESR. In silicon with $<$50~ppm \Si\ content, ensemble ESR measurements of electrons bound to \isotope{P}{31}\ donors
yielded $T_2$ $\approx$~2~s~\cite{Tyryshkin2012}. Again, in isotopically enriched Si, \onlinecite{Saeedi2015} demonstrated an ensemble nuclear spin coherence time of over 39 minutes. Hahn echo measurements of $T_2$ in electron spin qubits in isotopically purified silicon at fields greater than 100~mT gave comparable results, showing coherence times on the order of 1~ms in the small donor system (with larger hyperfine gradients)~\cite{Muhonen2014}, of order 1.2~ms in the somewhat larger SiMOS dot systems~\cite{Veldhorst2014}, and of order 30 $\mu$s to 1~ms in the larger Si/SiGe QD systems~\cite{Sigillito2019,Kawakami2014,Kerckhoff_quadrupolar_2020}.  \onlinecite{Stano2021} have compiled a thorough list of coherence times measured in semiconductor spin qubits to-date.  Some of these studies involve samples with micromagnets for EDSR, where the $T_2$ and $T_2^*$ values are not limited by nuclear spins at all, but rather by charge noise transduced to magnetic noise due the the micromagnet field gradient.
We address such effects in Sec.~\ref{sec:gradient_dephasing}.

\subsection{Phonon-mediated spin relaxation} \label{Subsec:Decoherence:SOT1}
As discussed in \refsec{SubSubSec:T1}, spin relaxation requires energy exchange with the environment.
For typical QD spin splittings, this often occurs via the emission of acoustic phonons coupled with a spin-mixing perturbation such as spin-orbit, hyperfine, or external magnetic gradient \cite{Hanson2007,Zwanenburg2013}. In polar semiconductors such as GaAs, the dominant phonon interaction is piezoelectric \cite{Khaetskii2001}. In nonpolar materials like Si, the deformation potential plays a key role~\cite{Tahan2014}. For single-phonon-mediated decay, the spin relaxation rate Eq. \ref{eq:T1BRW} can be expressed in Fermi golden rule form as
\be
\dfrac{1}{T_1} = \dfrac{2\pi}{\hbar}|\braket{\tilde{\uparrow}|H_p|\tilde{\downarrow}}|^2\rho(\Delta E),
\ee
where $\rho(\Delta E)$ is the density of modes (photon or phonon) at the level splitting $\Delta E$, equal to the Zeeman splitting for single-spin relaxation, and the electron-phonon interaction $H_p$ couples the spin states $\ket{\tilde{\uparrow}}$, $\ket{\tilde{\downarrow}}$, which are renormalized by the spin-mixing mechanism.

Evaluation of these rates for spin-orbit-mediated decay under the dipole approximation (valid for small energies) leads to characteristic scaling laws $1/T_1 \propto B^5$ and $B^7$, respectively, for piezoelectric-limited and deformation potential-limited one-phonon relaxation, in good agreement with single-spin $T_1$ measurements in GaAs \cite{hanson2003,fujisawa_allowed_2002} and Si \cite{xiao_measurement_2010}. The same microscopic interactions contribute to singlet-triplet decay in single QDs and DQDs; however, since the relevant spin splitting in those cases is usually exchange- rather than Zeeman-limited, and the excited state structure is strongly dependent on Coulomb interactions and confinement potential, the scaling and bias dependencies can change drastically \cite{Meunier2007,golovach_spin_2008,danon_spin-flip_2013}.

One recent development is the observation of spin relaxation ``hot spots'' when the spin splitting is resonant with another excited level. Hot spots are especially relevant in Si QDs where typical valley splittings of order 100~$\mu$eV can equal Zeeman energies at Tesla-scale magnetic fields. Similar to spin-orbit coupling, spin-valley coupling admixes the excited spin states with valley states of opposite spin, which then decay to the ground state via phonon or photon emission \cite{huang_spin_2014}. Valley relaxation is generally dominated by valley-orbit mixing due to interfacial disorder and is typically much faster than pure spin relaxation \cite{Tahan2014,penthorn_direct_2020}.
This leads to large enhancements in the single-spin relaxation rate when the spin and valley splittings are brought into resonance by tuning the magnetic field; relaxation suppression or ``cold spots'' due to the interplay of disorder are also possible \cite{Yang2013,Hosseinkhani2021}.
The Zeeman energy of relaxation hot spots can be used to directly measure valley splittings in Si QDs \cite{Yang2013,petit_spin_2018}. 
Spin-valley effects also play an important role in donor spin relaxation, as thoroughly discussed in \onlinecite{Zwanenburg2013,Tahan2014}; the weak interactions in these systems allow observations of spin lifetimes of up to 30 s in donor states \cite{watson_atomically_2017}.
Electric field-induced spin-orbit coupling can also significantly enhance the donor spin relaxation rate \cite{weber_spinorbit_2018}.

In general, spin lifetimes are shortened when spin-charge hybridization is enhanced. The spin-valley hotspots described above are one such example of this; another is the enhancement of interdot spin relaxation observed in GaAs DQDs at particular detunings where the excited spin state of one dot is resonant with an orbital energy in the other~\cite{srinivasa_simultaneous_2013}.
The directional dependence of SOC also leads to an anisotropic dependence of spin $T_1$ on the in-plane magnetic field orientation \cite{Scarlino2014}.
Furthermore, external magnetic gradients can provide a ``synthetic'' spin-orbit field that contributes to spin relaxation. In the dipole approximation limit, $1/T_1$ $\propto$ $B^5$ for deformation potential interactions due to a fixed external gradient. Experimentally, spin relaxation rates at high fields in micromagnet devices tend to increase more slowly~\cite{borjans_single-spin_2019,hollmann_large_2020}, possibly due to phonon bottleneck effects at these energies.

Hyperfine interactions provide yet another pathway for spin relaxation.
Coupling of an electron with local nuclear spins admixes spin states of different orbitals \cite{erlingsson_hyperfine-mediated_2002}, enabling relaxation via phonon or photon emission. The resulting single-spin relaxation rate scales as $B^3$ or $B^5$ for piezoelectric and deformation potential phonons, respectively. Hyperfine-induced relaxation in Si QDs is typically expected to be negligible due to the paucity of spinful nuclei~\cite{Tahan2014}. 
\onlinecite{camenzind_hyperfine-phonon_2017} observe long spin $T_1$ of around $57\pm 15$ s in a GaAs QD at $B = 0.6-0.7$ T, increasing as $B^3$ at low fields and insensitive to field orientation, strongly suggesting hyperfine-limited relaxation.
Hyperfine-induced relaxation can also lift Pauli spin blockade at low magnetic fields, as observed experimentally for (1,1) triplet decay in a GaAs DQD as a function of detuning \cite{Johnson2005}.

Single-phonon relaxation processes typically dominate at low temperatures, but two-phonon processes can become relevant at high temperatures. This leads to distinct temperature scalings which are observed in spin lifetime measurements above 200 mK in SiMOS \cite{petit_spin_2018} and Si/SiGe QDs \cite{borjans_single-spin_2019}. At small spin splittings, e.g., low magnetic fields for single spins or modest exchange splittings in singlet-triplet states, phonon-assisted decay is suppressed by the reduced density of states and (in Si) suppression of deformation potential coupling at long wavelengths. In such cases the dominant relaxation process may instead be mediated by charge noise, as described below. 

Overall, the long spin lifetimes in semiconductors mean that current spin qubit gate fidelities are rarely limited by $T_1$. In contrast, spin relaxation can lead to errors in spin readout when the readout time becomes comparable to $T_1$. The rich physics of spin relaxation rewards close study as it offers many insights into the microscopic physics of spin qubits.

\subsection{Charge noise} \label{Subsec:Decoherence:SOT2}
\label{sec:gradient_dephasing}
Charge noise can significantly limit the performance of spin qubits. In principle, a spin does not interact with fluctuating electric fields, but for all qubits we have discussed in this article, there is some form of spin-to-charge coupling, allowing charge noise to dephase, decohere, or otherwise reduce the operational fidelity of spin qubits.  Charge noise generally refers to random electric fields which occur at the spin location, which may be caused by fluctuating defect states in the device gate stack, by crystal deformations from phonons~\cite{Hu2011}, by spurious voltage noise transmitted through control gates, or by random charge motion from anywhere else in the device, such as the measurement channel.  Semiconductor charge noise processes typically have a $1/f$ noise spectral density~\cite{Dutta1981}, but white noise (e.g. Johnson-Nyquist noise) may also be present, usually at lower levels than $1/f$ noise.  Noise sources can to some extent be distinguished by how their spectral character translates to relaxation (via the filter function formalism discussed in Sec.~\ref{SubSec:Filter_Function}) and their temperature dependence~\cite{Beaudoin2015}. 
 
Although $1/f$ noise varies significantly from device to device, measurements in GaAs dots, SiMOS dots, Si donors near MOS gates, and Si/SiGe dots all see charge-noise induced energy fluctuations in the range of $A_\mu=0.1-10$ $\mu$eV/$\sqrt{\text{Hz}}$ at 1~Hz, meaning the chemical potential of the charge carrying the spin has noise spectral density $S_\mu(f)=A_\mu^2(1~\text{~Hz}/f)$~\cite{Freeman2016,Connors2021Spectroscopy,Petersson2010,mi_landau-zener_2018}. 

Charge noise can affect spin qubits via the large magnetic field gradients that enable EDSR for LD qubits (\ref{Subsec:Principles:SOC}).  The stray electric fields of charge noise translate directly to a fluctuating magnetic field, and in turn to fluctuations in both Zeeman splitting and in the transverse driving field, and therefore impacting all relaxation parameters $T_1, T_2^*, T_2, T_{1\rho}$ and $T_{2,R}$. For instance, \onlinecite{borjans_single-spin_2019} and \onlinecite{hollmann_large_2020} show the $T_1$ dependence of a Si/SiGe spin qubit in a large gradient is weakly field-dependent at low magnetic fields, strongly suggesting $1/f$ or Johnson noise-limited relaxation at these low energies. 

Charge noise as translated to spin by gradients or by spin-orbit effects~\cite{Huang2014} may also be observed in multipulse-sequence noise spectroscopy (Sec.~\ref{SubSec:Filter_Function}).  \onlinecite{Nakajima2020Coherence} examined a GaAs device,  \onlinecite{Kawakami2016} an isotopically natural Si/SiGe dot; \onlinecite{Yoneda2018} a 800 ppm $^{29}$Si/SiGe dot, and \onlinecite{Struck_npjQuantInfo_2020} a 60~ppm $^{29}$Si/SiGe dot. In all of these cases, the large micromagnet-induced gradient results in both $T_2^*$ and $T_2$ being limited by $1/f$ charge noise.  In contrast, noise spectroscopy performed on natural and 800~ppm Si/SiGe dots with no micromagnet~\cite{Kerckhoff_quadrupolar_2020} show $T_2^*$ and $T_2$ limited by hyperfine effects, although in Si these still have $1/f$ character as discussed in \refsec{Subsec:Decoherence:HFT2}.  \onlinecite{Chan2018} use noise spectroscopy in a SiMOS quantum device and find a predominantly $1/f$-charge-noise limited spectrum; in this case charge noise couples to the spin due to intrinsic spin-orbit or spin-valley effects. \onlinecite{petit_spin_2018} observe Johnson noise-limited spin $T_1$ at Zeeman energies below the valley splitting in a SiMOS device without a micromagnet. Hole qubits also feature large spin-orbit fields and therefore have $T_2^*$ and $T_2$ times limited by charge noise~\cite{hendrickx_single-hole_2020,Maurand_holes_2016}.

EO qubits in Si/SiGe may not suffer from SOI or field gradient effects, but are still susceptible to charge noise, since they utilize an exchange coupling $J$ that is a sensitive function of the wavefunction overlap between two spins.  Although fluctuations in the confinement potential may come from multiple sources, we may refer to it as though it arises from fluctuations in gate voltages $V_k$.  The exchange noise may therefore be written as $\delta J = \sum_{k} (\partial J / \partial V_k ) \delta V_k$, and hence for $1/f$ charge noise has a noise spectrum
from the noisy voltages,
\begin{equation}
	S_J(f) = \sum_{k} \left|\frac{\partial J}{\partial V_k}\right|^2 S_{V_k}(f).
\end{equation}
The partial derivatives ${\partial J}/{\partial V_k}$ quantify the \textit{sensitivity} to charge noise~\cite{Hu2006} and may be estimated through the Fermi-Hubbard \textit{ansatz}~\cite{Culcer2013}, Heitler-London/Hund-Milliken estimates~\cite{Culcer2009b}, or FCI calculations (Sec.~\ref{Subsec:2spins:FCI}).  They may also be measured to make a map of sensitivity to charge noise in bias space~\cite{Reed2016,Dial2013,Martins2016}, enabling an empirical search for operating regions of low charge-noise sensitivity (called ``sweet spots'').

Approximately, the simplest Fermi-Hubbard model \textit{ansatz} asserts that gates directly above the QDs impact the chemical potential $\mu_j$ of dot $j$ via a constant factor known as the lever arm $\alpha_V$, hence $\partial \mu_j/\partial V_k = e\alpha_V\delta_{jk}$. The dependence of tunnel couplings on gate voltages is more complex, but is typically assumed to be an exponential function of some linear sum of voltages, in which case $\partial{t_c}/\partial{V_k}\propto t_c$.  Under this model, one finds that in a DQD in the weak exchange limit, sensitivity to charge noise is maximized at high detuning and minimized at $\epsilon$ = 0; the latter condition means that the chemical potential of two dots are held equal, leaving only weaker tunnel-coupling noise~\cite{Taylor2007,Reed2016,Martins2016,Bertrand2015}.  We caution however that at high exchange and for simultaneous exchange across more than two dots, simple Fermi-Hubbard models are inaccurate at estimating charge noise sensitivity, as dot electrons merge into a regime in which exchange may be dominated by multi-dot orbital energies not parameterized by these models~\cite{pan_resonant_2020}; see Sec.~\ref{Subsec:2spins:FCI}.  

Recently, charge noise spectral densities in Si ST$_0$ qubits have confirmed a nearly $1/f$ spectrum over as many as 13 decades of frequency in both Si/SiGe~\cite{Connors2021Spectroscopy} and SiMOS~\cite{Jock2018}.  Moreover, temperature and fabrication dependencies of the $1/f$ noise amplitude point to fluctuations in materials or interfaces in the gate-stack, as opposed to noise emanating from the bulk of the semiconductor or instrumentation.  

In aggregate, the last 20 years of spin qubit research have indicated that, while material choices and judicious engineering of charge noise sensitivity may improve charge-noise-induced decoherence, the underlying sources of $1/f$ charge noise are unlikely to be completely removed from semiconductor devices (unlike hyperfine noise, which may be eliminated with sufficient isotopic enrichment).  The ease of control which comes from micromagnet-induced EDSR or RX qubits comes at the cost of persistent sensitivity to ever-present charge noise.  When only SOC is at play, as in hole qubits and high-field SiMOS systems, relaxation and decoherence due to charge noise may be lower, and it may be lowest in the nearly gradient-free and low spin-orbit environment of low-field ST$_0$ or EO qubits, but it is still activated during exchange pulsing and therefore provides some limit to control fidelity.  The balance of the speed and convenience of qubit control against sensitivity to charge noise remains a key design space for semiconductor spin qubits across multiple materials and modalities.

\section{Hybrid systems} \label{Sec:Hybrid}
The short-ranged nature of exchange coupling (see Sec.~\ref{Subsec:2spins:Exchange}) is most efficiently applied to implement two-qubit gates between nearest neighbor spin qubits. However, it has been shown experimentally that fully-connected quantum information processors can operate with higher fidelities as compared with systems which only provide nearest-neighbor coupling \cite{Linke2017}. Beyond the advantage of high-connectivity for quantum computing, the coupling of stationary qubits to mobile photonic qubits could form the basis of widespread quantum networks \cite{Kimble2008}. Some approaches for achieving long-range coupling of spin qubits are briefly outlined in Sec.~\ref{Subsec:2spins:LR}. This section is focused on one particularly promising approach, namely the development of hybrid devices consisting of semiconductor QDs embedded in a microwave cavity, to achieve long-range coupling of spin qubits and high-fidelity readout.

\subsection{Overview of superconducting circuit QED} \label{Subsec:Hybrid:SCQED}

Hybrid quantum systems consisting of QDs embedded in microwave cavities are an outgrowth of the field of circuit quantum electrodynamics (cQED). The main physical concepts associated with cQED were first explored by atomic physicists in the field of cavity quantum electrodynamics (cavity QED) \cite{Haroche2006,Miller2005,Walther2006,Mabuchi2002}. In cavity QED, a two-level atom with transition frequency $\omega_a$ is coupled to an optical cavity with a resonance frequency $\omega_c$. The photonic mode and atom interact through the electric dipole interaction $H_{\rm int}$ = $-e\vec{E}\cdot\vec{d}$, where $\vec{E}$ is the cavity electric field at the position of the atom and $\vec{d}$ is the dipole moment associated with the atomic transition. 

The Jaynes-Cummings Hamiltonian
$H_{\rm JC}$ = $\hbar \omega_c a^{\dagger}a + \hbar \omega_a \sigma^z/2 + \hbar g (a \sigma^+ + a^{\dagger}\sigma^-)$ 
describes the system dynamics
in cases where the rotating wave approximation is appropriate
\cite{Jaynes1963}. 
Here $a^{\dagger}$($a$) are the photon creation(annihilation) operators, $\sigma^z$ describes the state of the atom, and $\sigma^+$($\sigma^-$) are atomic raising(lowering) operators. When $\omega_a$ = $\omega_c$, the atom and cavity can exchange an excitation at a rate set by the vacuum Rabi frequency $g$. In the energy domain, the light-atom coupling manifests itself as the vacuum Rabi splitting between energy eigenstates formed as coherent superpositions that are part atom and part photon. It is directly observable in the cavity transmission in the strong coupling regime, where $g$ exceeds the cavity decay rate $\kappa$ and the atomic dephasing rate $\gamma$.

In the early 2000s, significant efforts were made to demonstrate cavity-QED physics in solid state systems. Strong-coupling physics was observed in systems consisting of self-assembled QDs embedded in a distributed Bragg reflector cavity \cite{Yoshie2004}, self-assembled QDs embedded in a photonic crystal cavity \cite{Reithmaier2004}, and a superconducting Cooper pair box \cite{Wallraff2004} or flux \cite{Chiorescu2004} qubit embedded in a microwave cavity. These seminal experiments demonstrated that a superconducting artificial atom could be coherently coupled to a microwave frequency photon in the strong-coupling regime with an interaction precisely described by the Jaynes-Cummings Hamiltonian \cite{Blais2004,Blais2007}. For a review of cQED physics with superconducting qubits, see \cite{Blais2020a,Blais2021}.

The energy scales associated with gate-defined QDs (charge transitions in a DQD and the Zeeman energy of a single spin in a moderate field $B$ = 0.25 T) are nicely matched with the energy of microwave frequency photons $f$ = 5 -- 15 GHz. Rapid developments in the cQED architecture led to growing interest in QD cQED and a number of theoretical proposals for physical implementations \cite{Burkard2006,Childress2004,Hu2012,Jin2012,Kerman2013,Guilherme2014,Russ2015b,Russ2016,Benito2016,Benito2017,Benito2019a}. Beyond coupling to charge through the electric dipole interaction, semiconductor QDs allow for cavity coupling to electron spins, long-range spin-spin interactions, and possibly even nuclear spin state readout. The main modes of interaction are described in the next section.

\subsection{Coherent interactions in quantum dot circuit QED} \label{Subsec:Hybrid:Interactions}

In this section we review the theory of charge-photon coupling, spin-photon coupling, and cavity-mediated spin-spin interactions in hybrid quantum systems consisting of semiconductor DQDs embedded in microwave cavities. The experimental signatures of coherent interactions in each of these cases are also presented.

\subsubsection{Charge-photon coupling}

The physics of a single electron confined in a semiconductor DQD is described by a charge qubit Hamiltonian $H_0$ = $(\epsilon/2) \tau^z + t_c \tau^x$, where $\epsilon$ is the DQD level detuning, $t_c$ is the interdot tunnel coupling, and the matrices $\tau^x$ and $\tau^z$ are Pauli matrices (see Appendix \ref{app:gates}) operating on the charge state of the qubit, i.e. $\tau^z |L\rangle$ = $\tau^z|(1,0)\rangle$ = $|(1,0)\rangle$
and $\tau^z |R\rangle$ = $\tau^z|(0,1)\rangle$ = $-|0,1)\rangle$. The cavity electric field $E_{\rm cav}$ = $E_0(a+a^{\dagger})$ couples to the charge dipole moment of an electron confined in a DQD through the interaction term $H_{\rm int} = \hbar g_c(a + a^{\dagger})\tau^z$. The charge-photon interaction strength $g_c = eE_0d$, where $d$ is the interdot spacing, and $E_0$ is the amplitude of the vacuum electric field fluctuations in the cavity, which characterizes the strength of the charge-photon interaction \cite{Burkard2020}. Diagonalizing $H_0$, transforming $H_{\rm int}$ into the eigenbasis of $H_0$, moving into a frame rotating at probe frequency $f_p$, and making the rotating-wave approximation yields the transverse coupling Hamiltonian $H = \frac{1}{2}\Omega\tau^z + \tilde{g}_c(a\tau^+ + a^{\dagger}\tau^-) + \Delta a^{\dagger}a$. Here, $\Omega = \sqrt{\epsilon^2 + 4 t_c^2}$ is the charge qubit transition energy, $\tilde{g}_c = 
g_0 t_c/\Omega$ is the coupling strength, and $\Delta$ = $2 \pi (f_c - f_p)$ is the detuning between the cavity resonance frequency $f_c=\omega_c/2\pi$ and the probe frequency. Note that from a practical perspective, $\Omega$ is first-order insensitive to charge noise at $\epsilon$ = 0. Conveniently, the coupling strength $\tilde{g}_c$ is maximal at the interdot charge transition ($\epsilon$ = 0) as well.  
Input-output theory \cite{Collett1984,Benito2017,Burkard2020} is used to calculate the cavity response. In the steady-state limit $\dot{a}$ = $\dot{\tau}_-$ = 0, the transmission amplitude through the cavity is 
\begin{align}
\label{eq:cavity_transmission}
A = \frac{-i\sqrt{\kappa_1 \kappa_2}}{\Delta - \frac{i \kappa}{2}+\tilde{g}_c\chi} = |A|e^{i\delta\phi},
\end{align}
with the electric susceptibility
$\chi=\tilde{g}_c/(-\Omega + 2\pi f_p+i\gamma/2)$ and the photon loss rates $\kappa_{1,2}$ at the cavity ports 1 and 2, where $\kappa=\kappa_1+\kappa_2+\kappa_i$
and $\kappa_i$ denotes the intrinsic photon loss rate. In terms of the resonator frequency and quality factor $Q_c$, $\kappa/(2\pi)=f_c/Q_c$.

\begin{figure}
    \centering
    \includegraphics[width=0.95\columnwidth]{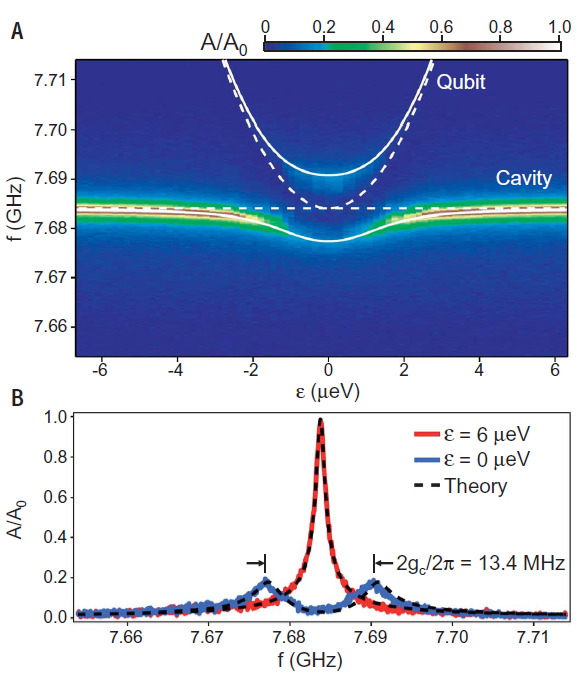}
    \caption{Cavity transmission for a charge qubit coupled to a superconducting microwave resonator, from~\cite{mi_strong_2017}. (a) As the double-dot detuning is swept across the tunneling transition, the charge qubit frequency comes into resonance with the cavity frequency. As a result of the strong coupling between the cavity and charge qubit, the system hybridizes, and two distinct transmission peaks separated by the vacuum Rabi splitting are observed. The eigenenergies of the uncoupled system are shown in dashed lines, and the eigenergies of the coupled system are shown in solid lines. (b) Cavity transmission at two different values of detuning, with theoretical predictions overlaid.}
    \label{fig:strongcoupling}
\end{figure}

Strong coupling between the qubit and cavity will occur when $\tilde{g}_c^2>(\gamma_c^2+(\kappa/2)^2)/2$, where $\gamma_c/(2 \pi)$ is the charge qubit decoherence rate. When $\gamma$ is dominated by inhomogeneous dephasing of the qubit, $\gamma/(2 \pi)=1/T_2^*$. In the strong coupling regime, the cavity resonance splits into two separate vacuum Rabi peaks separated by $2\tilde{g}_c$, as shown in Fig.~\ref{fig:strongcoupling}.
It is challenging to reach the strong coupling regime because in semiconductor systems the qubit decoherence rate $\gamma$ can be sizeable, e.g., several tens of MHz for GaAs. This can be overcome by increasing $\tilde{g}_c$ or by suppressing $\gamma$.  Both strategies have successfully been implemented to reach the strong coupling regime: an enhancement of $\tilde{g}_c \propto E_0 \propto\sqrt{Z}$ to a GaAs DQD was achieved by increasing the impedance $Z=\sqrt{L/C}$ of the resonator~\cite{Stockklauser2017}, while a reduction of $\gamma$ was possible using a Si DQD \cite{mi_strong_2017}.

\subsubsection{Spin-photon coupling}

For spin qubits, one is ultimately interested in coupling the spin to the cavity mode. While for optical cavities, SOC in the valence band of III-V semiconductors can enable spin-photon coupling \cite{Imamoglu1999}, the coupling to microwave photons requires mechanisms acting entirely in the conduction (valence) band for electrons (holes). Spin-charge hybridization using SOC or magnetic field gradients allows for a sizeable coupling between the electron spin and the cavity electric field~\cite{Burkard2020}. In particular, the coupling of a flopping-mode spin qubit via spin-charge hybridization using a magnetic field gradient $\Delta B^x=B_1^x-B_2^x$ in a Si DQD can be described by the additional term $(\Delta B^x/2)\sigma^x\tau^z$ in the single-electron Hamiltonian $H_0$ \cite{Benito2017}.  The direction of this gradient field is perpendicular to the homogeneous magnetic field $B^z=(B_1^z+B_2^z)/2$ described by the Zeeman term $(B^z/2)\sigma^z$. 

Transforming $H_{\rm int}$ into the eigenbasis of $H_0$, one obtains a coupling of the form $H_{\rm int}=g_c(a+a^\dagger)\sum_{n,m=0}^3 d_{nm}|n\rangle\langle m|$, where the sum represents the electric dipole operator in the spin-charge-hybridized DQD eigenbasis $|n\rangle$.  For microwave transmission through the cavity one finds again Eq.~\eqref{eq:cavity_transmission} with the susceptibility $\chi=\sum_{n=0}^3\sum_{j=1}^{3-n}d_{n,n+j}\chi_{n,n+j}$ and $\chi_{ij}$ follows from the stationary limit of the quantum master equation.
The relevant low-energy eigenstates of $H_0$ are ${|0\rangle\approx |-,\downarrow\rangle}$ and ${|1\rangle\approx \cos(\Phi/2)|-,\uparrow\rangle
+\sin(\Phi/2)|+,\downarrow\rangle}$
with the spin-orbit mixing angle $\Phi=\arctan(\Delta B^x/(2t_c-B^z))$ (in the symmetric case where $\epsilon=0$) and hybridized orbital states $|\pm\rangle=(|(1,0)\rangle\pm|(0,1)\rangle)/\sqrt{2}$.
The dipole transition matrix element for the predominantly spin-like transition between these two states is $d_{01}\approx -\sin(\Phi/2)$, whereas charge-like transitions to the next higher state are less important but can lead to an asymmetry in the vacuum Rabi peaks.
The resulting spin-photon coupling in this 
simplest two-level description and within the rotating-wave approximation can be described with a Jaynes-Cummings model
\begin{align}
\label{eq:JC-spin}
H = \hbar\Omega_s\sigma^z+\hbar\omega_c a^\dagger a+g_s \sigma^x (a+a^\dagger),
\end{align}
where $\Omega_s$ is the spin qubit transition frequency, and the spin-photon coupling $g_s \approx g_c|d_{01}|  \approx g_c|\sin(\Phi/2)|$.
A strength of this architecture is the electrical tunability of the spin-charge admixture via the inter-dot tunnel coupling $t_c$.

\begin{figure}
    \centering
    \includegraphics[width=0.95\columnwidth]{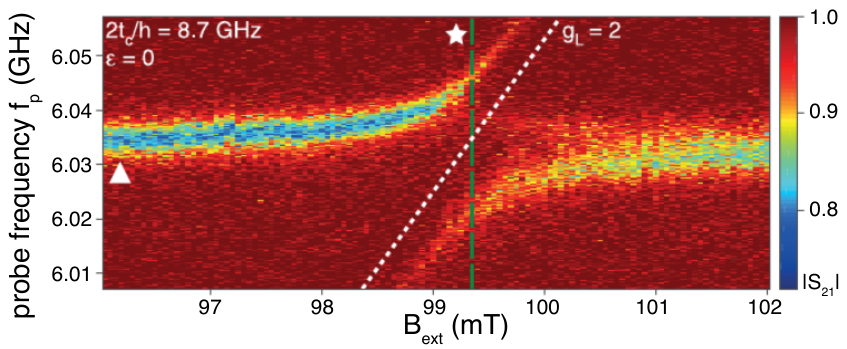}
    \caption{Cavity transmission for a single-spin qubit coupled to a superconducting microwave resonator, from~\cite{Samkharadze2018}. As the magnetic field is swept, the spin qubit comes into resonance with the cavity at about 6.03 GHz, and the qubit-cavity coupling splits the cavity resonance into two hybrid spin-photon modes. The characteristic vacuum Rabi splitting indicates the strong coupling regime.} 
    \label{fig:spinphoton}
\end{figure}

Strong spin-photon coupling will occur when $g_s > \gamma_s, \kappa$, where $\gamma_s$ is the spin deocherence rate. 
Remarkably, this condition is not identical with the strong coupling condition for charge, and in fact, the spin-photon system can be in the strong coupling regime while the charge-photon system is not. A key signature of strong coupling, split vacuum Rabi peaks, has been observed in microwave transmission through a superconducting Nb cavity with an embedded Si DQD \cite{mi_circuit_2017}. A similar experiment with NbTiN superconducting circuitry has also reached the strong coupling regime \cite{Samkharadze2018} [see Fig.~\ref{fig:spinphoton}]. The coupling of RX qubits to an electromagnetic cavity \cite{Russ2016} has been realized using a TQD in GaAs coupled to a NbTiN superconducting cavity \cite{Landig2018}.

\subsubsection{Cavity-mediated spin-spin interactions}

The coherent coupling Eq. \eqref{eq:JC-spin} of individual sub-micron scale  spin qubits to a single cavity mode extending over 100 $\mu$m or more lends itself to the pairwise coupling of spin qubits over distances much longer than their typical nearest-neighbor separation.  The exchange of virtual cavity photons in the dispersive limit gives rise to an effective coupling between spin qubits of the form 
\begin{align}\label{eq:cavity-spin-spin}
    H_I = \frac{g_s^2}{\Delta}\left( \sigma^+_{1}\sigma^-_{2}+\sigma^-_{1}\sigma^+_{2}\right),
\end{align}
with the detuning $\Delta=\Omega_s-\omega_c$ \cite{Benito2019a,Warren2019}.
The transverse (XY) coupling Eq. \eqref{eq:cavity-spin-spin} is known to generate 
the universal $\sqrt{\mathrm{iSWAP}}$ \cite{Imamoglu1999} and $\mathrm{iSWAP}$ gates \cite{Schuch2003}.  Cavity photon loss and qubit decoherence imply opposing requirements for the degree of spin-charge mixing, the optimum being defined by the ratio $\kappa/\gamma_c$ \cite{Benito2019b}. Fast and high-fidelity two-qubit gates in the presence of realistic charge noise have been supported by numerical calculations \cite{Warren2019}.

\begin{figure}
    \centering    \includegraphics[width=0.95\columnwidth]{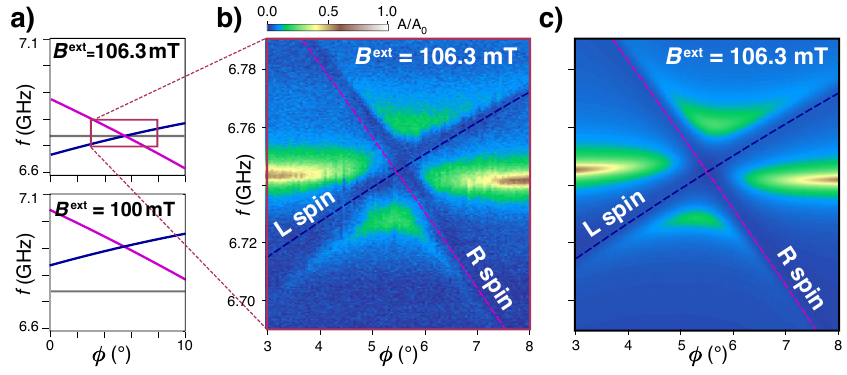}
    \caption{Resonant cavity-mediated spin-spin interactions, from~\cite{Borjans2020}. (a) Calculated spin resonance frequencies vs angle of the applied magnetic field, which changes the total magnetic field at the location of each qubit. At an external field of 106.3 mT and an angle of approximately 5.5$^\circ$, both qubits come into resonance with the cavity. (b) Measured cavity transmission vs angle, showing an avoided crossing between both qubits and the resonator at the expected angle.}
    \label{fig:spinspin}
\end{figure}

One challenge with experiments demonstrating cavity-mediated coupling between single spins involves bringing multiple spin qubits into resonance with each other and a cavity. For example, differences in qubit-micromagnet positioning of around ten nanometers, which are within typical fabrication tolerances, can easily detune two single-spin qubits from each other, even at the same value of the external magnetic field. 
To surmount this challenge, the micromagnets on different qubits can be fabricated at an angle with respect to each other~\cite{Astner_PRL,Borjans2020,harvey2021circuit}. By adjusting the angle and magnitude of the external magnetic field, the two spins can be brought into resonance with each other and the cavity. When two qubits, instead of just one, are tuned into resonance with the same cavity, an enhancement of the spin-photon coupling rate $g_s$ is observed (Fig.~\ref{fig:spinspin}), as reported for single electrons in Si DQDs coupled to both Nb~\cite{Borjans2020} and NbTiN superconducting resonators~\cite{harvey2021circuit}. Moreover, when both spins are detuned from the cavity but in resonance with each other, an avoided crossing between spins due to the cavity-mediated dispersive coupling can be observed~\cite{harvey2021circuit}. Microwave-photon-mediated coupling between charge qubits has also been demonstrated~\cite{vanWoerkom2018microwave}.

\subsection{Applications for readout} \label{Subsec:Hybrid:Readout}

Cavity coupled QDs can be readily probed be measuring the transmission through, or reflection from, the microwave cavity. Measurements are generally performed in the dispersive regime, where the detuning between the QD transition frequency and cavity photon is greater than the cavity linewidth, $|\omega_a - \omega_c| \gg \kappa$, where $\omega_a$ is the (charge or spin) qubit frequency. In this dispersive (i.e.\ off-resonant) regime, the Jaynes-Cummings Hamiltonian simplifies to the form 
$H \approx \hbar \left(\omega_c + \chi_d \sigma^z  \right)(a^{\dagger}a + 1/2) + \hbar \omega_a \sigma^z/2$
with the dispersive shift $\chi_d=\frac{g^2}{\omega_a - \omega_c}$. 
The first term in the Hamiltonian gives insight into the nature of the measurement. The bare cavity photon energy (energy in the absence of a qubit) $\hbar \omega_c$ is shifted by an amount $\chi_d$ that depends on the state of the qubit.

Detection of charge states using microwave photons has been demonstrated in GaAs, InAs, carbon nanotube and Si/SiGe DQDs
\cite{Frey2012,Petersson2012,Viennot2016,mi_strong_2017}. The dispersive shift can be detected by probing the cavity transmission amplitude $|A|$ or phase shift $\delta\phi$. Measurements of $\delta \phi$ as a function of the DQD gate voltages can be used to map out DQD charge stability diagrams and quantitatively extract the interdot tunnel coupling $t_c$ and the charge-qubit coupling rate $g_c$. High speed and high sensitivity real-time charge detection have benefited from the adoption of nearly quantum limited superconducting parametric amplifiers. Stehlik \textit{et al.} demonstrated ``video mode'' acquisition of DQD charge stability diagrams in 20 ms \cite{Stehlik2015}. It is also possible to use the cavity response at a single dot-to-lead charge transition for charge state readout with a very large signal-to-noise ratio of $>$450 and an integration time around 1 $\mu$s~\cite{Borjans_sensing_2021}

Cavity readout of spin states can be achieved using at least two different approaches. In the first approach, the Pauli exclusion principle is used to distinguish spin singlet and spin triplet states in a two-electron DQD. Pauli blocking is evident in the magnetic field dependence of the cavity response. With $B$ = 0, the spin singlet state is the ground state and tunneling from S(1,1) to S(2,0) leads to a large cavity response. In contrast, when $g \mu_B B > t_c$, the polarized spin triplet state T$_+$ (or T$_-$, depending on the sign of the $g$-factor) becomes the ground state near the charge transition. Due to Pauli blockade, charge tunneling from T$_+$(1,1) to T$_+$(2,0) is forbidden, and there is no cavity response. The magnetic field dependence of the interdot charge transition signal can thereby be used to determine the charge parity of a DQD interdot charge transition \cite{Schroer2012}. Control of two-electron spin states at a large DQD detuning, followed by cavity readout at zero detuning, has been used to distinguish singlet and triplet spin states in an InAs DQD \cite{Petersson2012} and later in a cavity-coupled Si/SiGe DQD \cite{zheng_rapid_2019}. Using an ancilla dot capacitively coupled to a singlet-triplet qubit has led to singlet-triplet spin state readout with a fidelity of 99.2\% \cite{Borjans_sensing_2021}.

\begin{figure}
    \centering
    \includegraphics[width=0.95\columnwidth]{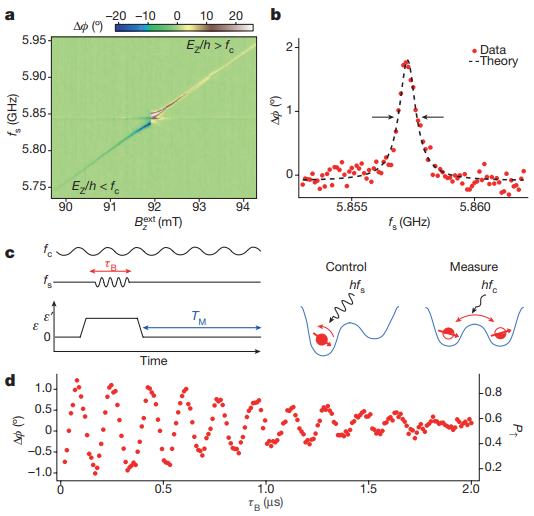}
    \caption{Cavity-mediated single-spin readout from~\cite{Mi2018}. (a) Cavity response vs magnetic field, showing the single-spin resonance frequency. (b) Electron-spin-resonance line at $B=$92.18 mT. (c) Pulse sequence to detect single-spin Rabi oscillations via the cavity resonance. (d) Rabi oscillations measured through the cavity dispersive shift. }
    \label{fig:cavity_readout}
\end{figure}

A second approach for spin state readout in the one-electron regime of a DQD utilizes spin-photon coupling and the dispersive interaction. For a spin interacting with a cavity photon, the dispersive shift is $\chi_d\sigma^z$. Dispersive readout of a single electron spin state using cQED was first demonstrated using a cavity-coupled Si/SiGe DQD \cite{Mi2018}. Figure~\ref{fig:cavity_readout} shows the cavity phase response as a function of magnetic field and microwave probe frequency. The spin-photon detuning dependence is clearly evident in the data, with the phase shift changing sign as the spin is taken through resonance with the cavity. The magnitude of the dispersive shift also decreases with detuning, as expected from the $1/\Delta$ dependence in the dispersive form of the Jaynes-Cummings Hamiltonian. Rabi oscillations of a single spin have been measured by probing the cavity with a microwave tone after the spin was driven with a microwave field. The signal-to-noise ratio of the dispersive readout of a single spin in a DQD coupled to a microwave cavity has been analyzed and optimized in \cite{DAnjou2019}.

\subsection{New avenues of research in cQED} \label{Subsec:Hybrid:Future}
\begin{figure}
    \centering
    \includegraphics[width=0.95\columnwidth]{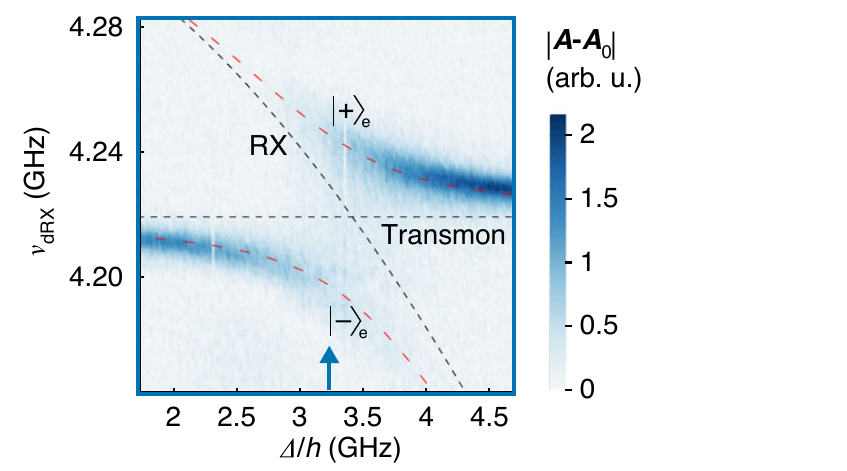}
    \caption{Cavity-mediated coupling between a triplet-dot resonant exchange qubit and a transmon superconducting qubit, from~\cite{landig2020virtual}. Here, a superconducting cavity is driven near its resonance frequency of about 5.6 GHz. The $y$ axis indicates the drive frequency of the resonance exchange qubit, and the $x$ axis corresponds to changes in the electrochemical potential of the middle dot, which changes the overall energy of the spin qubit. The color scale indicates the transmission through the cavity. The gray dashed lines indicate the eigenenergies of the system in the absence of coupling, and the red dashed lines indicate the eigenenergies of the system including the coupling}
    \label{fig:spintransmon}
\end{figure}
Cavity-coupled QDs have enormous potential for applications in quantum information science. In the span of just several years, coherent charge-photon and spin-photon interactions have been demonstrated, as well as evidence for long-range cavity mediated spin-spin interactions. Future research is likely to extend these results to spin-spin coupling in the dispersive limit, a time-domain demonstration of a cavity mediated two-qubit gate, and extensions to larger quantum networks. There is also the potential for cQED to probe the nuclear spin degree of freedom in dot-donor systems \cite{Mielke2021}.

Within the field of quantum information processing, hybrid systems employing cQED approaches may enable new forms of quantum information processors which could benefit from the advantages of different platforms. For example, recent work illustrates the feasibility of coupling spin qubits to superconducting qubits through microwave resonators, see Fig.~\ref{fig:spintransmon}~\cite{scarlino2019coherent,landig2020virtual}. A challenge for future hybrid systems such as these will be to ensure strong enough coupling rates to simultaneously capitalize on the benefits of the separate plaftorms while not introducing excess decoehrence. 

Hybrid quantum systems have had a major impact on the field of mesoscopic physics as well. For example, voltage biased DQDs have been shown to emit microwave photons \cite{Liu2014,Stockklauser_2015,Bruhat_2016}, and even enable the creation of a maser \cite{Liu2015}. Given the sensitivity with which charge state physics can be probed, signatures of electron-phonon coupling in suspended nanowire DQDs have been observed \cite{Hartke2018}. Kondo physics has been explored \cite{Desjardins2017} and there is potential to probe Majorana modes as well \cite{Dartiailh2017}. In Si/SiGe DQDs, cQED has proven to be very useful as a quantitative probe of valley splitting and intervalley coupling \cite{Burkard2016,mi_high-resolution_2017,borjans_probing_2021}. Looking ahead to the future, microwave spectroscopy may provide insight into a broader class of materials systems \cite{Curson_MIMS,Shim_valley_2019,Lee_phonons_arXiv} and qubit functionalities \cite{Larsen_gatemon_2015,deLange_gatemon_2015,Woerkom_2017}.

\section{Outlook} \label{Sec:Outlook}
Semiconductor spin qubits are uniquely positioned to benefit from the technologies that are available for classical semiconductor-based information processing devices. The most important observation is that no single roadblock stands in the way of reaching the types of yields now driving the industry of Si CMOS for classical information processing. Fidelities for both single-qubit and multi-qubit gates appear to be limited by processes with clear routes for reduction, such as judicious bias regimes for reducing sensitivity to charge noise and isotopic enhancement for reducing magnetic noise \cite{noiri_2Q_2021,xue_2Q_2021,mills_2Q_2021}. Readout visibilities and speeds have also improved substantially in the past few years \cite{mills_2Q_2021,madzik_arxiv,Borjans_sensing_2021}.

A clear advantage of semiconductor spin qubits is therefore the potential for their massive scaling and miniaturization. Due to their small size, semiconductor spin qubits have the distinction of having the most stringent demands on fabrication in comparison to superconducting qubits, trapped ion qubits, and photonic qubits.
As a result, the route to large arrays of spin qubits has been slower, as numerous problems have had to be overcome to more reliably yield qubit arrays \cite{Zajac2016,Ha_archive}.  In the past decade the progress not just for the most heroic of devices but also for the yielding of routine device arrays from many groups in many countries and using many different control strategies have indicated a clear positive trend. The number of demonstrations of coherent operation published worldwide has grown accordingly.

It seems too early to say which type of spin qubit (LD, ST$_0$, EO, etc.), spin qubit carrier (electron, hole, or nucleus), and material (Si, Ge, etc.), or which combination thereof, will end up being optimal for realizing a functioning large-scale quantum processor. While LD qubits offer efficient use of the available resources and high robustness against charge noise, ST$_0$ and EO qubits allow for baseband electrical control, in the case of EO qubits without the need for SOC or on-chip micromagnets. While Si offers extremely high coherence, Ge allows for spin-orbit engineering of electrically controlled qubit operations. The extremely long coherence time of nuclear spins can be contrasted with the readily available fast exchange coupling between electronic spins.


Even if fault-tolerant quantum computers are many years away, qubits serve as our most sensitive solid-state electrometers and magnetometers, and in the case of semiconductor spin qubits, they serve this role within the workhorse materials underpinning the most pervasive information processing technology in modern society.  Advances in the understanding of semiconductor device physics are at least one guaranteed outcome in the pursuit of future scalable quantum computers based on semiconductor spin qubits.

Until fault-tolerant quantum computation can be realized, computational demonstrations using noisy intermediate-scale quantum (NISQ) devices provide valuable proofs of principle \cite{Preskill2018}. Examples that can be tackled with noisy devices are simulations of condensed matter and quantum chemistry as well as optimization problems. Analog quantum simulations of condensed matter systems with three to four spin qubits have been demonstrated~\cite{Hensgens2017,Dehollain2020,vanDiepen2021}. On the level of two qubits, the variational quantum eigensolver method \cite{xue_2Q_2021} as well as small quantum algorithms~\cite{noiri_2Q_2021} have been implemented.

Ultimately, the utility of spin qubits, and in fact all other quantum computing platforms, lies in their ability to reach quantum fault-tolerance, since practical applications depend on a scale demanding lower-error operation than will be possible without quantum error correction.  It is not clear when we can declare any qubit is good enough for fault-tolerance, since many estimates of fault-tolerant thresholds, for example for the popular surface code~\cite{Fowler2012},
 make geometric layout and error-correlation assumptions which are certainly not consistent with semiconductor spin qubits as presently operated.
 Nearer-term approaches to error corrected logical qubits may nonetheless be pursued, even in strictly one-dimensional qubit arrays using the methodologies and geometries presently under study~\cite{Jones2016}, from which we may anticipate significant discovery, not only about the pathways to scalable quantum computers, but also to serendipitous advances in the understanding of the physics of solid-state devices.

\appendix \section{Spin Rotation Gates} 
\label{app:gates}

In this article we have discussed multiple encodings of spin-qubits in terms of spin-operators $\vec{S}$, which are typically represented as Pauli operators.  However, we reserve the notation of Pauli operators represented as Pauli matrices, $$\sigma^x=\begin{pmatrix} 0 & 1 \\ 1 & 0 \end{pmatrix},\quad
  \sigma^y=\begin{pmatrix} 0 & -i \\ i & 0 \end{pmatrix},\quad
  \sigma^z=\begin{pmatrix} 1 & 0 \\ 0 & -1 \end{pmatrix},$$ for the two encoded states of qubits $\ket{0}$ and $\ket{1}$.  Canonical quantum computing is accomplished by application of unitary qubit rotations generated by Pauli operators, i.e. single-qubit operations
\begin{equation}
	R_{\vec{n}}(\theta) = e^{-i\vec{n}\cdot\boldsymbol{\sigma}\theta/2}
\end{equation}
and two-qubit operations such as the controlled-$Z$ operation
\begin{equation}
	U_{\rm CZ} = e^{-i(\pi/4)(1-\sigma^z)\otimes(1-\sigma^z)}.
\end{equation}
Two-qubit gates for semiconductor spin qubits are generally drawn from two families, the controlled-rotations such as controlled-\textsc{NOT} and controlled-$Z$ which result from single-qubit rotations of the two-qubit unitary $\exp(-i\theta\sigma^z_1\sigma^z_2)$, with $\theta=\pi/4$ for a fully entangling gate; and fractional swaps, which result from single-qubit rotations of the two-qubit unitary $\exp(-i\theta\boldsymbol\sigma_1\cdot\boldsymbol\sigma_2)$, with a \textsc{swap} at $\theta=\pi/4$ and a fully entangling $\sqrt{\textsc{swap}}$ at $\theta=\pi/8$.

A $\pi$-pulse with unitary $R_{\vec{n}}(\pi)$ applies $-i\vec{n}\cdot\boldsymbol\sigma$, so if $\vec{n}$ is along $x,y$, or $z$ this is a Pauli operator with an overall phase.  
A Pauli-operator or $\pi$-pulse applies $\pm 1$ to the two eigenstates of a qubit in the associated basis.  
In the two-spin singlet-triplet basis, a $\pi$ pulse of the exchange operator, $U=\exp{-i\pi\vec{S}_1\cdot\vec{S}_2}$ applies a spin-swap, which from the antisymmetry of the spin-pair for the singlet and symmetry for the triplet states, applies a $-1$ phase to singlet and so is analogous to the Pauli operations for singlet-triplet and exchange-only systems.  
Exchange occurring for arbitrary duration generates a superposition of swapping and not swapping spins, so that  
\begin{equation}
\exp{-i\theta\vec{S}_1\cdot\vec{S}_2}\sim \exp{-i\sigma^z\theta/2}=R_{\vec{z}}(\theta).  
\end{equation}

Single qubits driven by RF signals (e.g. single-spin qubits controlled by ESR or EDSR, RX qubits, etc.) use a rotating frame for single-qubit control.  This means that the laboratory-frame Hamiltonian for qubit $j$ is
\begin{equation}
	H_j(t) = \frac{\hbar\omega_j}{2}\sigma^z 
		+ \hbar\Omega \cos(\omega t + \phi)\sigma^x, 
\end{equation}
where $\omega$ is the driving frequency, $\phi$ is the drive phase relative to a local oscillator, and $\Omega$ is proportional to the amplitude of the driving RF or microwave field.  In a rotating-frame analysis, we presume the driving frequency $\omega$ is close to the qubit resonant frequency $\omega_j$ and both of these are always much larger than the Rabi frequnecy $\Omega$.  Under these assumptions we transform $H_j$ and other operators to a frame rotating at the drive frequency $\omega$ and local oscillator phase for each qubit, 
\begin{multline}
	\tilde{H}_j(t) = e^{i\omega t\sigma^z/2}\biggl[H_j(t)-\frac{\hbar\omega}{2}\sigma^z \biggr]e^{-i\omega t\sigma^z/2}
\\
	=\frac{\hbar}{2}\biggl[(\omega_j-\omega)\sigma^z 
	+\Omega \{[1+\cos(2\omega t)][\sigma^x\cos\phi +\sigma^y\sin\phi]
\\
	          -\sin(2\omega t)[\sigma^x\sin\phi + \sigma^y\cos\phi]\}
   \biggr]
\end{multline}
The terms oscillating at frequency $2\omega$ with amplitude $\Omega$, when $\Omega \ll \omega$, are generally negligible; the lowest-order effect of these terms is the Bloch-Siegert shift\cite{Abragam1961} which amounts to a slight detuning of the resonance of order $(\Omega/\omega)^2$.  As such, these terms are generally dropped, resulting in the nominally time-independent rotating-frame Hamiltonian
\begin{equation}
	\tilde{H}_j=\frac{\hbar}{2}[\Omega(\sigma^x\cos\phi +\sigma^y\sin\phi)+\Delta\omega_j \sigma^z]
\end{equation}
for which $U=\exp(-i\tilde{H}_jt/\hbar)$ enables any single qubit rotation $R_{\vec{n}}(\theta)$ via control of the amplitude $\Omega$, phase $\phi$, and detuning $\Delta\omega$ of the drive frequency.  Since phase $\phi$ is relative to a local oscillator, a $z$-axis rotation is generally accomplished by a frame-shift, in which the phase of the local oscillator is updated without touching the qubit.

\begin{acknowledgements}
We thank John B. Carpenter of HRL Laboratories, LLC for assistance generating the figures in this manuscript. GB and JRP acknowledge the support of Army Research Office Grant No. W911NF-15-1-0149. GB also acknowledges funding from the European Union under Grant Agreement No. 951852 (Quantum Technology Flagship / QLSI) and German Research Foundation (Deutsche Forschungsgemeinschaft, DFG) under project number 450396347. JMN acknowledges support of Army Research Office Grant Nos. W911NF-16-1-0260 and W911NF-19-1-0167 and Office of Office of Naval Research Grant No. N00014-20-1-2424. The views and conclusions contained in this document are those of the authors and should not be interpreted as representing the official policies, either expressed or implied, of the Army Research Office or the U.S. Government. The U.S. Government is authorized to reproduce and distribute reprints for Government purposes notwithstanding any copyright notation herein.
\end{acknowledgements}

\bibliographystyle{apsrmp4-2}
\bibliography{RMP_master2_bib_v4}

\end{document}